\documentclass[11pt]{article}
\pdfoutput=1
\usepackage{jcapmod,parskip}
\usepackage{enumerate}
\usepackage{empheq}
\usepackage{booktabs}
\usepackage[english]{babel}
\usepackage[utf8]{inputenc}
\usepackage{amsmath,amssymb,amsbsy,amstext, amsthm, simplewick}
\usepackage{textgreek}
\usepackage{graphicx}
\usepackage{amsfonts}
\usepackage{upgreek}
 \usepackage{exscale,relsize}
 \usepackage[makeroom]{cancel}
\usepackage{float}
\usepackage{framed}
\usepackage{array}
\allowdisplaybreaks[1]
\usepackage{bm}
\usepackage{color}
\newcommand{\us}{\textcolor{black}}
\newcommand{\be}{\begin{equation}}
\newcommand{\ee}{\end{equation}}

\newcommand{\de}{\partial}
\newcommand{\nn}{\nonumber}
\newcommand{\ra}{\rangle}
\newcommand{\la}{\langle}
\makeatletter    
\begin{document}
\begin{titlepage}
\setcounter{page}{1} \baselineskip=15.5pt \thispagestyle{empty}
\bigskip\
\vspace{1cm}
\begin{center}
{\fontsize{20}{28}\selectfont  \sffamily \bfseries From matter to galaxies: General relativistic bias for the one-loop bispectrum.
}
\end{center}
\vspace{0.5cm}
\begin{center}
{\fontsize{13}{30}\selectfont Juan Calles, ${}^{\rm a}$ Lina Castiblanco,${}^{\rm a}$ Jorge Nore\~{n}a,${}^{\rm a}$ and Cl\'ement Stahl${}^{\rm a, b}$
}
\end{center}
\begin{center}
\vskip 8pt
\textsl{${}^{\rm a}\;$Instituto de F\'{\i}sica, Pontificia Universidad Cat\'{o}lica de Valpara\'{\i}so, Casilla 4950, Valpara\'{\i}so, Chile}\\
\textsl{${}^{\rm b}\;$Laboratoire Astroparticule et Cosmologie, Université Denis Diderot Paris 7, 75013 Paris, France}\\
\end{center}
\vspace{1.2cm}
\hrule \vspace{0.3cm}
\noindent {\sffamily \bfseries Abstract} \\[0.1cm]
We write down the Lagrangian bias expansion in general relativity up to 4th order in terms of operators describing the curvature of an early-time hypersurface for comoving observers. They can be easily expanded in synchronous or comoving gauges. This is necessary for the computation of the one-loop halo bispectrum, where relativistic effects can be degenerate with a primordial non-Gaussian signal. Since the bispectrum couples scales, an accurate prediction of the squeezed limit behavior needs to be both non-linear and relativistic. We then evolve the Lagrangian bias operators in time in comoving gauge, obtaining non-local operators analogous to what is known in the Newtonian limit. Finally, we show how to renormalize the bias expansion at an arbitrary time and find that this is crucial in order to cancel  unphysical $1/k^2$ divergences in the large-scale power spectrum and bispectrum that could be mistaken for a contamination to the non-Gaussian signal.

\vskip 10pt
\hrule
\vspace{0.6cm}
\end{titlepage}
\tableofcontents
\newpage


\section{Introduction}
The study of the Large Scale Structure of the Universe (LSS) is thriving, with next generation experiments (Euclid, LSST, SKA, SPHEREx \cite{Amendola:2016saw,Zhan:2017uwu,Bacon:2018dui,Dore:2014cca}) starting to collect data in the near future. Of particular relevance for our work, they are expected to be sensitive to a non-Gaussian signal of $f_{NL}=\mathcal{O}(1)$ \us{(where $f_{NL}$ is the amplitude of the primordial bispectrum divided by the amplitude power spectrum squared)}, opening the possibility of constraining fundamental physics with the LSS.
\us{More precisely, the observed three-point function in the squeezed limit is given by projection effects and can be thus trivially computed by a change of frame} \cite{Maldacena:2002vr, Creminelli:2004yq,Creminelli:2011rh,Creminelli:2013mca,Creminelli:2011sq}. Any deviation from this behavior would be a smoking gun for other degrees of freedom active during inflation\footnote{See \cite{Biagetti:2019bnp} for a review} such as several scalar fields \cite{Sugiyama:2011jt}, higher spins \cite{Arkani-Hamed:2015bza}, modified gravity \cite{Tahara:2018orv}, anisotropic inflation \cite{Emami:2015qjl} and the presence of an electromagnetic field~\cite{Chua:2018dqh}.

Very recently, we pointed out that the squeezed limit of the three-point correlation function may be contaminated by \emph{non-linear relativistic contributions} \cite{Castiblanco:2018qsd}. On large scales, the universe is linear, while Newtonian physics is a good approximation for the dynamics of small scales. But the bispectrum couples scales leading, in the squeezed limit, to a large non-linear relativistic signal \cite{Castiblanco:2018qsd}. This is particularly relevant since that limit is the most sensitive to the field spectrum during inflation. The main result of Ref.~\cite{Castiblanco:2018qsd} is the solution for the metric in the weak-field approximation.

In this work, we take a further step toward computing the observed bispectrum at one loop by considering  biased tracers (such as galaxies). Galaxy clustering is a complex non-linear problem that involves astrophysical processes that are not fully understood. A pragmatic  approach is to use an effective expansion: the small scale galaxy density field is smoothed out, in order to focus on the larger scales where the (unknown) physics is parametrized by (unknown) bias coefficients $b_{\mathcal{O}}$ that multiply gravitational operators $\mathcal{O}$ \cite{Desjacques:2016bnm}. Bias was historically developed in a Newtonian framework and was also generalized to GR~\cite{Yoo:2009au,Bonvin:2011bg,Challinor:2011bk,Bruni:2011ta,Baldauf:2011bh,Jeong:2011as,Yoo:2014vta}. Taking inspiration from \cite{Umeh:2019qyd} (see also \cite{Umeh:2019jqg}), in this paper we generalize their results to higher orders needed for the one-loop bispectrum. \us{We focus only on the issue of biasing up to fourth order in a relativistic context, which is necessary for the description of galaxy correlation functions. The correlation functions we compute and plot are only for illustrating the size and behavior of each term. In order to obtain the actual measurable quantity one should take into account the propagation of the photons from the source galaxies to the telescope in a perturbed universe. This has been done up to second order in \cite{Yoo:2014sfa, DiDio:2014lka, Clarkson:2018dwn} and recently up to third order for the redshift in \cite{DiDio:2018zmk}, see also \cite{DiDio:2020jvo} for an application. We leave the full fourth order calculation for future work.}

We structure our paper as follows: in section \ref{sec:DM}, we review our relativistic results on dark matter. In section \ref{sec:bias}, we describe the core of our work \us{and our first main result}: we write down a relativistic bias expansion to fourth order by using operators describing the curvature of the initial time hypersurface, and we evolve them in time using the continuity equation. \us{In section \ref{sec:rsl}, we review the computational tools required to obtain a bispectrum.} We then show in section \ref{sec:renorm} \us{our second main result:} how to extend the renormalization of the bias operators to the relativistic case. Finally, we plot the contributions of some of the operators for the one-loop power spectrum and bispectrum \us{in section \ref{sec:numResults}}, and conclude in section \ref{sec:ccl}. For clarity, we relegate some of the more technical calculations to the appendices.

\underline{\textbf{Notation}}
We use Greek letters (e.g.~$\mu, \nu$) for space-time indices that run from 0 to 3, and reserve latin letters (e.g.~$i,j$) for spatial indices that run from 1 to 3. Latin indices are written arbitrarily up or down as they differ only by powers of $a(\eta)$, which are always written down explicitly. We will indifferently write quantities in position space or in Fourier space. Our convention is  $$f(\bm{x}) = \int\frac{d^3 k}{(2\pi)^3}\, e^{i\bm{x}\cdot\bm{k}}f(\bm{k}) \equiv \int_{\bm{k}}e^{i\bm{k}\cdot\bm{x}}f(\bm{k})\,,$$
where we also introduced a short hand notation for integrals. An asterisk $*$ denotes a quantity evaluated at a very early time $\eta_\ast \rightarrow 0$.

\underline{\textbf{Approximations}}
\begin{itemize}
\item We take relativistic corrections to the Newtonian results to be small. This is quantified by taking all spatial scales to be small with respect to the Hubble radius $\epsilon \equiv H^2/k^2 \ll 1$, where $k$ is any of the Fourier modes involved. We work up to order $\epsilon$ and neglect terms of order $\epsilon^2$ and higher.

\item We will assume a matter-dominated Einstein-de Sitter universe throughout all its history in order to simplify the calculations. In this case, all quantities have a simple scaling with time that greatly helps with the bookkeeping.
While in Newtonian structure formation, the Einstein de Sitter approximation is accurate, the accuracy of this approximation has not been extensively studied in a relativistic setup. \\
\\
Relaxing this approximation involves solving our equations including a dark energy component (given for example by a cosmological constant), and accounting for the effects of the non-linear evolution of the plasma before matter domination, which leaves an imprint only on the initial conditions for our calculation since we start deep in the matter dominated era. \\
\\
The inclusion of dark energy is straightforward but tedious. 
Including the effects of radiation during the early evolution of the perturbations before matter domination can be done straightforwardly. As shown in our previous work \cite{Castiblanco:2018qsd}, the second order kernel for matter is completely fixed by these initial conditions while higher-order kernels are completely sourced by the subsequent gravitational evolution (to the order in $\epsilon$ to which we work). Thus, in order to include this, one only needs to replace our second order kernel for matter by the one computed by solving the second-order Boltzmann equations as done for example by \texttt{SONG} \cite{Pettinari:2013he}.

\item We neglect velocity bias. That is, we have assumed that the 4-velocity of galaxies equals that of the dark matter fluid. Adding the velocity bias would require separating the equations for the matter velocity and the galaxy velocity which would now include a new term to effectively account for this effect \cite{Desjacques:2016bnm}.

\item We assume that the primordial three-point function for the comoving gauge curvature perturbation $\zeta$ is exactly zero. If inflation is single-field slow-roll it will be given by Maldacena's result \cite{Maldacena:2002vr} which vanishes if we work to zeroth order in slow-roll. A non-zero primordial three-point function can be trivially included, similarly to the point above, by modifying the initial conditions, which only affects the second-order kernel for matter.

\item We neglect primordial tensor modes. They can again be straightforwardly included but would greatly complicate the algebra. We assume that transverse vector and tensor perturbations of the metric are only sourced by non-linear evolution.
\end{itemize}
\section{Dark matter perturbations self-gravitating in an expanding universe}
\label{sec:DM}
We now review our previous results for the dark matter density contrast. For more details on the physical setup, the reader can consult Ref.~\cite{Castiblanco:2018qsd}.  Our starting point is the perturbed FLRW metric
\begin{equation}
\label{eq:metric}
ds^2 = a(\eta)^2\left\{-(1+2\phi) d\eta^2+2 \omega_i dx^i d\eta + \left[(1-2\psi)\delta_{ij} + \gamma_{ij} \right]dx^i dx^j \right\}\,,
\end{equation}
where $a(\eta)$ is the background scale factor, $\eta$ is the conformal time, and $x^i$ are Cartesian comoving coordinates. The off-diagonal part of the metric is split into its transverse and longitudinal pieces $\omega_i = \de_i\omega + w_i$, with $\de_i w_i = 0$. The dark matter is taken to be a perfect irrotational fluid, with stress-energy
\begin{equation}
    T_{\mu\nu} = \bar{\rho}(1 + \delta) u_\mu u_\nu\,,
\end{equation}
where $\bar{\rho}(\eta)$ is the background density, $\delta(\eta,\textbf{x})$ is the dark matter density contrast, and $u_\mu = \de_\mu \varphi/\sqrt{X}$ is the matter 4-velocity, with $X=-\de_\mu \varphi \de^\mu\varphi$. Unless stated differently, we work in comoving gauge defined such that $\varphi = \eta$ and $\gamma_{ij}$ is transverse and traceless: $\partial^i\gamma_{ij}=\gamma_{i}^i=0$. One can show that the lapse can be set to $N = 1$ to all orders \cite{Yoo:2014vta}, though the shift is different from zero $N^i \neq 0$.\footnote{The lapse and the shift are defined, as usual, by writing the ADM decomposition of the metric
$$
ds^2 = a^2(\eta)\left[-N^2 d\eta^2 + h_{ij}(dx^i + N^i d\eta)(dx^j + N^j d\eta)\right]\,.
$$
}
The continuity and Euler equations that describe the evolution of the matter fluid follow from the conservation of this stress-energy tensor, and are
\begin{align}
& \nabla_{\mu} (\bar{\rho}(1+\delta) u^{\mu}) =0\,, \label{eq:NT1} \\
& u^{\mu}\nabla_{\mu} u^{\nu} =0\,. \label{eq:NT2}
\end{align}

Following the scheme described in section 3 of \cite{Castiblanco:2018qsd}, we adopt a \textit{weak-field approximation} which consists in taking metric fluctuations as small, but spatial derivatives large, which is a good approximation inside the Hubble radius. We then expand in the parameter $\epsilon \equiv H^2/\nabla^2 \ll 1$, which characterizes the smallness of the relativistic corrections. \us{Following~\cite{Castiblanco:2018qsd}, we take $\gamma_{ij}=\mathcal{O}(\epsilon^2)$, $\phi, \psi=\mathcal{O}(\epsilon)$, $\omega=\mathcal{O}(\epsilon)$, $w_i=\mathcal{O}(\epsilon^2)$.\footnote{It was shown in~\cite{Castiblanco:2018qsd} that the transverse vector and tensor modes are sourced at order $\epsilon^2$ if scalar perturbations are order $\epsilon$.}} We use the 4-velocity with an upper index that can be written as
\begin{equation}
u^{\mu}=\us{a^{-1}}\left(1, u^i\right)\,,
\end{equation}
which we split into a longitudinal piece $\theta \equiv \partial_i u^i$ and a transverse part $\de_i u^i_T = 0$. In the weak-field approximation, \us{we have $u^i=\mathcal{O}(\epsilon^{1/2})$, $u^i_T=\mathcal{O}(\epsilon^{3/2})$, $\delta=\mathcal{O}(1)$ and} equations \eqref{eq:NT1} and \eqref{eq:NT2} lead to the generalization of the continuity and Euler equation which we split into a dominant Newtonian part (written with a subscript N \us{and satisfying the Newtonian continuity and Euler equations}), and relativistic corrections (written with a subscript R) sourced by the Newtonian terms and satisfying:
\begin{align}
&\dot{\delta}_{R}+\theta_R= -\de_i(\delta_N u_R^i + \delta_R u_N^i) + \mathsf{S}_{\delta}[ \delta_N, \theta_N]\,, \label{eq:correc1} \\
 &\dot{\theta}_R+2 H \theta_R+ \frac{3}{2}H^2\delta_R= \de_j(u_N^i \de_j u_R^i + u_R^i \de_j u_N^i) + \mathsf{S}_{\theta}[\delta_N, \theta_N]\,. \label{eq:correc2}
\end{align}
An expression for the relativistic sources can be found in equations (C.4)-(C.5) of \cite{Castiblanco:2018qsd}.
We perform perturbation theory of these equations in the usual sense, \us{that is, expanding in powers of the linear matter density contrast evaluated at redshift zero $\delta_l$.} Thus, one can define the Newtonian perturbation kernels $F^N_n$ and $G^N_n$, obtained by solving the usual Newtonian continuity and Euler equations, along with their relativistic counterparts $F_n^R$, $G_n^R$, $\bm{G}^T_n$, and $F_n^\psi$ obtained by solving equations \eqref{eq:correc1} and \eqref{eq:correc2}:
\begin{multline}
    \delta(\eta,\bm{k}) = \sum_{n = 1}^\infty a^n\int_{\bm{k}_1...\bm{k}_n}\!\!\!\!\!\!\!\! (2\pi)^3\delta_D(\bm{k}-\bm{k}_{1...n})\Big[ F^N_n(\bm{k}_1,\dots,\bm{k}_n) \\ + a^2H^2 F_n^R(\bm{k}_1,\dots,\bm{k}_n)\Big]\delta_l(\bm{k}_1)\dots\delta_l(\bm{k}_n)\,,
    \label{eq:expansion1}
\end{multline}
\begin{multline}
    \theta(\eta,\bm{k}) = -H\sum_{n = 1}^\infty a^n\int_{\bm{k}_1...\bm{k}_n}\!\!\!\!\!\!\!\! (2\pi)^3\delta_D(\bm{k}-\bm{k}_{1...n})\Big[ G^N_n(\bm{k}_1,\dots,\bm{k}_n) \\ + a^2H^2 G_n^R(\bm{k}_1,\dots,\bm{k}_n)\Big]\delta_l(\bm{k}_1)\dots\delta_l(\bm{k}_n)\,,
    \label{eq:expansion2}
\end{multline}
\be
    \bm{u}_T(\eta,\bm{k}) = H^3 a^2\sum_{n = 1}^\infty a^n\int_{\bm{k}_1...\bm{k}_n}\!\!\!\!\!\!\!\! (2\pi)^3\delta_D(\bm{k}-\bm{k}_{1...n})  \bm{G}_n^T(\bm{k}_1,\dots,\bm{k}_n)\delta_l(\bm{k}_1)\dots\delta_l(\bm{k}_n)\,,
    \label{eq:expansion3}
\ee
\be
\psi(\eta,\bm{k}) = H^3 a^2\sum_{n = 1}^\infty a^n\int_{\bm{k}_1...\bm{k}_n}\!\!\!\!\!\!\!\! (2\pi)^3\delta_D(\bm{k}-\bm{k}_{1...n})  F_n^\psi(\bm{k}_1,\dots,\bm{k}_n)\delta_l(\bm{k}_1)\dots\delta_l(\bm{k}_n)\,,
\ee
\us{where $\delta_D$ refers to the Dirac-delta,} $\bm{k}_{1...n}\equiv \sum_{i=1}^n \bm{k}_i$. Under the weak field approximation, it is possible to obtain expressions for the relativistic kernels, which we reproduce in appendix \ref{sec:DMkern}.


\section{Relativistic Bias Expansion}
\label{sec:bias}

\subsection{Geometric approach to bias expansion}

Since halo formation is a local process, it should be described in the frame of reference of an observer moving with the halo's center of mass \cite{Baldauf:2011bh}. The bias expansion should only depend on the quantities that such an observer would measure such as the local curvature (corresponding to second derivatives of the gravitational potential in Newtonian physics).

\us{
 This description can be carried out by expanding around such an observer by working in Fermi coordinates \cite{Baldauf:2011bh, Pajer:2013ana}. The Conformal Fermi coordinates (CFC) \cite{Pajer:2013ana} separate dynamical effects from purely geometric (projection) effects by focusing on what a local observer measures. Fermi coordinates are defined around a geodesic, and all quantities are computed by Taylor expanding on the distance from this geodesic. This means that the effect of a slowly varying perturbation (with a weak dependence on this distance) on a small local patch can be easily included in this framework. Thus, they can be used to obtain for example the squeezed limit of the bispectrum. On the other hand, it is not practical for describing quickly varying (short-wavelength) perturbations, or large regions, such that other configurations of the bispectrum cannot be easily computed with this technique. Furthermore, since the calculations we perform are quite involved, they need to be compared with the existing literature. Most of that literature on explicit calculations of the CMB and LSS uses more traditional perturbation theory, and we thus choose to follow this path. Finally, let us clarify that we check that our results for the squeezed limit satisfy the LSS consistency relation (see Appendix B). The consistency relation is the manifestation of the fact that geometric effects are trivially computable through a coordinate transformation, which is equivalent to the CFC up to corrections quadratic in the distance to the geodesic, as we show in Appendix \ref{app:CFCconsistency}.}

A different approach was adopted recently in \cite{Umeh:2019jqg}, who write down the bias expansion in synchronous gauge, defined by the lapse and the shift being $N = 1$ and $N^i = 0$. This gauge is Lagrangian in the sense that the 4-velocity of fluid elements is equal to $u^\mu = \us{a^{-1}}(1,\bm{0})$, and is suitable for a Lagrangian bias expansion. They also pointed out (as remarked previously in \cite{Kehagias:2013rpa}) that the condition of making the bias expansion depend only on locally measurable quantities can be satisfied by requiring that there exist a local coordinate system where short scale physics is independent of the value and first gradient of a long wavelength gravitational potential. In synchronous gauge, the coordinate transformation that guarantees this takes a particularly simple form, given by dilations and special conformal transformations.

However, in the synchronous gauge the metric does not stay close to an unperturbed FLRW. This is because the spatial coordinate position of a fluid element is constant, and its physical displacement is contained in the metric, which always happens in a Lagrangian approach to perturbation theory, even in the Newtonian limit. Since the metric is greatly distorted with the gravitational evolution, this makes a weak field approximation impossible in this gauge.

It would be interesting to define a ``post-Lagrangian'' expansion around Newtonian Lagrangian perturbation theory, but we take a different approach. In order to describe biasing we use the comoving gauge, for which the comoving time of observers is equal to the coordinate time in matter domination, given that one can choose $N = 1$ (see \cite{Yoo:2014vta, Castiblanco:2018qsd}). Therefore, the constant-time hypersurfaces of the comoving gauge are the same as those in the synchronous gauge (they have the same slicing, the difference being in the threading). Furthermore, both gauges coincide at early times, that is $N^i = 0$ at $\eta_\ast \rightarrow 0$. We then follow a Lagrangian biasing prescription, and write down the bias expansion in terms of geometrical quantities describing those hypersurfaces of constant comoving time of observers at a very early time $\eta_\ast \rightarrow 0$. These quantities will be by definition gauge-independent once one chooses the hypersurface. Moreover, the coordinate transformation that eliminates the long-wavelength mode is also well known in this gauge \cite{Creminelli:2012ed}.

A similar approach was used to write down the effective theory of inflation \cite{Cheung:2007st}, and we follow their analysis with a few differences. Our first building blocks are the extrinsic curvature $K^\mu_{\phantom{\mu}\nu}$ of the constant-time hypersurfaces, and the matter density contrast $\delta$. This second quantity is geometrical in the sense that it can be written as proportional to the Einstein tensor contracted with the 4-velocity $G^{\mu\nu}u_\mu u_\nu$ using the Einstein equation. Since the galaxy number density transforms as a scalar under spatial diffeomorphisms, each term in the bias expansion is guaranteed to transform appropriately under such transformations. We ignore stochastic terms, which we expect behave as in the Newtonian case.

We thus write the Lagrangian galaxy number over-density \us{$\delta_g$} \us{in terms of combinations of these objects. Up to third order we write}
\be
\delta_g(\eta_*,\bm{x}) = \frac{b_1^*}{a_*} \delta + \frac{b_2^*}{a_*^2} \delta^2 + \frac{b_{s^2}^*}{a^2_*} S^i_{\phantom{i}j} S^j_{\phantom{j}i} + \frac{b_3^*}{a_*^3} \delta^3 + \frac{b_{\delta s^2}^*}{a_*^3} \delta S^i_{\phantom{i}j} S^j_{\phantom{j}i} + \frac{b_{s^3}^*}{a_*^3} S^i_{\phantom{i}j} S^j_{\phantom{j}k}S^k_{\phantom{k}i}  \,,
\label{biasExpansion3}
\ee
where $S^i_{\phantom{i}j} \equiv (K^\ell_{\phantom{\ell}\ell} \delta^i_j/3 - K^i_{\phantom{i}j} )/H^2$\us{, the arbitrary coefficients $b^*$ correspond to the bias parameters at the time $\eta_*$,} and all quantities are evaluated at $\eta_* \rightarrow 0$. \us{The bias coefficients are related to the usual Lagrangian bias coefficients by $b^{\mathcal{L}}_{\mathcal{O}} = b^*_{\mathcal{O}}/a^n$ where $n$ is the order at which the operator starts when expanded in perturbations.}

The extension of equation \eqref{biasExpansion3} to fourth order is straightforward, and is given in equation \eqref{biasExpansion}. Finally, since the coordinate transformation that eliminates the long mode involves only a spatial diffeomorphism, both sides of this expression transform in the same way, thus guaranteeing that the consistency relation is automatically satisfied, which we \us{explicitly} check in Appendix \ref{CRB}.

There are many other operators that can in principle be included, we discuss them in order:

\begin{itemize}

\item One can form spatial derivatives by projecting covariant derivatives orthogonal to the constant-time hypersurface. Higher spatial derivatives of the operators considered will be relevant for large enough halos. We expect them to work as in the Newtonian case, where each spatial derivative is accompanied by the Lagrangian size of the halo $R\de_i$, which is small for scales larger than $R$. It is straightforward to include this in our expansion, but we stick to the lowest derivative operators for simplicity.

\item One can apply a time derivative of any operator by acting on it with $u^\mu \nabla_\mu$. In synchronous gauge, or in comoving gauge at $\eta_i$, this is simply a time derivative $u^\mu \nabla_\mu = \de_\eta$. Since in matter domination each operator scales with powers of the growth factor, at each order one has  that $u^\mu \nabla_\mu \mathcal{O}^{(n)} = \de_\eta \mathcal{O}^{(n)}$ is simply proportional to $\mathcal{O}^{(n)}$ at the initial time (see section 2.5.2 of \cite{Desjacques:2016bnm} for further details).

\item Other contractions of the Ricci tensor will be different from the extrinsic curvature and $u^\mu u^\nu G_{\mu\nu}$. In general they will be independent operators, but at the lowest order in perturbation theory each operator will be expanded in terms of second derivatives of the metric fluctuations. At a given order $n$ the operators we chose at $\mathcal{O}^{(n)}$ give all possible combinations of second derivatives of the curvature fluctuation $(\de_i \de_j \zeta)^n$. Additional operators will be different from ours at higher orders in perturbation theory, but the different combinations that can appear are quite constrained. For example, at order $n + 1$ one can have additional combinations of second derivatives $(\de_i \de_j \zeta)^{n+1}$, but they will be degenerate with the operators starting at that order $\mathcal{O}^{(n+1)}$. Additionally, they can have relativistic corrections at a higher order, which to subleading order in the relativistic expansion are $\zeta(\de_i \de_j \zeta)^n$ or $\nabla\zeta.\nabla\zeta (\de_i \de_j \zeta)^n$, but they are fixed by the requirement of satisfying the consistency relation: A very specific combination of $(\de_i \de_j \zeta)^n$, $\zeta(\de_i \de_j \zeta)^n$ and $\nabla\zeta.\nabla\zeta (\de_i \de_j \zeta)^n$ is needed such that the unphysical correlation of the operator with a constant $\zeta$ or a constant gradient $\nabla\zeta$ disappears in the local frame (see Appendix~\ref{CRB}).\footnote{The dynamical evolution will however generate non-local terms that cannot be written as simple expansions in derivatives, as is remarked in \cite{McDonald:2009dh, Desjacques:2016bnm} and computed explicitly in section \ref{LagrangianEvolution}.}

\item Continuing with the previous point, one can in general have second derivatives of the tensor fluctuations $\de_i \de_j \gamma_{k\ell}$ \us{(e.g. in other components of the Riemann tensor)} but we take them to be sourced only by the scalar fluctuations such that the considerations of the previous point apply.

\end{itemize}

\subsection{Explicit Lagrangian expansion and dynamical evolution}\label{LagrangianEvolution}

We follow a Lagrangian framework for the bias expansion as detailed in the previous section. That is, we write the galaxy number density contrast in terms of geometrical quantities describing an early time hypersurface. These can be explicitly written in synchronous or comoving gauge to obtain the relativistic Lagrangian bias prescription. In order to obtain the later time Eulerian density contrast, we work in comoving gauge and evolve the galaxy number density as follows:

\begin{itemize}
    \item We assume conservation of the fraction of  Lagrangian volume that will eventually collapse into halos by the time of observation,
        \begin{equation}
           \nabla_{\mu}(u^{\mu} u^{\nu} \rho_g)=0\,,
        \end{equation}
where we have also assumed that the 4-velocity of galaxies equals that of the dark matter fluid, i.e. we ignore velocity bias. The solution to this equation in the Newtonian limit recovers the standard relation between Lagrangian and Eulerian bias coefficients (see section 2.3 of \cite{Desjacques:2016bnm}).

    Using the perturbed metric \eqref{eq:metric} and the background equations, we find
        \begin{equation}\label{eq:dyna}
            \dot{\delta_g}+\theta= -\de_i(\delta_g u^i) + \mathcal{S}_{\delta}[\delta_g, \theta_N]\,,
        \end{equation}
    where $\mathcal{S}_{\delta}$ is the same source as in equation \eqref{eq:correc1} but evaluated for the biased tracer. We reproduce it here for completeness
    \be
     \mathcal{S}_{\delta}[\delta_g, \theta_N] = 3(1 + \delta_g) \dot{\psi} + 3(1 + \delta_g) u^i \de_i \psi.
    \ee

    \item We set adiabatic initial conditions at a very early time $\eta_* \rightarrow 0$. Explicitly, up to fourth order we write
        \begin{equation}
        \label{biasExpansion}
            \us{\delta_ g(\eta_*) = \sum _ {n=1}^{4} \frac{b_n^*}{n!a_*^n} \delta^n + \sum_{n=2}^4 \frac{b_{s^n}^*}{a_*^n} (S^n) + \frac{b_{\delta s^2}^*}{a_*^3} (S^2) \delta  + \frac{b_{\delta^2 s^2}^*}{a_*^4} (S^2)\delta^2  + \frac{b_{\delta s^3}^*}{a_*^4} (S^3)\delta  + \frac{b_{(s^2)^2}^*}{a_*^4}(S^2)^2\,,}
        \end{equation}
      where all quantities are evaluated at the initial time $\eta_\ast$. Here, $(S^n)$ stands for the trace of $n$ operators, e.g., $(S^n) = \text{tr} (S \cdots S)$, and $S^i_{\phantom{i} j}$ is the traceless part of the extrinsic curvature, which reduces to the usual expression in the Newtonian limit $S^i_{\phantom{i} j} = (k_i k_j/k^2 -\delta_{ij}/3) \delta_\ell(\bm{k})$. In the perturbative expression for the bias expansion, we keep only the terms which don't vanish in the limit $\eta \rightarrow 0$.

\end{itemize}

For future reference, after some work the operator $S^i_{\phantom{i} j}$ can be explicitly written as
\be\label{eq:Sij}
S^i_{\phantom{i} j} = \frac{1}{2}(\de_i u^j + \de_j u^i) - \frac{1}{3} \de_k u^k \delta^i_j + \mathcal{O}(\epsilon^2)\,.
\ee

Following this prescription allows us to consistently write solutions order by order.
The explicit solution for the relativistic contribution to the galaxy number density contrast will be written in terms of bias Kernels given in Appendices~\ref{app:NBkernels} and~\ref{app:RBkernels}.

\subsubsection{First order}
The integration of \eqref{eq:dyna} gives:
\begin{equation}
\delta_g^{(1)} =a(\eta) \delta_{\ell}(\bm{k})+C_1(\bm{k}),
\end{equation}
where $C_1$ is a constant in time, which we set such that it satisfies the first-order initial conditions \eqref{biasExpansion}:
\begin{equation}
C_1= b_1^*\delta_{\ell}(\bm{k})\,,
\end{equation}
to get
\begin{equation}
\label{eq:final1}
\delta_g^{(1)}(\bm{k})= a(\eta) \delta_\ell(\bm{k}) \left(1+ \frac{\us{b_1^*}}{a(\eta)}\right)\,.
\end{equation}

\subsubsection{Second order}
At second order, the Lagrangian bias expansion \eqref{biasExpansion} is given by
\begin{equation}
\delta_g^{(2)}(\bm{k}, \eta_\ast) = \int_{\bm{k}_1, \bm{k}_2}\!\!\!\!\!(2\pi)^3\delta_D(\bm{k} - \bm{k}_{12}) \left[ b_1^* a^3_* H^2_* F_2^R(\bm{k}_1,\bm{k}_2) + \frac{1}{2} b_2^* + b^*_{s^2}s^2(\bm{k}_1,\bm{k}_2)\right] \delta_\ell(\bm{k}_1)\delta_\ell(\bm{k}_2)\,,
\end{equation}
where we have taken the limit $a_* \rightarrow 0$ such that the Newtonian quadratic kernel vanishes. This sets the initial conditions for the evolution of $\delta_g$, and the first term reflects the fact that initial conditions for $\delta$ must be set at second order. We have also defined
\be
s^2(\bm{k}_1,\bm{k}_2)= \frac{(\bm{k}_1\cdot\bm{k}_2)^2}{k_1^2k_2^2}-\frac{1}{3}.
\ee
The integration of equation \eqref{eq:dyna} gives
\begin{equation}
\delta_g^{(2)}(\bm{k},\eta)= \int_{\bm{k}_1, \bm{k}_2}\!\!\!\!\!(2\pi)^3\delta_D(\bm{k} - \bm{k}_{12}) \left\{\left[a^2F^N_2(\bm{k}_1,\bm{k}_2) + a b_1^* \alpha(\bm{k}_1,\bm{k}_2)\right]\delta_\ell(\bm{k}_1)\delta_\ell(\bm{k}_2)+C_2(\bm{k}_1,\bm{k}_2)\right\}\,,
\end{equation}
where the second term is usually rewritten as $(F_2^N - (2/7)s^2 - 2/21)\delta_\ell^2$. Fixing the integration constant then gives
\begin{multline}
\delta_g^{(2)}(\bm{k},\eta) = a^2 \int_{\bm{k}_1, \bm{k}_2}\!\!\!\!\!(2\pi)^3\delta_D(\bm{k} - \bm{k}_{12}) \bigg[\left(1 + \frac{b_1^*}{a}\right)F_2(\bm{k}_1,\bm{k}_2) +\frac{1}{2}\left(\frac{b_2^*}{a^2} - \frac{4}{21}\frac{b_1^*}{a}\right) \\+ \left(\frac{b^*_{s^2}}{a^2} - \frac{2}{7}\frac{b_1^*}{a}\right)s^2(\bm{k}_1, \bm{k}_2)\bigg]\delta_\ell(\bm{k}_1)\delta_\ell(\bm{k}_2)\,,
\end{multline}
where we denote $F_2 = F_2^N + a^2H^2 F_2^R$ for brevity. 

\subsubsection{Third order}
At third and higher orders, it won't be possible to neatly regroup the different terms into the geometric operators at a given time. This may be a gauge issue, but it doesn't spoil the result, merely complicating the bookkeeping.

At third order, the Lagrangian bias expansion is given by
\begin{multline}
\delta_g^{(3)}(\bm{k}, \eta_*) = \int_{\bm{k}_1, \bm{k}_2, \bm{k}_3}\!\!\!\!\!(2\pi)^3\delta_D(\bm{k} - \bm{k}_{123})  \bigg[ \frac{b_2^*}{a^2} a^5 H^2 F_2^R(\bm{k}_1, \bm{k}_2) + \frac{b_{s^2}^*}{a^2} a^5 H^2 M_3^{s^2,R}(\bm{k}_1, \bm{k}_2, \bm{k}_3) \\+ \frac{1}{6} \frac{b_3^*}{a^3}a^3 + \frac{b_{\delta s^2}^*}{a^3}a^3s^2(\bm{k}_1,\bm{k}_2) + \frac{b_{s^3}^*}{a^3}a^3s^3(\bm{k}_1,\bm{k}_2,\bm{k}_3)\bigg]\delta_\ell(\bm{k}_1)\delta_\ell(\bm{k}_2)\delta_\ell(\bm{k}_3)\,,
\end{multline}
where all the terms are constant in time, we define
$$
s^3(\bm{k}_1,\bm{k}_2,\bm{k}_3) = \frac{\bm{k}_1.\bm{k}_2\ \bm{k}_1.\bm{k}_3\ \bm{k}_2.\bm{k}_3}{k_1^2 k_2^2 k_3^2} - \left(\frac{(\bm{k}_1.\bm{k}_2)^2}{3 k_1^2 k_2^2} + 2\,\text{perms.}\right) + \frac{2}{9}\,,
$$
and $M_3^{s^2,R}$ is the cubic piece of the relativistic correction to the operator $S^i_{\phantom{i}j}S^j_{\phantom{j}i}$, given in equation~\eqref{eq:M3Rbk} and obtained from equation~\eqref{eq:Sij}. Note that there is no term proportional to $b_1^*$ since its cubic piece goes to zero as $\eta \rightarrow 0$. Integration of equation \eqref{eq:dyna} then gives
\begin{align}
&\delta^{(3)}_g(\bm{k}, \eta) =  \int_{\bm{k}_1, \bm{k}_2,\bm{k}_3}\!\!\!\!\!(2\pi)^3\delta_D(\bm{k} - \bm{k}_{123}) \bigg\{a^3 F_3(\bm{k}_1,\bm{k}_2,\bm{k}_3) + \frac{1}{2} a^2b_1^*\alpha(\bm{k}_1 + \bm{k}_2, \bm{k}_3)G^N_2(\bm{k}_1,\bm{k}_2) \nn \\
&+ \alpha(\bm{k}_3,\bm{k}_1 + \bm{k}_2) \left[ \frac{1}{2}a^2 b_1^*F_2^N(\bm{k}_1,\bm{k}_2) +\frac{1}{2}\left(ab_2^* - \frac{2}{21}a^2b_1^*\right) + \left(ab^*_{s^2} - \frac{1}{7}a^2b_1^*\right)s^2(\bm{k}_1, \bm{k}_2)\right] \nn \\
&+a^4 H^2 b_1^*\bigg[\alpha(\bm{k}_3,\bm{k}_1 + \bm{k}_2)F_2^R(\bm{k}_1,\bm{k}_2) + \alpha(\bm{k}_1 + \bm{k}_2, \bm{k}_3) G_2^R(\bm{k}_1,\bm{k}_2)  + 3F_2^\psi(\bm{k}_1, \bm{k}_2) \nn \\
&\phantom{+a^4 H^2 b_1^*\bigg[}-i\bm{k}_3\cdot \bm{G}_2^T(\bm{k}_1,\bm{k}_2)+ \frac{15}{2} \frac{\bm{k}_1.\bm{k}_2}{k_1^2 k_2^2}\bigg] \bigg\}\delta_\ell(\bm{k}_1)\delta_\ell(\bm{k}_2)\delta_\ell(\bm{k}_3) + \delta^{(3)}_g(\bm{k}_1, \bm{k}_2, \bm{k}_3,\eta^\ast)\,.
\end{align}

We don't attempt to rewrite this expression in terms of simple quantities since not much insight is gained by doing so. Even in the Newtonian limit, non-local terms appear in the evolved galaxy density contrast (see \cite{McDonald:2009dh, Desjacques:2016bnm}), and we expect the same to hold for the relativistic terms. This means that it won't be possible to neatly write them in terms of geometrical quantities of the hypersurface of constant time $\eta$.

\subsubsection{Fourth order}

At fourth order, the calculation proceeds in an analogous fashion, and we obtain schematically
\be
\delta_g^{(4)} = a^4\left(F_4(\bm{k}_1, \bm{k}_2,\bm{k}_3,\bm{k}_4) + \sum b^\mathcal{L}_{\mathcal{O}} M_4^{\mathcal{O}}(\bm{k}_1, \bm{k}_2,\bm{k}_3,\bm{k}_4)\right) \delta_\ell(\bm{k}_1)\delta_\ell(\bm{k}_2)\delta_\ell(\bm{k}_3)\delta_\ell(\bm{k}_4)\,,
\ee
where we used the notation $b_{\mathcal{O}}^\mathcal{L} = b^*_\mathcal{O}/a^n$ with $n$ the order at which a given operator starts, e.g. $b_2^\mathcal{L} = b^*_2/a^2$ and $b_{\delta s^2}^\mathcal{L} = b^*_{\delta s^2}/a^3$. The kernels are given by $M_4^\mathcal{O} = M_4^{\mathcal{O},N} + a^2 H^2 M_4^{\mathcal{O},R}$, and $M_4^{\mathcal{O},N}$ and $M_4^{\mathcal{O},R}$ can be found in Appendixes~\ref{app:NBkernels} and~\ref{app:RBkernels}.

\us{
Following this notation, which is useful to compute correlation functions, we write the galaxy density contrast as
\begin{equation}
\delta_g(\bm{k},\eta) = \delta(\bm{k},\eta) + \sum_{n=1}^{\infty}a^n \int_{\bm{k}_1\cdots\bm{k}_n} (2\pi)^3 \delta_{D}(\bm{k}-\bm{k}_{1\cdots n})\sum_{\mathcal{O}} b_{\mathcal{O}}^{\mathcal{L}} M_n^{\mathcal{O}}(\bm{k}_1,\cdots,\bm{k}_n,\eta) \delta_\ell(\bm{k}_1)\cdots\delta_\ell(\bm{k}_n)\,, \label{eq.delta_h}
\end{equation}
where $\delta(\bm{k},\eta)$ has been calculated in section \ref{sec:DM} and the second part contains all the terms proportional to the bias parameters, explicitly $M_n^\mathcal{O}(\bm{k},\eta) = M_n^{\mathcal{O},N}(\bm{k}) + a^2 H^2 M_n^{\mathcal{O},R}(\bm{k})$. Expressions for  $M_n^{\mathcal{O},N}$ and $M_n^{\mathcal{O},R}$ are given in Appendixes~\ref{app:NBkernels} and~\ref{app:RBkernels} respectively.
}
\us{
\section{Galaxy correlation functions}\label{sec:rsl}
In order to have a better understanding of the interplay between the bias operators, and the evolution of dark matter, we now explicitly compute the two and three point correlation function of the galaxy density contrast $\delta_g$. We start by defining the galaxy  power spectrum and the galaxy bispectrum in the standard way \footnote{Here and in the rest of the paper $\langle \dots \rangle$ denotes the average over a large region of space with a fundamental Fourier mode $k_f$. Assuming ergodicity, this is close to the ensemble average over many realization of the stochastic field $\delta_g(\bm{k},\eta)$ when $k_f \rightarrow 0$.}
\begin{align}
    & \langle \delta_g(\bm{k}_1,\eta) \delta_g(\bm{k}_2,\eta)\rangle= (2\pi)^3 \delta_D(\bm{k}_1+\bm{k}_2) P_{g}(k_1,\eta)\,, \\
    & \langle \delta_g(\bm{k}_1,\eta) \delta_g(\bm{k}_2,\eta)\delta_g(\bm{k}_3,\eta)\rangle=(2\pi)^3 \delta_D(\bm{k}_1+\bm{k}_2+\bm{k}_3) B_{g}(\bm{k}_1,\bm{k}_2,\bm{k}_3,\eta)\,.
\end{align}
For this section only we will use the notation $\delta_m$ for the matter density field given by equation \eqref{eq:expansion1} and $\delta_g$ defined in \eqref{eq.delta_h}.
In order to organize our expressions, we define 
\begin{equation}
\delta_b \equiv \delta_g - \delta_m\,,
\end{equation}
such that we can separate the contribution to each expression that comes from the fact that galaxies are a biased tracer.
}
\us{
\subsection{Galaxy power spectrum}
To leading-order (LO), or tree-level, there is no relativistic correction as we see from \eqref{eq:final1} and we recover the Newtonian result \cite{Desjacques:2016bnm}.
To next-to-leading-order (or 1-loop) corrections to the galaxy power spectrum can be written as
\begin{equation}
	P_g(\bm{k},\eta) = P_g^{LO}(\bm{k},\eta) + P^g_{1-loop}(\bm{k},\eta)\,,  \label{Pg}
\end{equation}
where $P^g_{1-loop}$ is the 1-loop galaxy power spectrum, which can be split into three main contributions:
\begin{equation}
	P^g_{1-loop}(\bm{k},\eta)= P_{1-loop}^{mm}(\bm{k},t)  + 2P_{1-loop}^{bm}(\bm{k},t) + P_{1-loop}^{bb}(\bm{k},t)\,. 
\end{equation}
The matter-matter power spectrum at 1-loop $P_{1-loop}^{mm}$ has two contributions $P_{13}^{mm}$ and $P_{22}^{mm}$ given by
\begin{equation}
	P_{13}^{mm}(\bm{k},\eta) =    6 a^4(\eta) P_{L}(k)\int_{\bm{q}} P_{L}(q)F_3^N(\bm{q},-\bm{q},\bm{k}) + 
    6 H_0^2a^3(\eta) P_{L}(k)\int_{\bm{q}} P_{L}(q)F_3^R(\bm{q},-\bm{q},\bm{k}),
\end{equation}
\begin{multline}
    P_{22}^{mm}(\bm{k},\eta) =  2 a^4(\eta) \int_{\bm{q}}  \left[F_2^N(\bm{q},\bm{k}-\bm{q})\right]^2P_L(q) P_L(|\bm{k}-\bm{q}|)  \\
     +4  H_0^2a^3(\eta) \int_{\bm{q}}  F_2^N(\bm{q},\bm{k}-\bm{q}) F_2^R(\bm{q},\bm{k}-\bm{q})P_L(q) P_L(|\bm{k}-\bm{q}|).
\end{multline} 
The 1-loop correlation between the matter density contrast and the bias operators is denoted $P_{1-loop}^{bm}$ and given by
\begin{multline}
	P_{1-loop}^{bm}(\bm{k},\eta) \equiv  b_1^\mathcal{L}P_{13}^{mm}(\bm{k},\eta) + 3 a^4(\eta) P_{L}(k)\sum b_\mathcal{O}^\mathcal{L}\int_{\bm{q}} P_{L}(q)M_3^{\mathcal{O}}(\bm{q},-\bm{q},\bm{k})   \\
     + 2 a^4(\eta)\sum b_\mathcal{O}^\mathcal{L} \int_{\bm{q}} F_2(\bm{q},\bm{k}-\bm{q}) M_2^{\mathcal{O}}(\bm{q},\bm{k}-\bm{q}) P_L(q) P_L(|\bm{k}-\bm{q}|) .
\end{multline}
And finally the 1-loop auto-correlation between the bias operators $P_{1-loop}^{bb}$ reads
\begin{multline}
	P_{1-loop}^{bb}(\bm{k},\eta) \equiv 6 a^4(\eta) P_{L}(k)\sum _{\mathcal{O}}b_1^\mathcal{L} b_\mathcal{O}^\mathcal{L} \int_{\bm{q}}P_{L}(q) M_3^{\mathcal{O}}(\bm{q},-\bm{q},\bm{k})\\
	+2a^4(\eta)\sum _{\mathcal{O},\mathcal{O}'}b_\mathcal{O}^\mathcal{L}b_\mathcal{O'}^\mathcal{L} \int_{\bm{q}} M_2^{\mathcal{O}}(\bm{k}-\bm{q}, \bm{q})M_2^{\mathcal{O'}}(\bm{k}-\bm{q},\bm{q})P_L(q)P_L(|\bm{k}-\bm{q}|).
\end{multline}
In section~\ref{sec:numResults} we show the results obtained from numerically integrating these expressions.
}

\us{
\subsection{Galaxy bispectrum}
To next-to-leading-order (or 1-loop) the galaxy bispectrum can be written as
\begin{equation}
	B _{g}(\bm{k}_1,\bm{k}_2,\bm{k}_3,\eta) = B_g^{LO}(\bm{k}_1,\bm{k}_2,\bm{k}_3,\eta) + B^g_{1-loop}(\bm{k}_1,\bm{k}_2,\bm{k}_3,\eta)\,.
\label{Bg} 
\end{equation}
Similarly to the power spectrum, one can split the galaxy bispectrum at 1-loop into pieces that involve the bias operators and pieces that don't.}

\us{
At tree level, the galaxy bispectrum can be decomposed into the following contributions
\begin{align}
	B_{g}^{LO}(\bm{k}_1,\bm{k}_2,\bm{k}_3,\eta) & = B^{mmm}_{211}(\bm{k}_1,\bm{k}_2,\bm{k}_3,\eta) + B^{bmm}_{211}(\bm{k}_1,\bm{k}_2,\bm{k}_3,\eta)+2B^{mmb}_{211}(\bm{k}_1,\bm{k}_2,\bm{k}_3,\eta)\nonumber\\ 
	&\phantom{=}+ 2B^{bbm}_{211}(\bm{k}_1,\bm{k}_2,\bm{k}_3,\eta) +B^{mbb}_{211}(\bm{k}_1,\bm{k}_2,\bm{k}_3,\eta)+   B^{bbb}_{211}(\bm{k}_1,\bm{k}_2,\bm{k}_3,\eta)\nonumber\\
	&\phantom{=}+2\,\text{cyclic permutations}
\end{align}
The first one is the matter-matter-matter bispectrum which has a Newtonian and a relativistic part:
\begin{multline}
	B^{mmm}_{211}(\bm{k}_1,\bm{k}_2,\bm{k}_3,\eta) =   2a^4(\eta) F_2^N(\bm{k}_1,\bm{k}_2) P_L(k_1) P_L(k_2) \\
       +4 a^3(t)H_0^2 F_2^R(\bm{k}_1,\bm{k}_2) P_L(k_1) P_L(k_2) \,.
\end{multline}
We denote them $B^N_{tree}$ and $B^R_{tree}$.}

\us{
Next we have the correlations between the bias operators and the matter density contrast:
\begin{align}
B^{bmm}_{211}(\bm{k}_1,\bm{k}_2,\bm{k}_3,\eta) &= 2 a^4(\eta) \sum_{\mathcal{O}} b_\mathcal{O}^\mathcal{L}M_2^{\mathcal{O}}(\bm{k}_2,\bm{k}_3) P_L(k_2) P_L(k_3)\,,\\
B^{mmb}_{211}(\bm{k}_1,\bm{k}_2,\bm{k}_3,\eta)&=2 a^4(\eta)b_1^\mathcal{L}F_2(\bm{k}_2,\bm{k}_3)P_L(k_2) P_L(k_3)\,,\\
B^{bbm}_{211}(\bm{k}_1,\bm{k}_2,\bm{k}_3,\eta)& =   2a^4(\eta) \sum_{\mathcal{O}} b_1^\mathcal{L} b_\mathcal{O}^\mathcal{L}M_2^{\mathcal{O}}(\bm{k}_2,\bm{k}_3) P_L(k_2) P_L(k_3) \,,\\
B^{mbb}_{211}(\bm{k}_1,\bm{k}_2,\bm{k}_3,\eta)&=2a^4(\eta) \left(b_1^\mathcal{L}\right)^2 F_2(\bm{k}_2,\bm{k}_3)P_L(k_2) P_L(k_3)\,. \\
B^{bbb}_{211}(\bm{k}_1,\bm{k}_2,\bm{k}_3,\eta) & = 2 a^4(\eta) \sum_{\mathcal{O}} \left(b_1^\mathcal{L}\right)^2 b_\mathcal{O}^\mathcal{L}M_2^{\mathcal{O}}(\bm{k}_1,\bm{k}_2) P_L(k_1) P_L(k_2) \,.
\end{align}
 For the sake of brevity, we refer the reader to Appendix~\ref{app:G_BS} for explicit expressions for the 1-loop calculation. 
}


\section{Renormalization of bias parameters}\label{sec:renorm}

In order for the bias expansion to be well defined, the operators should be appropriately renormalized such that the expectation value of $\delta_g$ vanishes, and its correlation with long-wavelength perturbations behaves as expected. We do this at the initial time $\eta_*$ take these as renormalized initial conditions for the dynamical evolution. Since this evolution is non-linear, one then needs to renormalize the result at each time $\eta$.

To be more explicit, let's study how this works for the operator proportional to $b_2^*$. At initial time, the bare operator is given by
\begin{multline}
\frac{1}{2 a_*^2}b_2^* \delta^2(\bm{k}, \eta_*) = \frac{1}{2}b_2^* \int_{\bm{q}_1,\bm{q}_2} \delta_D(\bm{k} - \bm{q}_1 - \bm{q}_2) \delta_\ell(\bm{q}_1) \delta_\ell(\bm{q}_2) \\+  b_2^* a_*^3 H_*^2 \int_{\bm{q}_1,\bm{q}_2,\bm{q}_3} \delta_D(\bm{k} - \bm{q}_1 - \bm{q}_2 - \bm{q}_3) F_2^R(\bm{q}_1, \bm{q}_2) \delta_\ell(\bm{q}_1) \delta_\ell(\bm{q}_2)  \delta_\ell(\bm{q}_3)\,.
\end{multline}
Following \cite{Assassi:2014fva}, we compute the expectation value of this operator, which is
\be
\frac{1}{2 a_*^2}b_2^*\left\la \delta^2 \right\ra = \frac{1}{2}b_2^* \int_{\bm{q}} P(q) = \frac{1}{2}b_2^* \sigma^2(\Lambda)\,.
\ee
\us{where $\sigma^2(\Lambda)=\int_{\bm{q}} P(q)$ and} $\Lambda$ is the UV cutoff of the integral, or equivalently the scale over which the density contrast is smoothed. This needs to be subtracted from the bare operator in order to guarantee that the full operator have zero expectation value. Next, we correlate with a long-wavelength perturbation, which gives
\begin{align}
\frac{1}{2 a_*^2}b_2^*\lim_{k \rightarrow 0} \left\la \delta_\ell(\bm{k})  \delta^2(-\bm{k}) \right\ra' &= b_2^* a_*^3 H_*^2 \lim_{k \rightarrow 0} P(k) \int_{\bm{q}} \left(F_2^R(\bm{k},\bm{q}) + F_2^R(\bm{k},-\bm{q})\right) P(q) \nn \\
&= \left( -\frac{5}{k^2} \sigma^2(\Lambda)  - 5\sigma^2_{-2}(\Lambda)\right)b_2^* a_*^3 H_*^2 P(k)\,.
\end{align}
where $\sigma^2_{-2} = \int_{\bm{q}} P(q)/q^2$, and the prime denotes the fact that we factor out the Dirac delta of momentum conservation. In order to cancel the unphysical cutoff dependence coming from this result, we need to write \us{the renormalized operator \footnote{\us{Renormalized operators are denoted as $[\mathcal{O}]_{\Lambda}$}}} 
\be
\left[\frac{1}{2 a_*^2}b_2^* \delta^2\right]_{\Lambda} = \frac{1}{2 a_*^2}b_2^* \delta^2 - \frac{1}{2}b_2^* \sigma^2 + 5b_2^\ast a_*^3 H_*^2 \sigma^2_{-2} \delta_\ell  + 2 b_2^\ast \sigma^2 \zeta\,.
\label{renormalizedL}
\ee
\us{where $\zeta(\bm{k})= \frac{5}{2 k^2}\delta_\ell(\bm{k})$ is the curvature perturbation in comoving gauge. The third term of this expression, proportional to $\sigma^2_{-2}$, represents a relativistic correction to the standard renormalization.}

\us{Let us take a moment to analyze the last term of expression \eqref{renormalizedL}, which would seem to suggest that we need to include an unphysical ``non-Gaussian scale-dependent bias'' term in our expansion. {\bf This is not the case} as remarked in \cite{dePutter:2015vga}, as it merely represents the fact that we are working in a set of coordinates that don't represent what a local observer measures. Specifically, the cutoff scale has been fixed to be some coordinate value $\Lambda$, while it should correspond, physically, to a scale defined by a local observer $\Lambda_{ph}$. It was shown by \cite{dePutter:2015vga} that  $\sigma^2(\Lambda_{ph}) = (1 - 4\zeta) \sigma^2(\Lambda)$. Their argument is as follows: The physical length-scale for an observer in the presence of a very large wavelength perturbation $\zeta_L$ is given by $a(\eta)\Lambda(1 + \zeta_L)$, such that the comoving physical cutoff scale is modified by the presence of the long-wavelength perturbation. They write the variance as
\begin{align}
\sigma^2(\Lambda_{ph}) &= \int_{\bm{q}} |W(q\Lambda_{ph})|^2 P(q) \nonumber \\
&= (1 + \zeta_L)^{-3} \int_{\tilde{\bm{q}}} |W(\tilde{q}\Lambda)|^2 P\left(\tilde{q}(1 - \zeta_L)\right) \nonumber \\
&= (1 - 4\zeta_L) \sigma^2(\Lambda)\,. 
\end{align}
Thus, the problematic last term combines with the second term to form $ -\frac{1}{2}b_2^*\sigma^2(\Lambda_{ph})$.\footnote{We assumed scale invariance for this calculation. A deviation from scale invariance induces an additional term in the rescaling of the short-mode variance, but it also induces a primordial contribution to the three-point function in these coordinates (given by Maldacena's consistency relation) which gives an additional contribution to eq.~\eqref{renormalizedL}, and the two again cancel.} Moreover, the same is true for all bias operators as we explicitly check in Appendix~\ref{app:ernormOps}. Note that, written in terms of the global coordinates, this counter-term proportional to $\zeta$ should be kept and it cancels a $1/k^2$ divergence in the power spectrum.}

We now want to take this renormalized operator as initial conditions, and evolve it using equation \eqref{eq:dyna}. For this, let us write the evolution equation for a given operator, which is obtained by defining $\delta_g = \delta + \mathcal{O}$ in equation \eqref{eq:dyna}, such that
\begin{multline}
\label{evolO}
\dot{\mathcal{O}}(\bm{k}) = \int_{\bm{q}_1, \bm{q}_2} (2\pi)^3 \delta_D(\bm{k} - \bm{q}_{12})\left[-\alpha(\bm{q}_1, \bm{q}_2)
\theta(\bm{q}_1)\mathcal{O}(\bm{q}_2) + 3 \mathcal{O}(\bm{q}_1)\dot{\psi}(\bm{q}_2) - i \bm{q}_2\cdot\bm{u}_T(\bm{q}_1) \mathcal{O}(\bm{q}_2)\right] \\
+ 3\int_{\bm{q}_1, \bm{q}_2, \bm{q}_3} (2\pi)^3 \delta_D(\bm{k} - \bm{q}_{123}) \frac{\bm{q}_1 \cdot\bm{q}_3}{q_1^2}\theta(\bm{q}_1)\mathcal{O}(\bm{q}_2) \psi(\bm{q}_3)\,.
\end{multline}
Since this equation is linear in $\mathcal{O}$, the evolution of the initial renormalized operator is given by the superposition of the solution for the bare operator presented in the previous section, and the solution for which the initial conditions are only given by the counter-terms in equation~\eqref{renormalizedL}. Since this evolution is non-linear, it will induce a further cutoff dependence that needs to be subtracted at each order. In particular,
\be
\lim_{k_1\rightarrow 0} \lim_{k_2\rightarrow 0}\left\langle \delta_\ell(\bm{k}_1) \delta_\ell(\bm{k}_2) \left[\frac{1}{2a_*^2}b_*^2\delta^2\right](\bm{k},\eta)\right\rangle' = -\frac{115}{6} a^4 H^2 b_2^*\sigma^2_{-2}P(k_1) P(k_2),
\ee
where $[b_2^\ast \delta^2/2a_*^2]$ is the operator evolved from expression \eqref{renormalizedL} using equation \eqref{evolO}. \us{This operator was renormalized at the initial conditions but contains an additional cutoff dependence induced by the non-linear time evolution so it is no longer properly renormalized.} We add an additional counter-term to the evolved operator in order to cancel this cutoff dependence. In this way, we obtain for this second piece
\be
M_0^{\delta^2, c.t.}(\bm{k}) = -\frac{1}{2}\frac{b_2^*}{a^2} a^2 \sigma^2 (2\pi)^3\delta_D(\bm{k})\,,
\ee
\be
M_1^{\delta^2, c.t.}(\bm{k}) = -\frac{1}{2}\frac{b_2^*}{a^2} a^3 \sigma^2 + 5\frac{b_2^\ast}{a^2} a^5 H^2 \sigma^2_{-2} + 5 \frac{b_2^*}{a^2}a^5H^2 \sigma^2 \frac{1}{k^2} \,,
\ee
\begin{align}
M_2^{\delta^2, c.t.}(\bm{k}_1, \bm{k}_2) &= -\frac{1}{4}\frac{b_2^\ast}{a^2}a^4\sigma^2 \left(G_2^N(\bm{k}_1, \bm{k}_2) + \alpha(\bm{k}_1, \bm{k}_2) \right) + \frac{b_2^\ast}{a^2} a^6 H^2 \sigma^2_{-2}\left( 5\alpha(\bm{k}_1, \bm{k}_2) +\frac{115}{12}\right) \nn\\
&\phantom{=}  -\frac{1}{2}\frac{b_2^\ast}{a^2}a^6 H^2\sigma^2 \left(G_2^R(\bm{k}_1, \bm{k}_2) + 3 F_2^\psi(\bm{k}_1, \bm{k}_2) +\frac{15}{2} \frac{\bm{k}_1.\bm{k}_2}{k_1^2 k_2^2} - \frac{10}{k_2^2}\alpha(\bm{k}_1,\bm{k}_2) \right).
\end{align}
One can check that, after a reshuffling of terms, this procedure recovers the standard renormalized bias expansion in the Newtonian limit.
The renormalization of the other operators used proceeds in a similar fashion, and is performed in Appendix~\ref{app:ernormOps}.

\us{
In order to have well behaved correlation functions at long wavelengths it is necessary to add the corresponding contribution from the counterterms. Therefore, the power spectrum \eqref{Pg} and bispectrum \eqref{Bg} are renormalized by adding 
\begin{equation}
    P_g^{c.t.}(k)=2\sum_{\mathcal{O}} b_\mathcal{O}^\mathcal{L} M_1^{\mathcal{O}, c.t.}(\bm{k})P_L(k)\,,\label{C.t.Pg}
\end{equation}{}
\begin{align}
    B_g^{c.t.}(\bm{k}_1,\bm{k}_2,\bm{k}_3) = & 2 \sum_{\mathcal{O}} b_{\mathcal{O}}^\mathcal{L}M_2^{\mathcal{O}, c.t.}(\bm{k}_2,\bm{k}_3)P_{L}(k_2)P_{L}(k_3)+2\,\text{permutations}\nonumber\\&+2 \sum_{\mathcal{O}} b_{\mathcal{O}}^\mathcal{L}M_1^{\mathcal{O},c.t.}(\bm{k}_1)F_2(\bm{k}_2,\bm{k}_3)P_{L}(k_2)P_{L}(k_3)+2\,\text{permutations}
\label{C.t.Bg}
\end{align}
respectively.
}
\us{
\section{Numerical computation of galaxy correlation functions}
\label{sec:numResults}
}
\us{To illustrate our findings we numerically compute the power spectrum \eqref{Pg} and bispectrum \eqref{Bg} together with the counter-term contributions \eqref{C.t.Pg} and \eqref{C.t.Bg} generated by the second-order operators proportional to $b_2^*$ and $b_{s^2}^*$. The numerical integration is performed using a code written in C which uses the Cuba library \cite{Hahn:2004fe}. For each plot, we set all bias parameters to zero except for the one being studied. For the linear power spectrum used in producing the plots we used the following parameters: $H_0 = 67.6$ km/s/Mpc, $\Omega_m = 0.31$, $n_s=1$, $\mathcal{A}_s=2.2 \times 10^{-9}$  and $f_{NL}=1$. The superscript $N$ refers to results using only the standard Newtonian kernels, while the superscript $R$ refers to results that involve the relativistic kernels. Explicit expressions for the integrals are given in Section~\ref{sec:rsl} and Appendix~\ref{app:G_BS}.\\}

\us{We show these plots only to illustrate the relative importance of the relativistic contributions to biasing. The quantities we compute do not represent an actual observable, since for that one needs to take into account how that observable is defined, that is, number of galaxies observed at a given redshift and position in the sky. This requires solving for photon propagation in a perturbed universe, which has been done up to second order in perturbation theory (see for example \cite{Yoo:2014sfa, DiDio:2014lka, Clarkson:2018dwn}), and was recently done ignoring lensing up to third order in the weak field approximation in \cite{DiDio:2018zmk}. We leave the full fourth-order calculation for future work. We find that all relativistic effects we have computed for the first time are degenerate with a primordial signal of $f_{NL} \sim \mathcal{O}(1)$, which is the same order of magnitude as the effects coming from photon propagation at tree level. A complete one-loop calculation is needed in order to account for these effects in the observed bispectrum and separate them from a primordial signal. }

\begin{figure}[!htb]
        \centering
        \begin{tabular}[t]{cc}
        \includegraphics[width=0.47\textwidth]{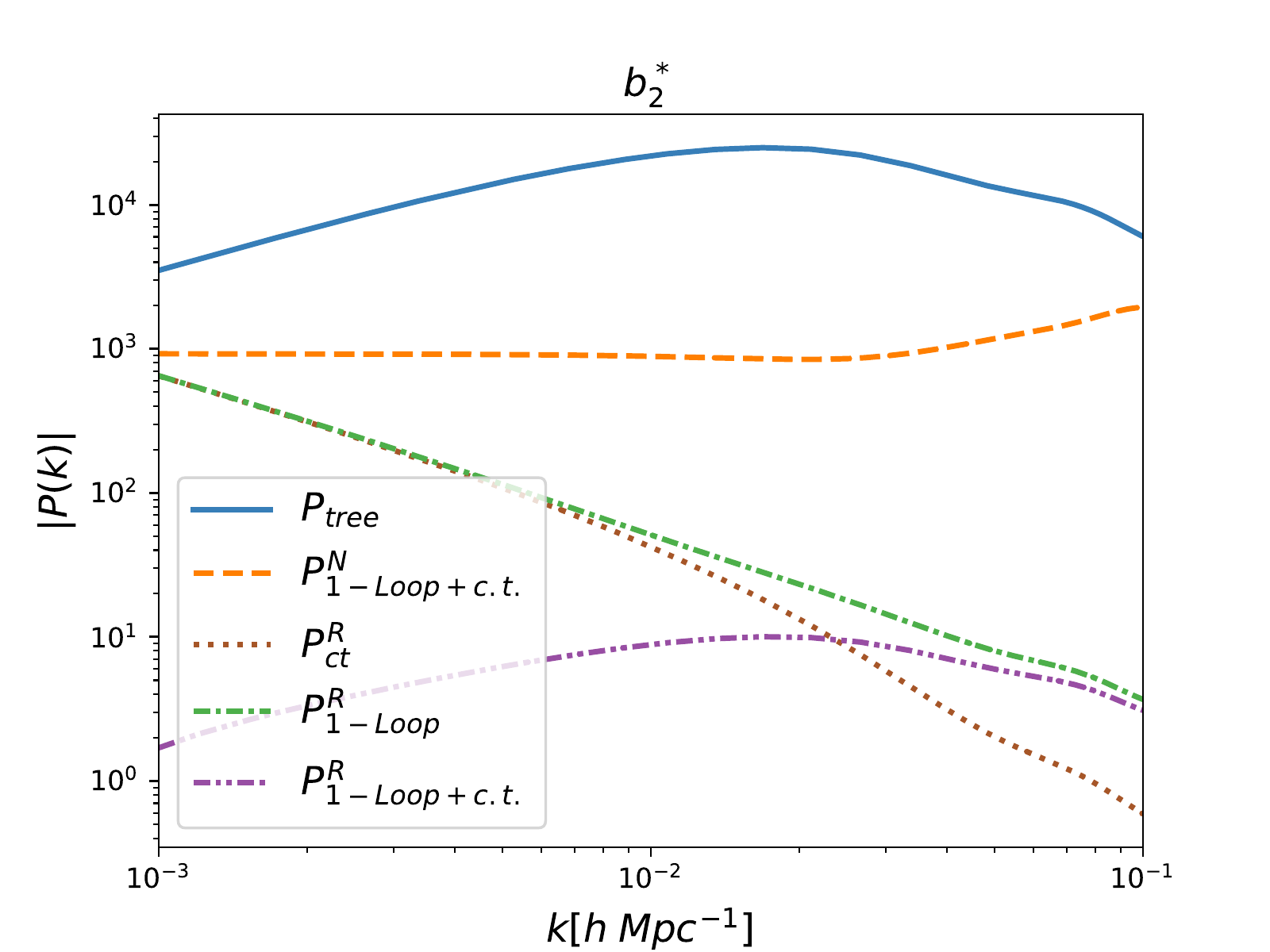} &
        \includegraphics[width=0.47\textwidth]{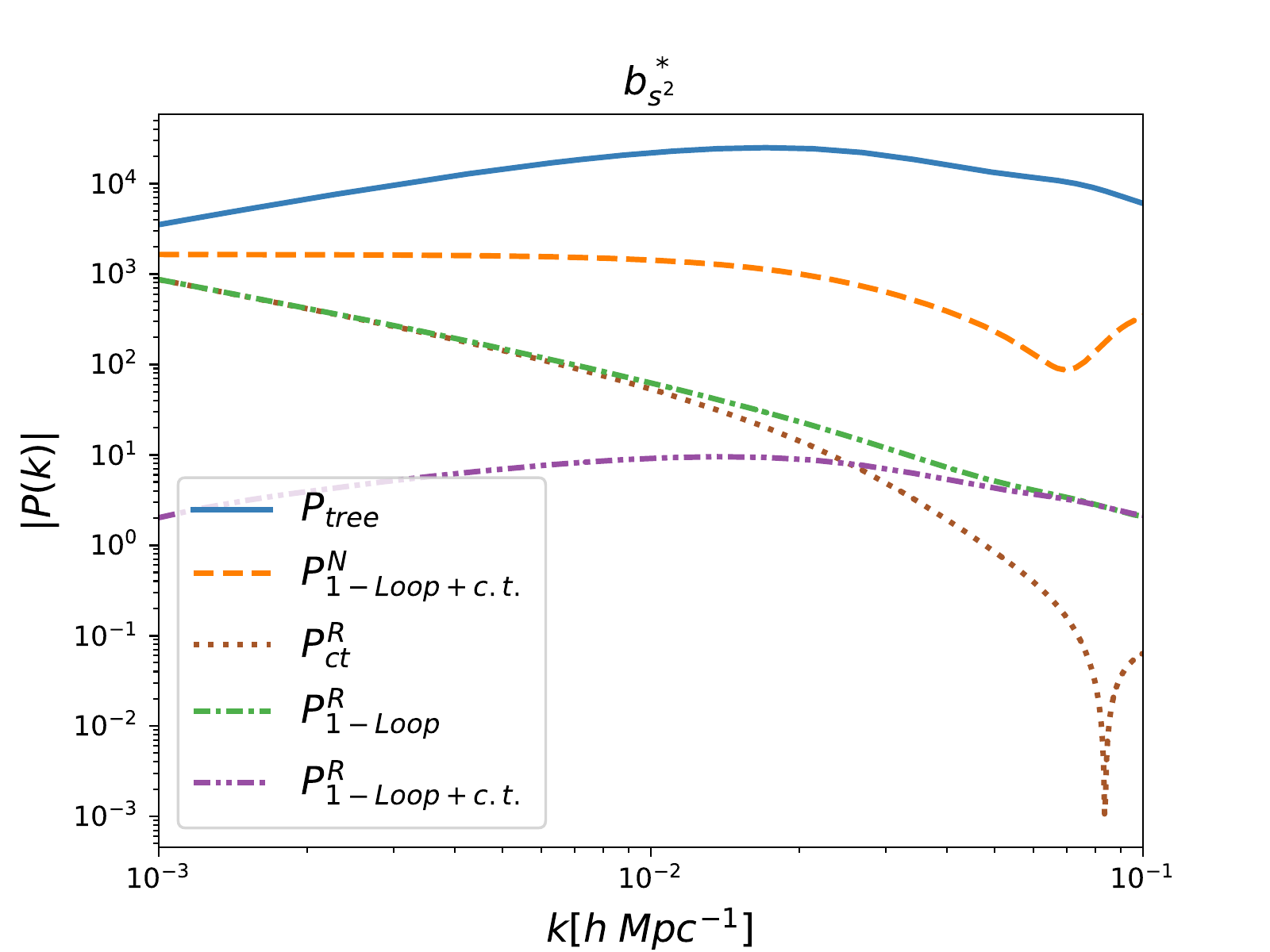}
        \end{tabular}
        \caption{\label{fig:PS} \us{Comparison between the tree-level power spectrum and various contributions to the one-loop power spectrum} computed from the operators proportional to $b_2^*$ and $b_{s^2}^*$ \us{(left and right panels respectively)}. For each plot, we set all bias parameters to zero except for the one being studied. \us{We separated the Newtonian (N) and Relativistic (R) contributions as well as the contributions from the counter-terms (ct).} Notice that the unphysical $1/k^2$ behavior of the loop is cancelled by the counter-terms. All quantities are evaluated at redshift $z = 0$.}
\end{figure}

\begin{figure}[!htb]
        \centering
        \begin{tabular}[t]{cc}
        \includegraphics[width=0.48\textwidth]{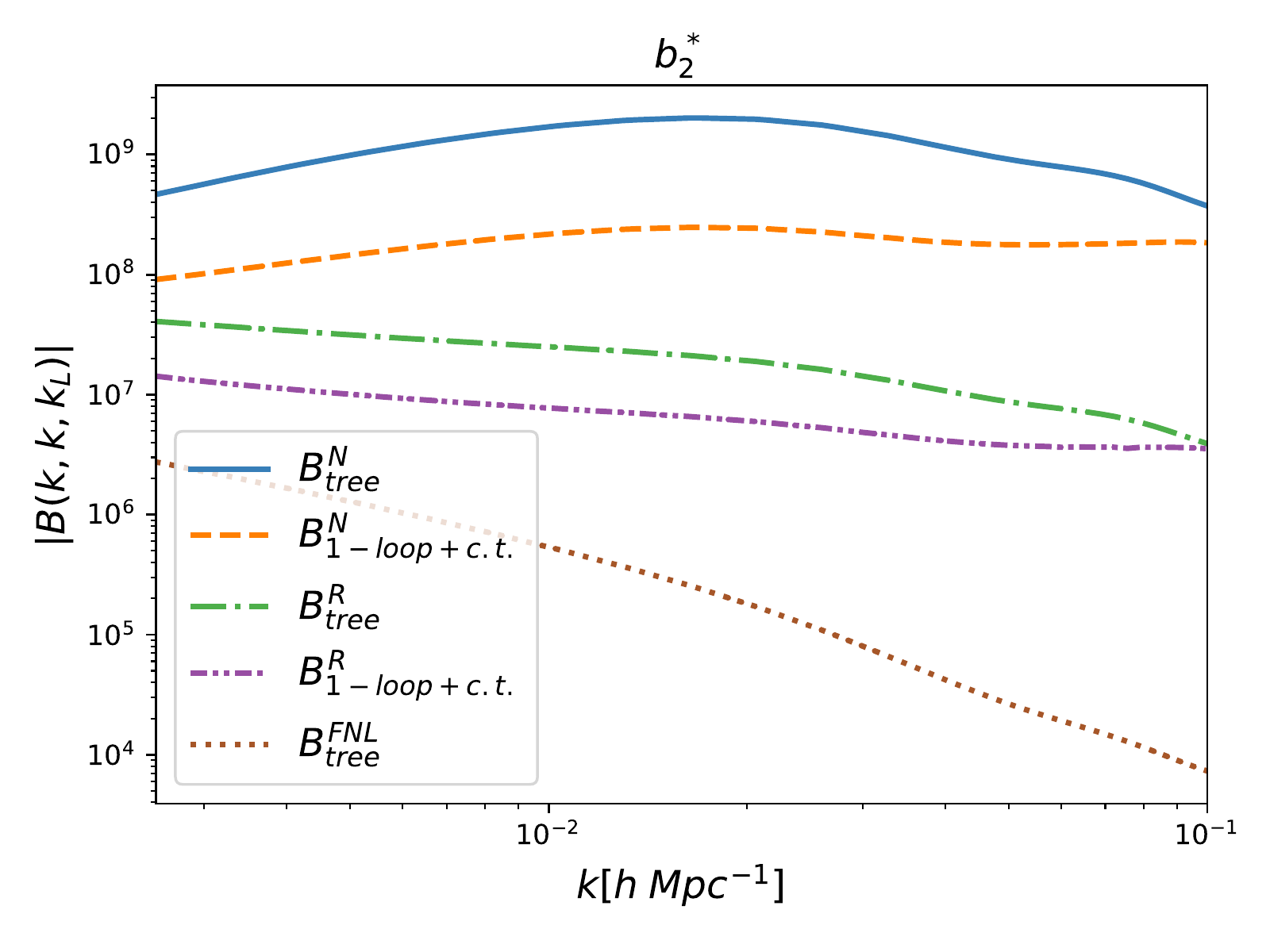} &
        \includegraphics[width=0.48\textwidth]{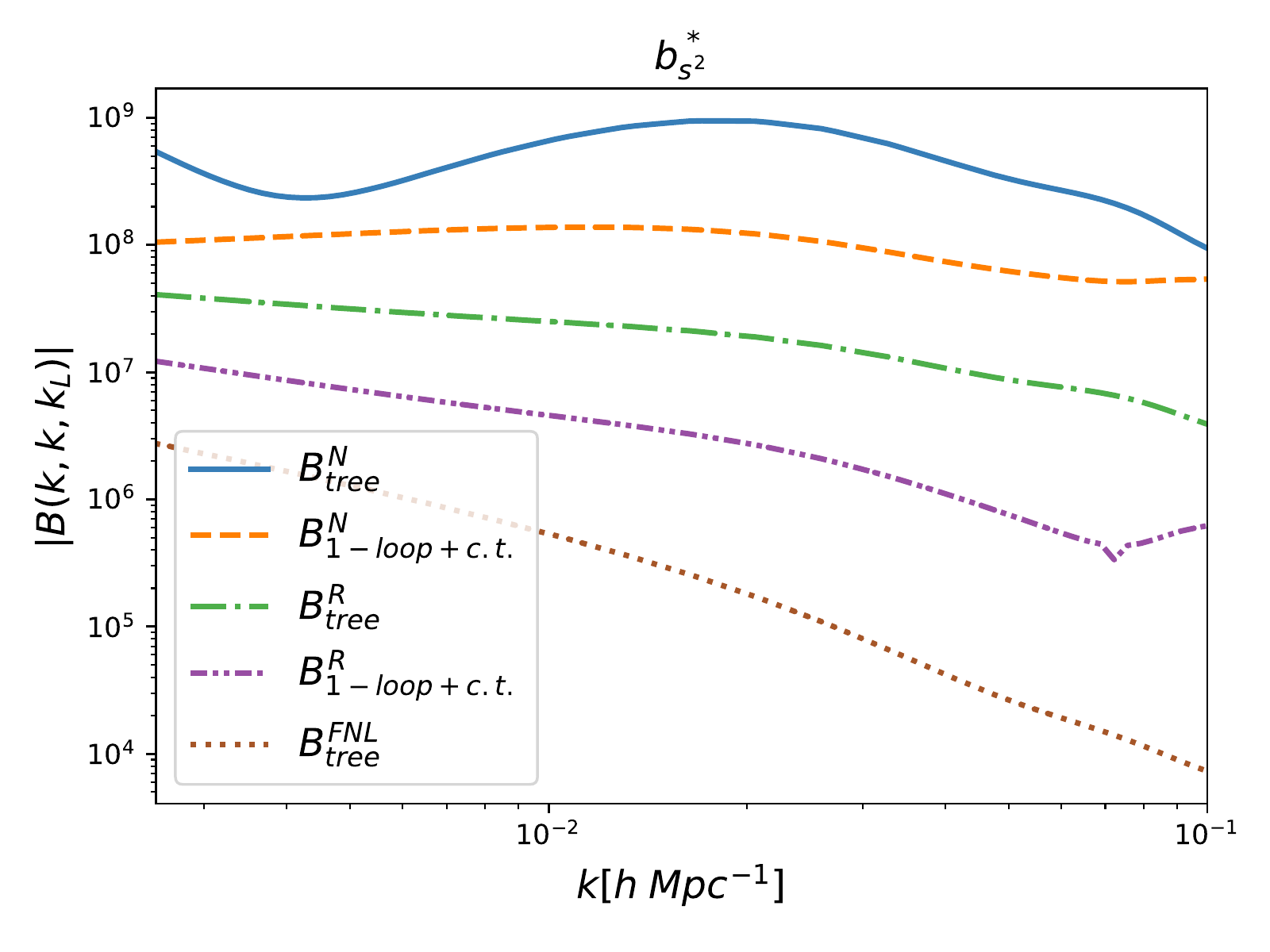}
        \end{tabular}
        \caption{\label{fig:klong} \us{Comparison between the tree-level bispectrum, the one-loop bispectrum, and a primordial bispectrum signal with local non-Gaussianity of $f_{NL} = 1$. We separated the Newtonian (N) and Relativistic (R) contributions. All lines are computed from the operators proportional to $b_2^*$ and $b_{s^2}^*$ (left and right panel). For each plot, we set all bias parameters to zero except for the one being studied. We fixed $k_L = 0.005\,h\,\text{Mpc}^{-1}$ and $k$ is varied. All quantities are evaluated at redshift $z = 0$.}}
\end{figure}

\begin{figure}[!htb]
        \centering
        \begin{tabular}[t]{cc}
        \includegraphics[width=0.48\textwidth]{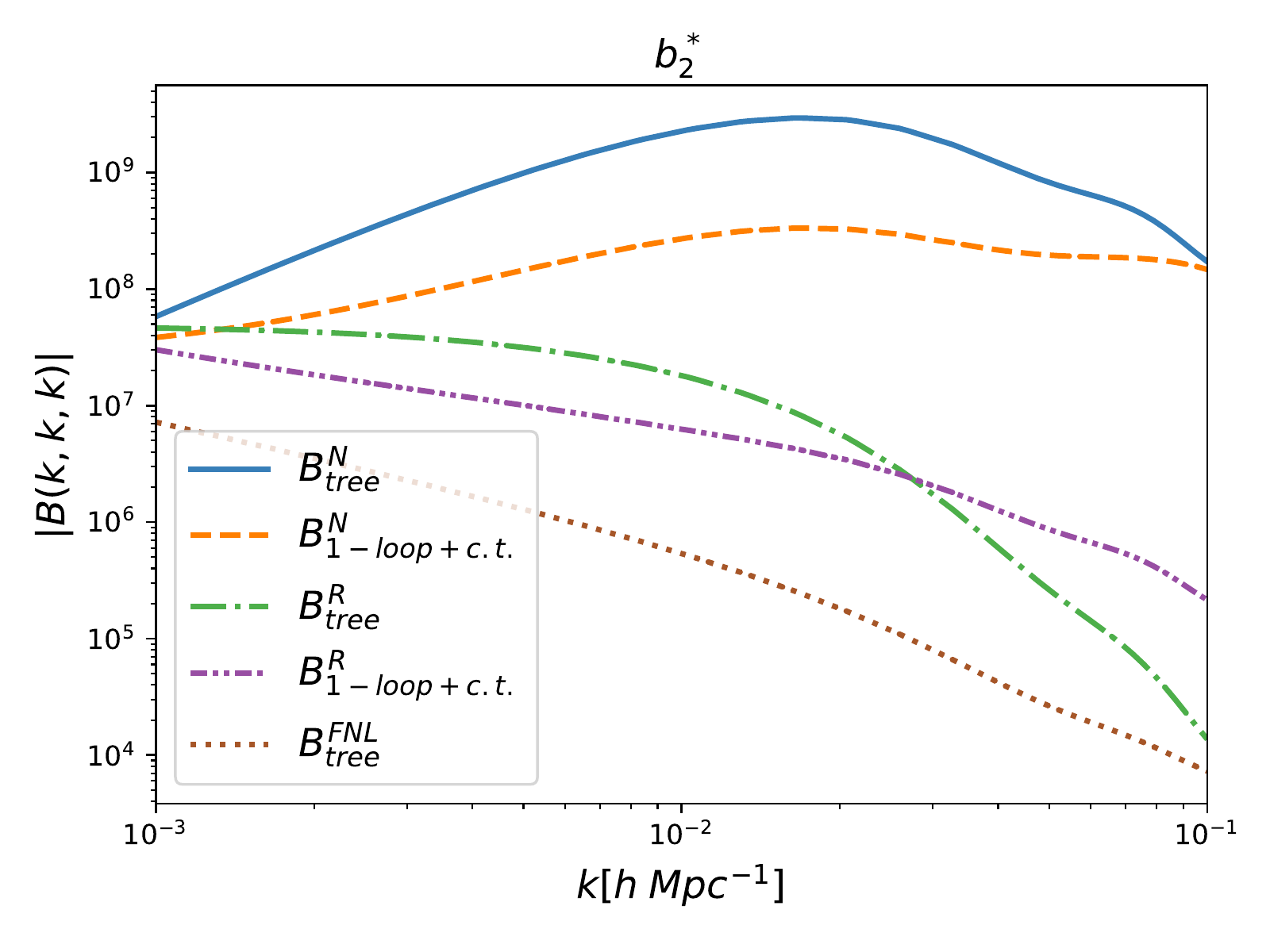} &
        \includegraphics[width=0.48\textwidth]{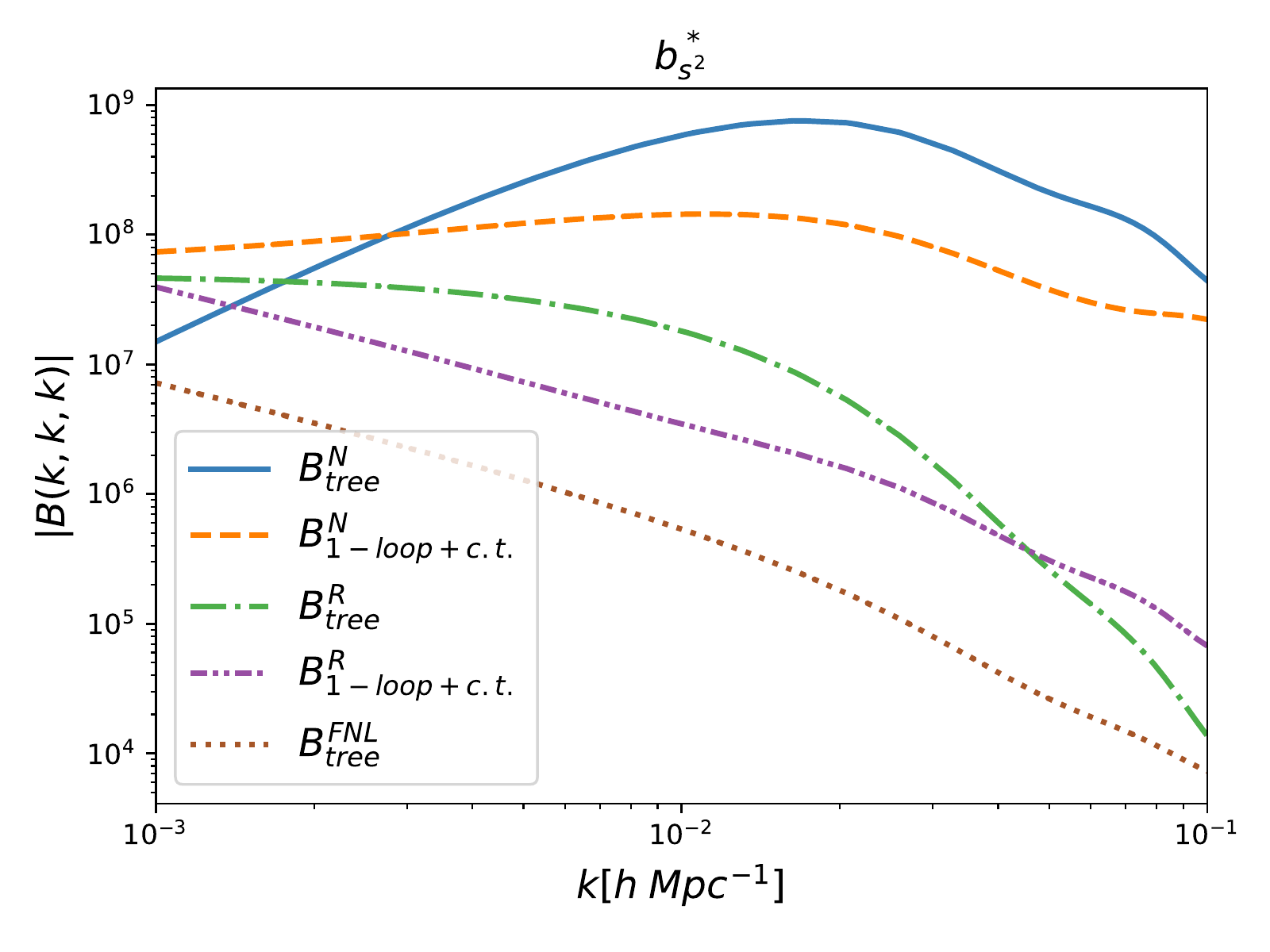}
        \end{tabular}
        \caption{\label{fig:equilateral} \us{Comparison between the tree-level bispectrum, the one-loop bispectrum in the equilateral configuration and a primordial signal with local non-Gaussianity of $f_{NL} = 1$. We separated the Newtonian (N) and Relativistic (R) contributions. All lines are computed from the operators proportional to $b_2^*$ and $b_{s^2}^*$ (left and right panel). For each plot, we set all bias parameters to zero except for the one being studied. All quantities are evaluated at redshift $z = 0$.}}
\end{figure}

\us{The one-loop power spectrum is plotted in figure~\ref{fig:PS}, where the tree-level, loop contributions and counter-terms are plotted separately. As expected, the integrals involved in the loops are sensitive to the UV cutoff chosen, but this dependence is cancelled out by the counter-terms. Interestingly, the relativistic terms in the loops induce a $1/k^2$ contribution that could be interpreted as a local-type non-Gaussianity. However, this contribution is unphysical and is exactly cancelled by the counter-terms, see section \ref{sec:renorm} and Ref.~\cite{dePutter:2015vga} for details about the renormalization of the operators. The renormalized relativistic corrections are negligible with respect to the Newtonian results. This is expected since relativistic corrections are only important on large scales (where they are cancelled by the counter-terms) while loop corrections are sizeable only on small scales. However, since the bispectrum couples scales, the effect of one-loop relativistic corrections can be sizeable for certain bispectrum configurations.}
\\
\begin{figure}[!htb]
        \centering
        \begin{tabular}[t]{cc}
        \includegraphics[width=0.49\textwidth]{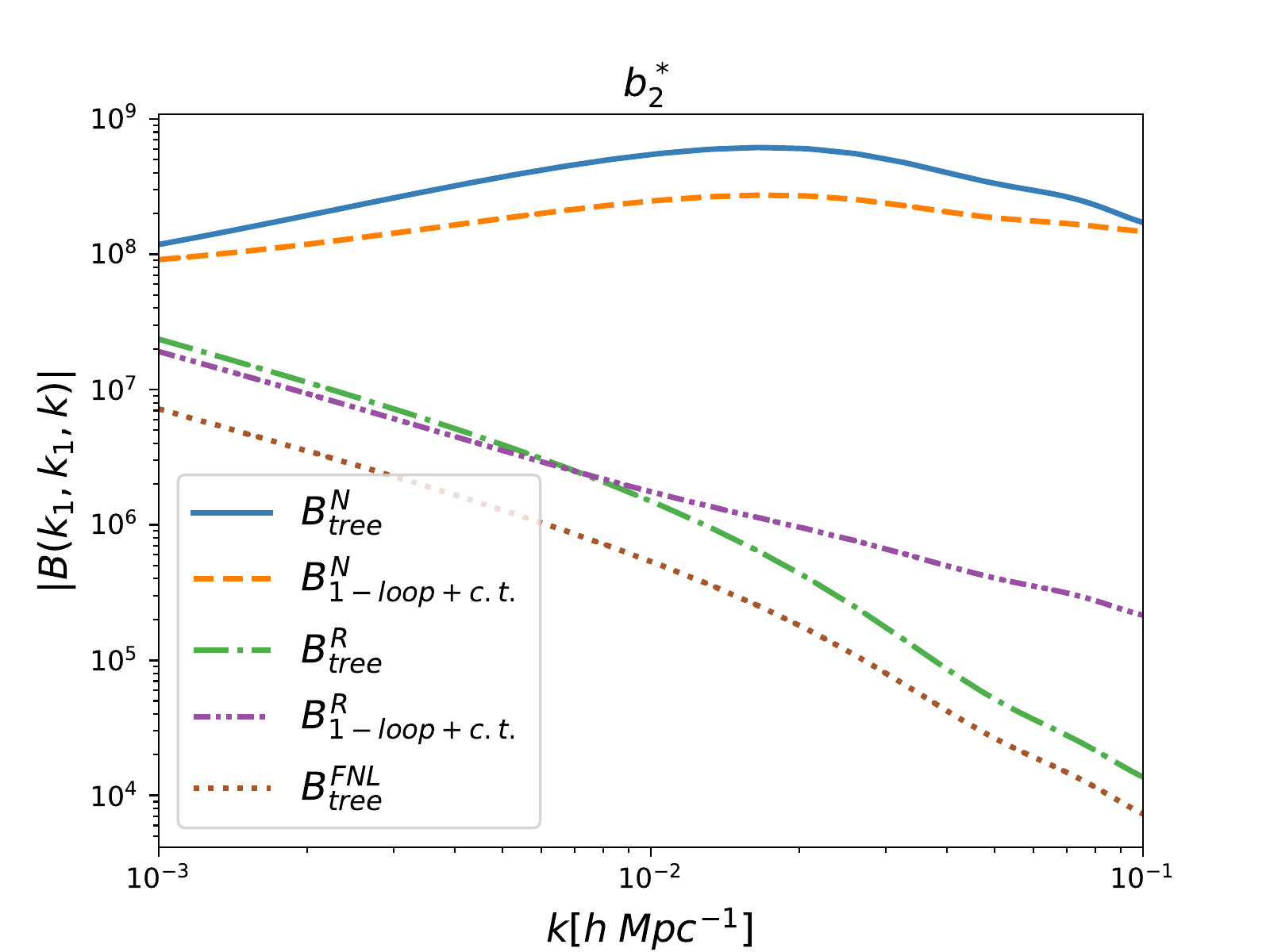} &
        \includegraphics[width=0.49\textwidth]{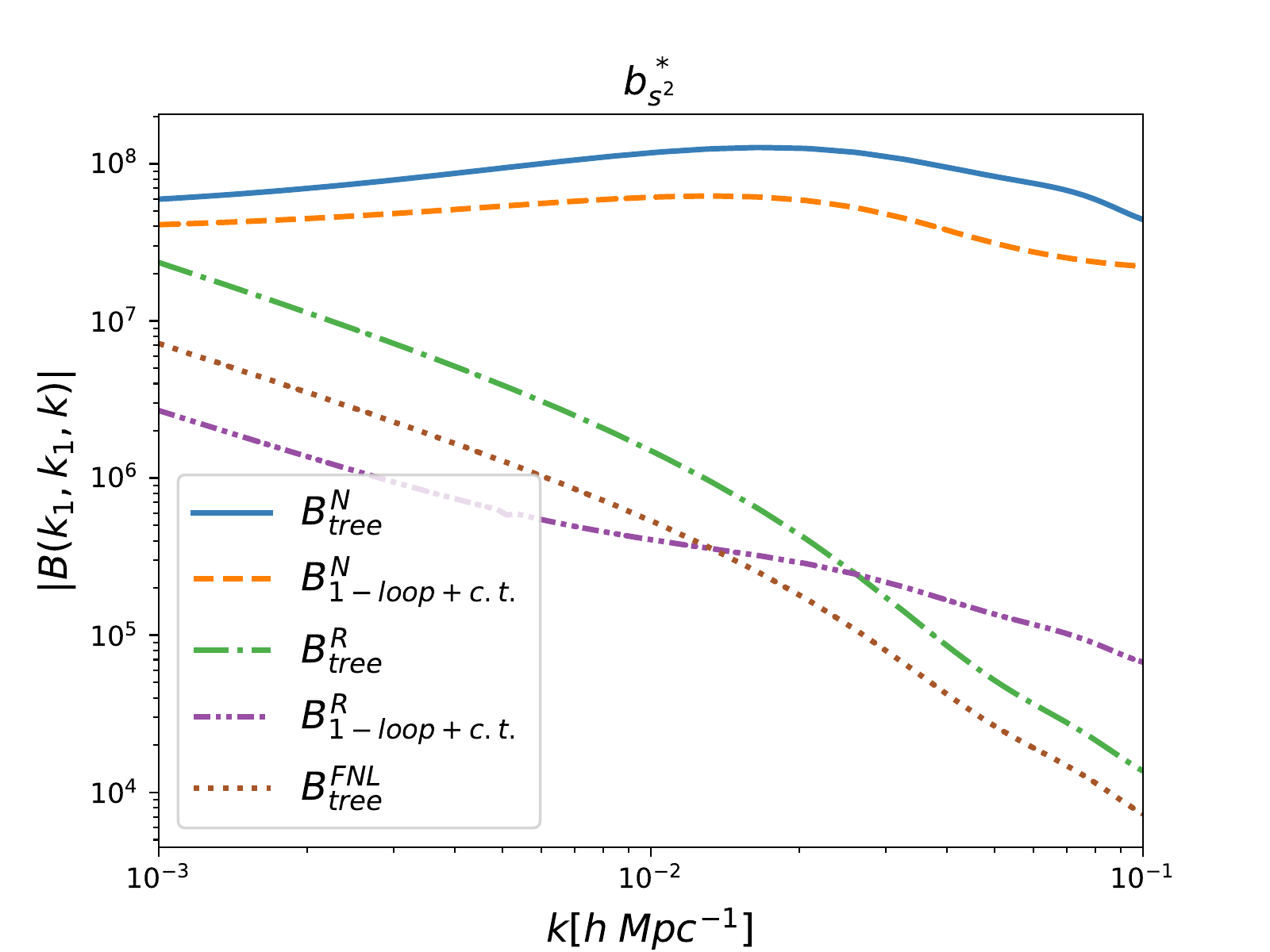}
        \end{tabular}
        \caption{\label{fig:squeezed} \us{Comparison between the tree-level bispectrum, the one-loop bispectrum, and a primordial bispectrum signal with local non-Gaussianity of $f_{NL} = 1$. We separated the Newtonian (N) and Relativistic (R) contributions. All lines are computed from the operators proportional to $b_2^*$ and $b_{s^2}^*$ (left and right panel).} For each plot, we set all bias parameters to zero except for the one being studied. Each contribution is plotted separately. For these plots $k_1 = 0.1\,h\,\text{Mpc}^{-1}$ is held fixed and $k$ is varied. The squeezed limit is approached as $k$ becomes smaller, which illustrates the behavior of the bispectrum as a function of the squeezed momentum. All quantities are evaluated at redshift $z = 0$.}
\end{figure}
\us{In figure~\ref{fig:squeezed}, we plot the behavior of the bispectrum as it approaches the squeezed limit. We plot $B(k_1, k_1, k)$ where $k_1 = 0.1\,h\,\text{Mpc}^{-1}$ and $k$ is varied. The squeezed limit is attained as $k$ goes to zero which happens towards the left of the figure, and we can clearly see the functional dependence of the bispectrum on the soft momentum. Since the bispectrum is generated by both the dark matter non-linearity and the bias expansion, this plot shows that there is a $1/k^2$ behavior as it approaches that limit, similar to what happens for the dark matter case. \us{This $1/k^2$ behavior is the same as that induced by a primordial non-Gaussian signal of the local type, and for comparison we show a line corresponding to $f_{NL} = 1$ which shows a similar behavior as a function of $k$ towards the squeezed limit. As remarked in \cite{Kehagias:2015tda,Castiblanco:2018qsd}, this $1/k^2$ behavior must be present such that it can be cancelled by the coordinate transformation that goes to that local frame, and in that sense it is a geometric projection effect. We also show the full bispectrum in figure~\ref{fig:B} keeping only one of the momenta fixed at $0.1\,h\,\text{Mpc}^{-1}$, where the growth towards the squeezed limit can also be seen.}
\begin{figure}[!htb]
        \centering
        \begin{tabular}[t]{cc}
        \includegraphics[width=0.49\textwidth]{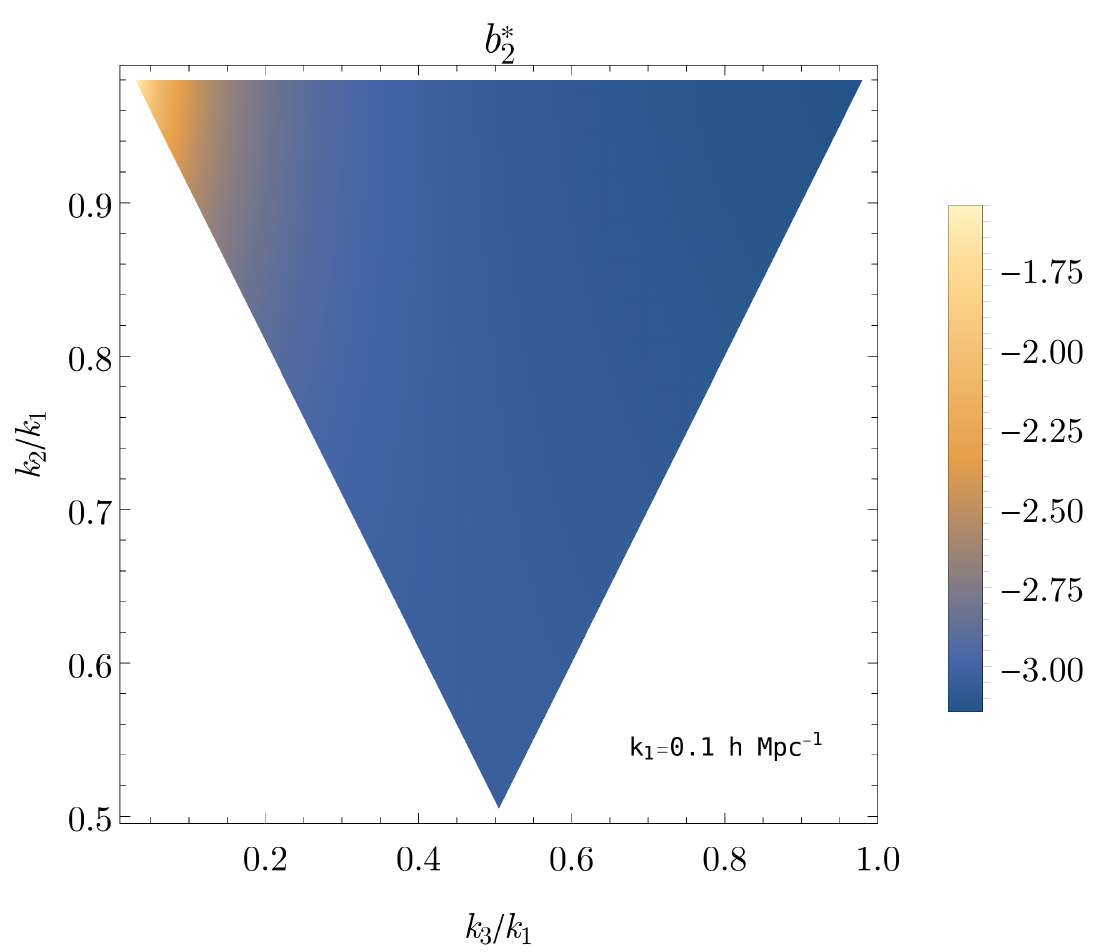} &
        \includegraphics[width=0.49\textwidth]{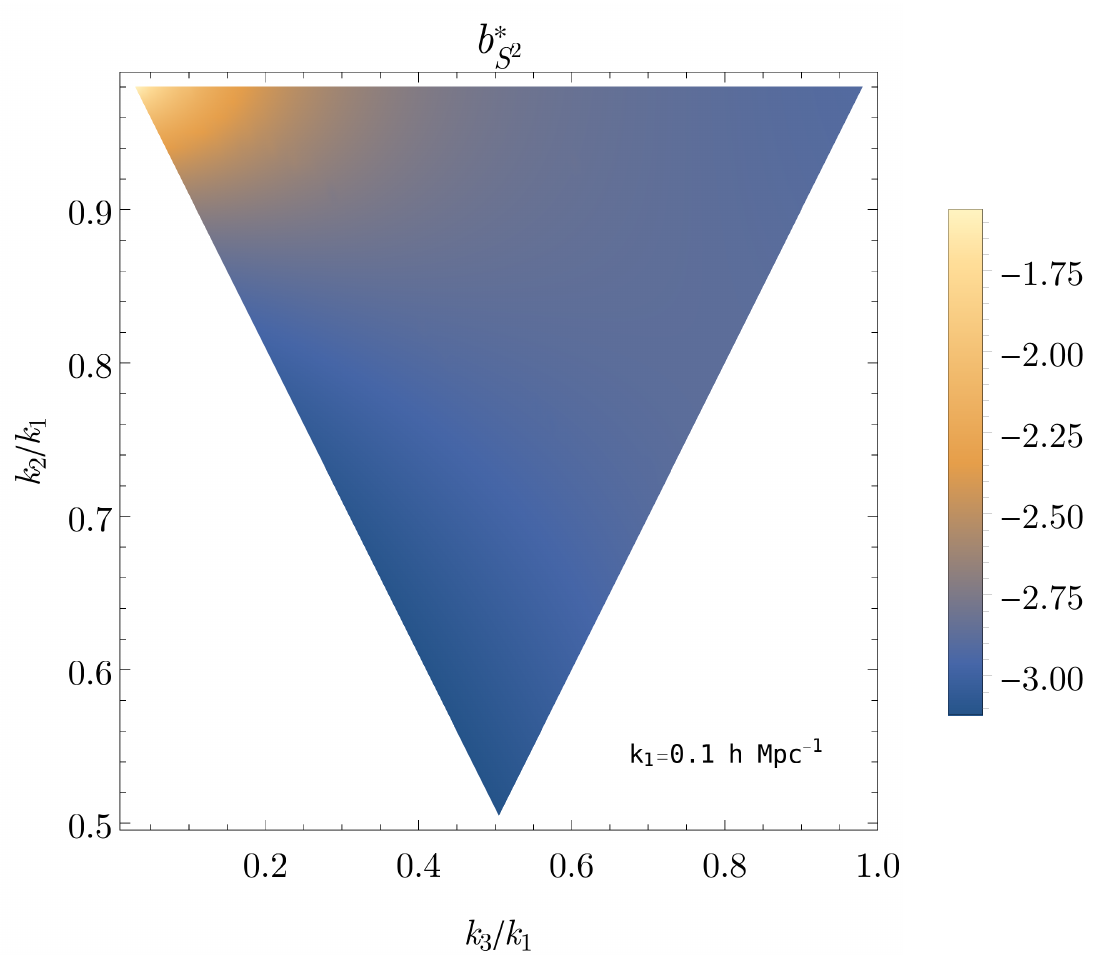}
        \end{tabular}
        \caption{\label{fig:B} Logarithm \us{base $10$} of the ratio between the relativistic one-loop bispectrum to its Newtonian counter-part, computed from the operators proportional to $b_2^*$ and $b_{s^2}^*$. For each plot, we set all bias parameters to zero except for the one being studied. Here, one of the momenta was fixed to $k_1 = 0.1\,h\,\text{Mpc}^{-1}$. All quantities are evaluated at redshift $z = 0$.}
\end{figure}
}
\us{
We also plot the bispectrum for other configurations, which are usually plotted in the literature. In figure \ref{fig:klong} we plot $B(k_L, k, k)$ for which $k$ is varied. This also illustrates the behavior of the bispectrum in the squeezed limit as $k$ becomes large, though it does not show the behavior of the bispectrum with the small (soft) momentum $k_L$ as the limit is approached, but rather shows it as a function of the momenta becoming large (hard). We observe that the behavior of a primordial non-Gaussian signal of the local type is a somewhat different function of the hard momenta, such that it is not completely degenerate with the relativistic corrections we computed. Note however that an observation of an effect going as $1/k_L^2$ would be confused with a violation of the consistency relation and considered a smoking gun of primordial non-Gaussianity, regardless of the behavior as a function of the hard momenta. In figure \ref{fig:equilateral} we plot the bispectrum for equilateral configurations $B(k,k,k)$. We see that, as expected, relativistic corrections are comparable to the Newtonian terms only at large scales for equilateral configurations. Finally, we plot the cross-spectra involving one galaxy number density contrast correlated with matter density contrast fields in Appendix~\ref{sec:newplts}.}
\section{Conclusions}
\label{sec:ccl}
We wrote a generalization of the Lagrangian bias prescription in the relativistic case \eqref{biasExpansion}, written in terms of operators describing the curvature of the initial time hypersurface. We explicitly wrote these operators in comoving gauge (which coincides with the synchronous gauge at the initial time hypersurface) and evolved them in time to obtain an Eulerian result, see section~\ref{sec:bias}. \us{Together with the renormalization of the relativistic operators, the relativistic bias expansion in the comoving gauge is the main result of this paper. We found that the relativistic loop corrections to the bias operators induce a signal in the squeezed bispectrum which has an amplitude and a behavior with the soft momentum similar to a primordial non-Gaussian signal with $f_{NL} \sim \mathcal{O}(1)$, and should thus be taken into account. We checked that this behavior is produced by geometric projection effects in the sense that our results satisfy the relativistic LSS consistency relation. However, the amplitude of this effect is comparable to the amplitude of projection terms coming from photon propagation (see e.g. \cite{Yoo:2014sfa, DiDio:2014lka, Clarkson:2018dwn}), and if they are ignored when computing the bispectrum to compare with data, one may artificially believe to have found a primordial signal.}

\us{We also checked that, as has already been pointed out in the literature (see e.g. \cite{dePutter:2015vga}), there is no $1/k^2$ behavior in the power spectrum of the galaxy number density contrast in comoving gauge at large scales induced by the relativistic bias expansion. The $1/k^2$ behavior appears in the power spectrum computed using non-renormalized operators, but is cancelled by a counter-term. This counter-term has a $1/k^2$ form, which seems naively incompatible with the equivalence principle (or the consistency relation). However, it has the exact coefficient that neatly combines with other terms in the expansion making it equivalent to changing the cutoff in the integrals from a coordinate cutoff to the cutoff defined in a local frame (as discussed in \cite{dePutter:2015vga}). This happens for all bias operators where the change in the cutoff is expected to have such an effect, and in all of them the coefficient of the counter-term is precisely the one needed to account for this.}

It would be interesting to write this expansion in Poisson gauge for a couple of reasons: First it is a popular gauge chosen for instance to implement the N-body code \texttt{gevolution} \cite{Adamek:2016zes}. Second, the UV behavior of the non-linear loop integrals is better modeled in the Poisson gauge with EFT techniques. Indeed, in synchronous and comoving gauges the inverse of the smoothed large scale metric is not the same as the smoothed inverse metric, see Appendix B of \cite{Baumann:2010tm}. However, the Poisson gauge does not coincide with the synchronous gauge at the initial time hypersurface, making this task an arduous one to fourth order since it requires a gauge transformation.

It would also be interesting to write down the bias expansion at formation time instead of the far past. Our bias expansion can be straightforwardly used in that case, but writing the geometric operators explicitly in comoving gauge is again difficult due to the fact that this gauge does not coincide with the synchronous gauge at an arbitrary time.

\section*{Acknowledgments}
We thank R. Gannouji for initial collaboration and many interesting discussions. It is also a pleasure to thank V.~Desjacques for a useful conversation. J.C.~is supported by ``Beca Postgrado PUCV 2019,'' L.C.~is supported by CONICYT scholarship Nr. 21190484, J.N.~is supported by Fondecyt Regular Nr. 1171466, and C.S.~is supported by Fondecyt Grant 3170557.
\appendix

\section{Explicit expressions of the large scale structure kernels}
\subsection{For dark matter}
\label{sec:DMkern}
The relativistic  kernels for dark matter were computed in \cite{Castiblanco:2018qsd}, we list them here since we use them to write the galaxy bias kernels.
\begin{align}
F_1^R(\bm{k})  &= 0\,, \\
\us{F_1^\psi(\bm{k})}&\us{=-\frac{5}{2 k^2}\,,} \\
F_{2}^{R}(\bm{k}_1,\bm{k}_2) & =  -\frac{5}{2}\frac{(k_1^2+k_2^2)}{k_1^2k_2^2}+\frac{5}{4}\frac{\bm{k}_1\cdot\bm{k}_2}{k_1^2k_2^2}\,, \label{F2RSync} \\
F_{2}^\psi(\bm{k}_1,\bm{k}_2) & =  \frac{1}{4 k_{12}^{2}}\left[1-6F_2^N(\bm{k}_1,\bm{k}_2)-4G_2^N(\bm{k}_1,\bm{k}_2 )-\frac{(\bm{k}_1\cdot\bm{k}_2)^2}{k_1^2k_2^2}\right]\,,\\
G_{2}^{R}(\bm{k}_1,\bm{k}_2) & =   F_{2}^{R}(\bm{k}_1,\bm{k}_2)-3F_{2}^{\psi}(\bm{k}_1,\bm{k}_2)-\frac{15}{2}\frac{\bm{k}_1\cdot\bm{k}_2}{k_1^2k_2^2}\,, \\ \label{G2RSync}
\bm{G}^{T}_{2}(\bm{k}_1,\bm{k}_2) & = i \left[ 4\frac{\bm{k}_{12}}{k^2_{12}} F_{2}^{\psi}(\bm{k}_1,\bm{k}_2)+ \frac{5}{k_2^2}\left(-\frac{\bm{k}_1}{k_1^2} + \frac{\bm{k}_{12}}{k_{12}^2}\frac{\bm{k}_{12}\cdot\bm{k}_1}{k_1^2} + \frac{\bm{k}_1}{k_1^2} \frac{k_2^2}{k_{12}^2} + \frac{\bm{k}_1\cdot\bm{k}_2}{k_1^2}\frac{\bm{k}_2}{k_{12}^2}\right) \right]\,,\\
F_{3}^R(\bm{k}_1,\bm{k}_2,\bm{k}_3) & = \frac{5}{7}\alpha(\bm{k_1},\bm{k}_{23})F_2^R(\bm{k}_2,\bm{k}_3)+G_2^R(\bm{k}_2,\bm{k}_3)\left(\frac{5}{7}\alpha(\bm{k}_{23},\bm{k}_1)+\frac{4}{7}\beta(\bm{k}_1,\bm{k}_{23})\right)\nonumber \\
& +F_{2}^{\psi}(\bm{k}_2,\bm{k}_3)\left(\frac{19}{7}+\frac{6}{7}\frac{\bm{k}_1\cdot\bm{k}_{23}}{k_1^2}\right)+\frac{95}{14}\frac{(\bm{k}_1\cdot\bm{k}_3)}{k_{1}^2 k_3^2}\nonumber \\
&-i\frac{1}{7}\bm{G}^{T}_{2}(\bm{k}_1,\bm{k}_3)\cdot\bm{k}_2\left(7+4\frac{\bm{k}_{13}\cdot\bm{k}_2}{k_2^2}\right)+\frac{15}{7}\frac{(\bm{k}_1\cdot\bm{k}_2)(\bm{k}_2\cdot\bm{k}_3)}{k_{1}^2 k_{2}^2k_3^2}\,,\\ \label{F3R}
F_{3}^{\psi}(\bm{k}_1,\bm{k}_2,\bm{k}_3) & =\frac{1}{2k_{123}^2}\left[G_2^N(\bm{k}_1,\bm{k}_{3})\left(1-\frac{(\bm{k}_{13}\cdot\bm{k}_{2})^2}{k_{13}^2\bm{k}^2_{2}}\right)-3F_3^N\bm{k}_1,\bm{k}_2\bm{k}_3)-2G_3^N\bm{k}_1,\bm{k}_2\bm{k}_3) \right]\,, \\
G_{3}^R(\bm{k}_1,\bm{k}_2,\bm{k}_3) & =  2F_{3}^{R}(\bm{k}_1,\bm{k}_2,\bm{k}_3) -6F_{3}^{\psi}(\bm{k}_1,\bm{k}_2,\bm{k}_3)-\alpha(\bm{k}_1,\bm{k}_{23})F_{2}^{R}(\bm{k}_2,\bm{k}_3)\nonumber\\
&-\alpha(\bm{k}_{13},\bm{k}_2)G_{2}^{R}(\bm{k}_1,\bm{k}_3)-\frac{15}{2}\frac{\bm{k}_1\cdot\bm{k}_3}{k_1^2k_3^2}-\frac{15}{2}G_2^N\bm{k}_{1},\bm{k}_3)\frac{\bm{k}_{13}\cdot\bm{k}_2}{k_{13}^2k_2^2}\nonumber \\
& -3F_{2}^{\psi}(\bm{k}_2,\bm{k}3)\left(1-\frac{\bm{k}_1\cdot\bm{k}_{23}}{k_1^2}\right)+i\bm{k}_2\cdot\bm{G}^T_{2}\,,\\
\bm{G}^T_{3}(\bm{k}_1,\bm{k}_2,\bm{k}_3) &  = i \left[ \frac{8\bm{k}_{123}}{k_{123}^2}F_3^{\psi}(\bm{k}_1,\bm{k}_2,\bm{k}_3)\right. \nonumber \\
&+\frac{5}{k_2^2}G_2^N(\bm{k}_1,\bm{k}_3)\left(-\frac{\bm{k}_{13}}{k_{13}^2}+\frac{\bm{k}_{123}}{k_{123}^2}\frac{\bm{k}_{123}\cdot\bm{k}_{13}}{k_{13}^2}+\frac{\bm{k}_{13}}{k_{13}^2}\frac{k_2^2}{k_{123}^2}+\frac{\bm{k}_{13}\cdot\bm{k}_2}{k_{123}^2}\right)\nonumber\\
&\left.-2 F_2^\psi(\bm{k}_2,\bm{k}_3)\left(-\frac{\bm{k}_1}{k_1^2}+\frac{\bm{k}_{123}}{k_{123}^2}\frac{\bm{k}_{123}\cdot\bm{k}_1}{k_1^2}+\frac{\bm{k}_1}{k_1^2}\frac{\bm{k}_{23}^2}{k_{123}^2}+\frac{\bm{k}_1\cdot\bm{k}_{23}}{k_1^2}\frac{\bm{k}_{23}}{k_{123}^2}\right)\right]\,,\\
F_4^R(\bm{k}_1,\bm{k}_2,\bm{k}_3,\bm{k}_4) & =  \frac{7}{18}\alpha\left(\bm{k}_1,\bm{k}_{234}\right)F_3^R\left(\bm{k}_2,\bm{k}_3,\bm{k}_4\right) + \frac{7}{18} \alpha\left(\bm{k}_{12},\bm{k}_{34}\right)G_2^N\left(\bm{k}_1,\bm{k}_2\right) F_2^R\left(\bm{k}_3,\bm{k}_4 \right)\nonumber \\
&+\frac{7}{18}\alpha\left(\bm{k}_{134},\bm{k}_2\right)G_3^R\left(\bm{k}_1,\bm{k}_3,\bm{k}_4\right)+\frac{7}{18} \alpha\left(\bm{k}_{12},\bm{k}_{34}\right)G_2^R\left(\bm{k}_1,\bm{k}_2\right) F_2^N\left(\bm{k}_3,\bm{k}_4 \right)\nonumber\\
&+\frac{2}{9}\beta\left(\bm{k}_{134},\bm{k}_2\right)G_3^R\left(\bm{k}_1,\bm{k}_3,\bm{k}_4\right)+\frac{2}{9}\beta\left(\bm{k}_{12},\bm{k}_{34}\right)G_2^N\left(\bm{k}_1,\bm{k}_2\right)G_2^R\left(\bm{k}_3,\bm{k}_4\right)\nonumber\\
&+\frac{1}{18}F_{2}^{\psi}(\bm{k}_3,\bm{k}_4)\left[9F_2^N\left(\bm{k}_1,\bm{k}_2\right)+G_2^N\left(\bm{k}_1,\bm{k}_2\right)\left(4+6\frac{\bm{k}_{12}\cdot\bm{k}_{34}}{k_{12}^2}\right)\right] \nonumber\\
&+ \frac{5}{36}G_2^N\left(\bm{k}_1,\bm{k}_4\right)\left(29\frac{\bm{k}_{14}\cdot\bm{k}_3}{k_{14}^2k_3^2}+12\frac{(\bm{k}_{14}\cdot\bm{k}_2)(\bm{k}_2\cdot\bm{k}_3)}{\bm{k}^2_{14}k_3^2k_2^2}\right)\nonumber\\
& -\frac{1}{18}F_{2}^{\psi}(\bm{k}_3,\bm{k}_4)\left(\frac{(\bm{k}_1\cdot\bm{k}_{34})}{k_1^2}+6\frac{(\bm{k}_1\cdot\bm{k}_2)(\bm{k}_2\cdot\bm{k}_{34})}{k_1^2k_2^2}\right)\nonumber\\
&-i\frac{1}{18}\bm{G}_{3}^T(\bm{k}_1,\bm{k}_3,\bm{k}_4)\cdot\bm{k}_2\left(9+4\frac{\bm{k}_{134}\cdot\bm{k}_2}{k_2^2}\right)+\frac{105}{36}\frac{\bm{k}_1\cdot\bm{k}_3}{k_1^2k_3^2}F_2^N\left(\bm{k}_2,\bm{k}_4\right)\nonumber\\
&-i\frac{1}{18}\bm{G}_{2R}^{T}(\bm{k}_1,\bm{k}_3)\cdot\bm{k}_{24}\left(7F_2^N(\bm{k}_2,\bm{k}_4)+2G_2^N(\bm{k}_2,\bm{k}_4)\frac{\bm{k}_{13}\cdot\bm{k}_{24}}{k_{24}^2}\right).
\end{align}

\subsection{Newtonian galaxy bias kernels}\label{app:NBkernels}

\begin{align}
M_2^{\delta,N}(\bm{k}_1,\bm{k}_2)&=\;F_2^N(\bm{k}_1,\bm{k}_2)+\frac{4}{21}-\frac{2}{7}s^2(\bm{k}_1,\bm{k}_2)\,, \\
M_2^{\delta^2,N}(\bm{k}_1,\bm{k}_2)&=\;\frac{1}{2}\,, \\
M_2^{s^2,N}(\bm{k}_1,\bm{k}_2)& =\; s^2(\bm{k}_1,\bm{k}_2)\,,\\
M_3^{\delta,N}(\bm{k}_1,\bm{k}_2,\bm{k}_3)&=\; \frac{1}{2} \left[G_2^N(\bm{k}_1,\bm{k}_2) \alpha(\bm{k}_{12},\bm{k}_3)+F_2^N(\bm{k}_1,\bm{k}_2) \alpha(\bm{k}_3,\bm{k}_{12})\right.\nonumber\\
&\;\left.+\left(\frac{4}{21}-\frac{2}{7} s^2(\bm{k}_1,\bm{k}_2)\right) \alpha(\bm{k}_3,\bm{k}_{12})\right]\,,\\
M_3^{\delta^2,N}(\bm{k}_1,\bm{k}_2,\bm{k}_3)&=\; \frac{1}{2}\alpha(\bm{k}_1,\bm{k}_{23})\,,\\
M_3^{s^2,N}(\bm{k}_1,\bm{k}_2,\bm{k}_3)&=\; s^2(\bm{k}_2,\bm{k}_3)\alpha(\bm{k}_1,\bm{k}_{23})\,,\\
M_3^{\delta^3,N}(\bm{k}_1,\bm{k}_2,\bm{k}_3)&=\;\frac{1}{6}\,,\\
M_3^{s^3,N}(\bm{k}_1,\bm{k}_2,\bm{k}_3)&=\;s^3(\bm{k}_1,\bm{k}_2,\bm{k}_{3})\,,\\
M_3^{\delta s^2,N}(\bm{k}_1,\bm{k}_2,\bm{k}_3)&=\; \frac{1}{3}\left[s^2(\bm{k}_1,\bm{k}_2)+s^2(\bm{k}_1,\bm{k}_3)+s^2(\bm{k}_2,\bm{k}_3)\right]\,,\\
M_4^{\delta,N}(\bm{k}_1,\bm{k}_2,\bm{k}_3,\bm{k}_4)&=\; \frac{1}{3}
M_3^{\delta,N}(\bm{k}_2,\bm{k}_3,\bm{k}_4) \alpha(\bm{k}_1,\bm{k}_{234})+\frac{1}{3} M_2^{\delta,N}(\bm{k}_2,\bm{k}_4)
G_2^N(\bm{k}_1,\bm{k}_3) \alpha(\bm{k}_{13},\bm{k}_{24})\,,\\
M_4^{\delta^2,N}(\bm{k}_1,\bm{k}_2,\bm{k}_3,\bm{k}_4)&=\; \frac{1}{2}\left[
M_3^{\delta^2,N}(\bm{k}_2,\bm{k}_3,\bm{k}_4) \alpha(\bm{k}_1,\bm{k}_{234})+ M_2^{\delta^2,N}(\bm{k}_2,\bm{k}_4)
G_2^N(\bm{k}_1,\bm{k}_3) \alpha(\bm{k}_{13},\bm{k}_{24})\right] \,, \\
M_4^{s^2,N}(\bm{k}_1,\bm{k}_2,\bm{k}_3,\bm{k}_4)&=\; \frac{1}{2}\left[
M_3^{s^2,N}(\bm{k}_2,\bm{k}_3,\bm{k}_4) \alpha(\bm{k}_1,\bm{k}_{234})+M_2^{s^2,N}(\bm{k}_2,\bm{k}_4)
G_2^N(\bm{k}_1,\bm{k}_3) \alpha(\bm{k}_{13},\bm{k}_{24})\right]\,,\\
M_4^{\delta^3,N}(\bm{k}_1,\bm{k}_2,\bm{k}_3,\bm{k}_4)&=\; M_3^{\delta^3,N}(\bm{k}_2,\bm{k}_3,\bm{k}_4) \alpha(\bm{k}_1,\bm{k}_{234})\,,\\
M_4^{s^3,N}(\bm{k}_1,\bm{k}_2,\bm{k}_3,\bm{k}_4)&=\; M_3^{s^3,N}(\bm{k}_2,\bm{k}_3,\bm{k}_4) \alpha(\bm{k}_1,\bm{k}_{234})\,,\\
M_4^{\delta s^2,N}(\bm{k}_1,\bm{k}_2,\bm{k}_3,\bm{k}_4)&=\; M_3^{\delta s^2,N}(\bm{k}_2,\bm{k}_3,\bm{k}_4) \alpha(\bm{k}_1,\bm{k}_{234})\,,\\
M_4^{\delta^4,N}(\bm{k}_1,\bm{k}_2,\bm{k}_3,\bm{k}_4)&=\; \frac{1}{24} \,,\\
M_4^{\delta^2 s^2,N}(\bm{k}_1,\bm{k}_2,\bm{k}_3,\bm{k}_4)&=\; s^2(\bm{k}_1,\bm{k}_2)\,,\\
M_4^{\delta s^3,N}(\bm{k}_1,\bm{k}_2,\bm{k}_3,\bm{k}_4)&=\; s^3(\bm{k}_1,\bm{k}_2,\bm{k}_3) \,,\\
M_4^{(s^2)^2,N}(\bm{k}_1,\bm{k}_2,\bm{k}_3,\bm{k}_4)&=\;s^2(\bm{k}_1,\bm{k}_2)s^2(\bm{k}_3,\bm{k}_4) \,,\\
M_4^{s^4,N}(\bm{k}_1,\bm{k}_2,\bm{k}_3,\bm{k}_4)&=\;\frac{\bm{k}_1.\bm{k}_2 \bm{k}_1.\bm{k}_4 \bm{k}_2.\bm{k}_3 \bm{k}_3.\bm{k}_4}{k_1^2 k_2^2 k_3^2 k_4^2} - \frac{4}{3}\frac{\bm{k}_2.\bm{k}_4 \bm{k}_2.\bm{k}_3 \bm{k}_3.\bm{k}_4}{k_2^2 k_3^2 k_4^2} + \frac{2}{3}\frac{(\bm{k}_3.\bm{k}_4)^2}{k_3^2 k_4^2}- \frac{1}{9}\,.
\end{align}

\subsection{Relativistic galaxy bias kernels}\label{app:RBkernels}

The non-symmetrized relativistic galaxy bias kernels are
\begin{align}
M_2{}^{\delta,R}(\bm{k}_1,\bm{k}_2)=& F_{2}^R(\bm{k}_1,\bm{k}_2) \,, \label{eq:M2Rb1} \\
M_3{}^{\delta,R}(\bm{k}_1,\bm{k}_2,\bm{k}_3) = & \; \frac{15}{2} \frac{\bm{k}_1 \cdot \bm{k}_3}{ k_1 ^2 k_3 ^2} -i \bm{k}_2 \cdot \bm{G}_{2}^T(\bm{k}_1,\bm{k}_3)
+ 3  F_2^{ \psi} (\bm{k}_2,\bm{k}_3)  \label{eq:M3Rb1}\\
&  + F_{2}^R(\bm{k}_2,\bm{k}_3) \alpha (\bm{k}_1,\bm{k}_{23})+G_2^R(\bm{k}_2,\bm{k}_3)\alpha (\bm{k}_{23},\bm{k}_1) \,,\nonumber \\
M_3{}^{\delta^2,R}(\bm{k}_1,\bm{k}_2,\bm{k}_3) = &\; \frac{1}{3}\left[F_{2}^R(\bm{k}_1,\bm{k}_2)+F_{2}^R(\bm{k}_1,\bm{k}_3)+F_{2}^R(\bm{k}_2,\bm{k}_3)\right] \label{eq:M3Rb2} \,, \\
M_3{}^{s^2,R}(\bm{k}_1,\bm{k}_2,\bm{k}_3)  = &\; 2 \left[ G_2^{R}(\bm{k}_2,\bm{k}_3)s^2(\bm{k}_1,\bm{k}_{23})  -i  \bm{k}_2 \cdot \bm{G}_{2}^T(\bm{k}_1,\bm{k}_3) \frac{( \bm{k}_{13})\cdot \bm{k}_2 }{k_2^2} \right] \label{eq:M3Rbk} \,, \\
M_4^{\delta,R}(\bm{k}_1,\bm{k}_2,\bm{k}_3,\bm{k}_4) = & \; \frac{1}{2} \left[-i  \frac{4}{21} \bm{k}_{24}\cdot \bm{G}_{2}^T(\bm{k}_1,\bm{k}_3) +\bm{k}_2 \cdot \bm{G}_{3}^T(\bm{k}_1,\bm{k}_3,\bm{k}_4)  +\frac{10}{7} \frac{\bm{k}_1 \cdot \bm{k}_3}{k_1^2 k_2^2} \right.  \nonumber \\
& -i \bm{k}_{24} \cdot \bm{G}_{2}^T(\bm{k}_1,\bm{k}_3)F_2^N(\bm{k}_2,\bm{k}_4) + \frac{15}{2} F_2 (\bm{k}_2,\bm{k}_4) \frac{\bm{k}_1 \cdot \bm{k}_3}{k_1^2 k_3^2} \nonumber\\
& + \frac{4}{7}F_2^{ \psi} (\bm{k}_2,\bm{k}_4) + 3 F_2^N(\bm{k}_1,\bm{k}_3) F_2^{ \psi} (\bm{k}_2,\bm{k}_4) - 3 F_2^{ \psi} (\bm{k}_3,\bm{k}_4) \frac{ \bm{k}_1 \cdot\bm{k}_{34}}{k_1^2} \nonumber\\
& + 6 F_3^{ \psi} (\bm{k}_2, \bm{k}_3,\bm{k}_4) + \frac{15}{2} G_2^N(\bm{k}_1,\bm{k}_4) \frac{ \bm{k}_{14}\cdot \bm{k}_3}{(\bm{k}_1 + \bm{k}_4)^2 k_3^2}- \frac{6}{7} F_2^{ \psi}(\bm{k}_2,\bm{k}_4) s^2(\bm{k}_1,\bm{k}_3) \nonumber\\
& +i\frac{2}{7} ( \bm{k}_2 + \bm{k}_4) \cdot \bm{G}_{2}^T(\bm{k}_1, \bm{k}_3) s^2(\bm{k}_2,\bm{k}_4) - \frac{15}{7} s^2(\bm{k}_2,\bm{k}_4) \frac{ \bm{k}_1 \cdot \bm{k}_3}{k_1^2 k_3^2}   \nonumber\\
& + M_3{}^{\delta,R} (\bm{k}_2,\bm{k}_3,\bm{k}_4) \alpha (\bm{k}_1,\bm{k}_{234}) + \frac{4}{21} G_2^R(\bm{k}_2,\bm{k}_4) \alpha (\bm{k}_{24},\bm{k}_{13}) \nonumber\\
&  + F_2^N(\bm{k}_1,\bm{k}_3) G_2^R(\bm{k}_2,\bm{k}_4) \alpha (\bm{k}_{24}, \bm{k}_{13}) + G_3^R (\bm{k}_2,\bm{k}_3,\bm{k}_4) \alpha (\bm{k}_{234}, \bm{k}_1) \nonumber\\
&  - \frac{2}{7} G_2^R(\bm{k}_2,\bm{k}_4) s^2(\bm{k}_1,\bm{k}_3)\alpha (\bm{k}_{24}, \bm{k}_{13})   \nonumber\\
&  \left. + G_2^N(\bm{k}_1,\bm{k}_3)M_2{}^{\delta,R}(\bm{k}_2,\bm{k}_4) \alpha (\bm{k}_{13} , \bm{k}_{24})  \right]\,, \\
M_4^{\delta^2,R}(\bm{k}_1,\bm{k}_2,\bm{k}_3,\bm{k}_4)  = & \;\frac{1}{2}\Bigg[-i\bm{k}_{24}\cdot \bm{G}_{2}^T(\bm{k}_1,\bm{k}_3) +\frac{15}{2} \frac{\bm{k}_1 \cdot \bm{k}_3}{k_1^2 k_ 3^2} +3 F_2^{\psi} (\bm{k}_2,\bm{k}_4)\nonumber  \\
& + G_2^R (\bm{k}_2,\bm{k}_3) \alpha (\bm{k}_{23},\bm{k}_{14})  +M_3{}^{\delta^2,R}(\bm{k}_2,\bm{k}_3,\bm{k}_4) \alpha (\bm{k}_1,\bm{k}_{234})\Bigg] \,,\\
M_4^{s^2,R}(\bm{k}_1,\bm{k}_2,\bm{k}_3,\bm{k}_4)  = & \; 3 F_2 ^{\psi} (\bm{k}_2,\bm{k}_4) s^2(\bm{k}_1,\bm{k}_3) -i \bm{k}_{24}\cdot\bm{G}_{2}^T (\bm{k}_1,\bm{k}_3)s^2(\bm{k}_2,\bm{k}_4) \nonumber \\
& + \frac{15}{2} s^2(\bm{k}_2,\bm{k}_4) \frac{\bm{k}_1\cdot \bm{k}_3}{k_1^2 k_3^2}+ G_2^R(\bm{k}_2,\bm{k}_4) s^2(\bm{k}_1,\bm{k}_3)^2 \alpha (\bm{k}_{24},\bm{k}_{13}) \nonumber\\
& +  M_3{}^{s^2,R}(\bm{k}_2,\bm{k}_3,\bm{k}_4) \alpha(\bm{k}_1,\bm{k}_{234}) \,,\\
M_4{}^{\delta^3,R}(\bm{k}_1,\bm{k}_2,\bm{k}_3,\bm{k}_4) =&\frac{1}{2}F_2^R (\bm{k}_1,\bm{k}_2) \label{eq:M4Rb3} \,, \\
M_4{}^{s^3,R}(\bm{k}_1,\bm{k}_2,\bm{k}_3,\bm{k}_4) =&3s^3(\bm{k}_2,\bm{k}_3,\bm{k}_{14}) G_2^R(
\bm{k}_1,\bm{k}_4)\nonumber\\
&-i3 \left(-\frac{2}{3}\frac{\bm{k}_3\cdot \bm{k}_{14}}{k_3^2}+\frac{\bm{k}_2\cdot \bm{k}_3\bm{k}_{14}\cdot \bm{k}_2}{k_2^2 k_3^2}\right)\bm{k}_3\cdot\bm{G}_{2}^T(\bm{k}_1,\bm{k}_4) \label{eq:M4Rbs3}\,,\\
M_4{}^{\delta s^2,R}(\bm{k}_1,\bm{k}_2,\bm{k}_3,\bm{k}_4) =&i\bm{k}_4\cdot\bm{G}_{2}^T(
\bm{k}_2,\bm{k}_3)\frac{2 \bm{k}_4\cdot \bm{k}_{23}}{\bm{k}_4\cdot \bm{k}_4}+
F_2^R(\bm{k}_1,\bm{k}_4)
s^2(\bm{k}_2,\bm{k}_3)\nonumber\\
&+2 G_2^R(\bm{k}_2,\bm{k}_3)
s^2(\bm{k}_{23},\bm{k}_4) \label{eq:M4Rbds2}\,.
\end{align}

\section{Consistency relation for bias evolution}\label{CRB}

We check our calculation of the bias evolution kernels by verifying that they satisfy the consistency relations. Consistency relations are based on the fact that the effect of  a long  mode on the short modes is equivalent to a  linear coordinate transformation \cite{Creminelli:2013mca}. Under a coordinate transformation  $N$ point correlation functions are related to  $N-1$ correlation function, which is the consistency check that all the kernels must satisfy.  We correlate de galaxy density with matter density perturbations.

The  coordinate transformation in comoving gauge is given by
\begin{align}
\tilde{\eta} & = \eta\,, \label{eq:coordtran} \\
\tilde{x}^{i} & = (1 + \zeta)x^i +\frac{1}{2}\left(2x^ix^j\partial_j\zeta -  x^2 \partial_i\zeta\right) - \frac{1}{5} \eta^2 \partial_i\zeta \,,\label{coordinatetr}
\end{align}
where $\zeta$ is the curvature perturbation, set at the initial conditions (e.g. by inflation). \us{The first term in \eqref{coordinatetr} corresponds to dilations while the second terms is due to special conformal transformations. The last term is added to preserve the gauge conditions.}

Since time does not transform, the galaxy bias solution satisfies the same relation as the dark matter density solution.  Under the coordinate transformation, the N-point function in Fourier space is
\begin{align}
\left\langle \delta(\eta,\bm{q})\delta_{g}(\eta_1,\bm{k}_1)\cdots\delta(\eta_n,\bm{k}_n)\right\rangle'_{q\rightarrow{0}} & =-\frac{5H_0^2}{2q^2}a(\eta)P(q)\left[ 3(n-1)+\sum_a \bm{k}_a\cdot\bm{\partial}_{k_a}\right.\nonumber \\
& \left.+\frac{1}{2}q^iD_i-\frac{1}{5}\sum_a \bm{q}\cdot\bm{k}_a\eta^2_a\right]\left\langle\delta_{g}(\eta_1,\bm{k}_1)\cdots\delta(\eta_n,\bm{k}_n)\right\rangle', \label{CRS}
\end{align}
where we used the relation \us{$\zeta(\bm{k}) =\frac{5}{2k^2}\delta_l(\bm{k})$}, and
\begin{equation}
q^iD_i\equiv\sum_{a=1}^n\left[6\bm{q}\cdot\bm{\partial}_{k_a}-\bm{q}\cdot\bm{k}_a\bm{\partial}^2_{k_a}+2\bm{k}_a\cdot\bm{\partial}_{k_a}(\bm{q}\cdot\bm{\partial}_{k_a})\right].
\end{equation}
For simplicity we compute the consistency relation at equal times. In this case the last term in equation \eqref{CRS} is zero at all orders.  Since the relativistic solution for the bias evolution starts to contribute to second order, we begin by checking the consistency relation for $n=2$. Thus,  the l.h.s   of  equation \eqref{CRS} at equal times for $n=2$ is
\begin{align}
\left\langle \delta(\bm{q})\delta(\bm{k}_1)\delta(\bm{k}_2)\right\rangle'_{q\rightarrow{0}} & =\sum_\mathcal{O}2H_0^2b_{\mathcal{O}}^\mathcal{L}\left[M_{2}^{\mathcal{O},R}(-\bm{q},\bm{k}_1 + \bm{q})P(|\bm{k}_1 + \bm{q}|) + M_{2}^{\mathcal{O},R}(-\bm{q},-\bm{k}_1)P(k_1)\right]P(q) \nonumber\\
&=\sum_\mathcal{O} 2H_0^2 b_\mathcal{O}^\mathcal{L} \left[M_{2}^{\mathcal{O},R}(-\bm{q},\bm{k}_1)\left(1+\frac{\bm{q}\cdot\bm{k}_1}{k_1^2}\right)+M_{2}^{\mathcal{O},R}(-\bm{q},-\bm{k}_1)\right]P(q)P(k_1)\, \label{CR2L}
\end{align}
and for  r.h.s it gives
\begin{align}
\left\langle \delta(\bm{q})\delta_{g}(\bm{k}_1)\delta(\bm{k}_2)\right\rangle'_{q\rightarrow{0}} & =-\frac{5}{2q^2}\left[3+\sum_a \bm{k}_a\cdot\bm{\partial}_{k_a}+\frac{1}{2} q^iD_i\right]b_1^\mathcal{L} M_{1}^{b_1}(\bm{k}_1)P(k_1).
\end{align}

For $n=3$ the l.h.s is
\begin{align}
\left\langle\delta(\bm{q})\delta_{g}(\bm{k}_1)\delta(\bm{k}_2)\delta(\bm{k}_3)\right\rangle'_{q\rightarrow{0}} =&  \langle \delta^{(1)}(\bm{q})\delta^{(3)}_{g}(\bm{k}_1)\delta^{(1)}(\bm{k}_2)\delta^{(1)}(\bm{k}_3)\rangle + 2 \text{ perm} \nonumber \\
&\left\langle\delta^{(1)}(\bm{q})\delta^{(2)}_{g}(\bm{k}_1)\delta^{(2)}(\bm{k}_2)\delta^{(1)}(\bm{k}_3)\right\rangle + 2\text{ perm}.
\end{align}
As we did in \cite{Castiblanco:2018qsd}, it is sufficient  to compare only a combination of the momenta. We will compare the terms on both sides which are proportional to $P(k_2)P(k_3)$. The relation that we use is given by
\begin{align}
\left\langle\delta({\bm{q}})\delta_{g}(\bm{k}_1)\delta(\bm{k}_2)\delta_{g}(\bm{k}_3)\right\rangle'_{q\rightarrow{0}} \supset & P(k_2)P(k_3)\sum_i b_\mathcal{O}^\mathcal{L}\left[ 6M_{3}^{\mathcal{O},R}(-\bm{q},-\bm{k}_2,-\bm{k}_3)\right. \nonumber \\
&+4F_2^R(-\bm{q},\bm{q}+\bm{k}_2)M_{2}^{\mathcal{O},N}(-\bm{q}-\bm{k}_2,-\bm{k}_3)\left(1+\frac{\bm{q}\cdot\bm{k}_2}{k_{2}^{2}}\right) + \bm{k}_2\leftrightarrow\bm{k}_3 \nonumber \\
&\left.+4F_2^N(-\bm{q},\bm{q}+\bm{k}_2)M_{2R}^{\mathcal{O}}(-\bm{q}-\bm{k}_2,-\bm{k}_3)\left(1+\frac{\bm{q}\cdot\bm{k}_2}{k_{2}^{2}}\right)+ \bm{k}_2\leftrightarrow\bm{k}_3\right]\,, \label{CR3L}
\end{align}
and the r.h.s is
\begin{align}
\left\langle \delta(\bm{q})\delta_{g}(\bm{k}_1)\delta(\bm{k}_2)\delta_{g}(\bm{k}_3)\right\rangle'_{q\rightarrow{0}} & =-\frac{5}{2q^2}\left[6+\sum_a \bm{k}_a\cdot\bm{\partial}_{k_a}+\frac{1}{2} q^iD_i\right]\sum_\mathcal{O} b_\mathcal{O}^\mathcal{L} M_{2}^{\mathcal{O},N}(\bm{k}_2,\bm{k}_3)P(k_2)P(k_3)\,.\label{CR3R}
\end{align}

For $n=4$  we take the combination proportional to $P(k_2)P(k_3)P(k_4)$ on both sides of \ref{CRS}. The l.h.s reads
\begin{align}
\left\langle \delta(\bm{q})\delta_{g}(\bm{k}_1)\delta(\bm{k}_2)\delta(\bm{k}_3)\delta(\bm{k}_4)\right\rangle'_{q\rightarrow{0}}  = & 24 \sum_i b_\mathcal{O}^\mathcal{L} M_{4}^{\mathcal{O},R}(-\bm{q},-\bm{k}_2,-\bm{k}_3,-\bm{k}_4) \nonumber \\
&+	12\sum_\mathcal{O} b_\mathcal{O}^\mathcal{L} \left[F_2^R(-\bm{q} ,\bm{k}_2+\bm{q})\
M_{3}^{\mathcal{O},N}(-\bm{k}_2,-\bm{q} ,-\bm{k}_3,-\bm{k}_4)\right.\nonumber\\
&+F_2^R(-\bm{q} ,\bm{k}_3+\bm{q}) M_{3}^{\mathcal{O},N}(-\bm{k}_3,-\bm{q} ,-\bm{k}_2,-\bm{k}_4)\nonumber\\
&\left.+F_2^R(-\bm{q}
,\bm{k}_4+\bm{q}) M_{3}^{\mathcal{O},N}(-\bm{k}_4,-\bm{q}  ,-\bm{k}_3,-\bm{k}_2)\right]\,,\label{CR4L}
\end{align}
and the r.h.s is
\begin{multline}
\left\langle \delta(\bm{q})\delta_{g}(\bm{k}_1)\delta(\bm{k}_2)\delta(\bm{k}_3)\delta(\bm{k}_4)\right\rangle'_{q\rightarrow{0}} = \\-\frac{5}{2q^2}\left[9+\sum_a \bm{k}_a\cdot\bm{\partial}_{k_a}+\frac{1}{2} q^iD_i\right]\sum_\mathcal{O} b_\mathcal{O}^\mathcal{L} M_{3}^{\mathcal{O},N}(\bm{k}_2,\bm{k}_3,\bm{k}_4)P(k_2)P(k_3)P(k_4).\label{CR4R}
\end{multline}

As an example, we show here that the operator $S^2$ satisfies the consistency relation. We take all the bias parameters equal to zero but  $b_{s^2}^*$ and we check the relativistic evolution for the  operator $S^2$ which starts to contribute to third order. For n=3,  the limit $q\rightarrow0$ in equation \eqref{CR3L}, and the corresponding derivatives in equation \eqref{CR3R} give the same result, therefore the consistency relation is satisfied at this order. The result is given by
\begin{align}
\left\langle\delta(\bm{q})\delta_g(\bm{k}_1)\delta(\bm{k}_2)\delta(\bm{k}_3)\right\rangle' & = - \frac{1}{q^2}\left[40s^2(\bm{k}_2,\bm{k}_3)+\frac{25}{3}\left(\frac{\bm{k}_2\cdot\bm{q}}{k^2}+\frac{\bm{k}_3\cdot\bm{q}}{k^3}\right)-30\frac{\bm{k}_2\cdot\bm{k}_3}{k_2^2k_3^2}\bm{q}\cdot\left(\bm{k}_{23}\right)\right.\nonumber\\
&\left.+5\frac{(\bm{k}_2\cdot\bm{k}_3)^2}{k_2^2k_3^2}\left(\frac{\bm{k}_2\cdot\bm{q}}{k_2^2}+\frac{\bm{k}_3\cdot\bm{q}}{k_3^2}\right)\right]  P(k_2)P(k_3)\,.
\end{align}
For $n=4$ we only check the leading contribution to the consistency relation, given by the dilation transformation. The result for equations \eqref{CR4L} and \eqref{CR4R} in this case is
\begin{align}
\left\langle\delta(\bm{q})\delta_g(\bm{k}_1)\delta(\bm{k}_2)\delta(\bm{k}_3)\delta(\bm{k}_4)\right\rangle' & = - \frac{1}{q^2}\left[60s^2(\bm{k}_2,\bm{k}_3)+20\bm{k}_2\cdot\bm{k}_3\left(\frac{1}{k_2^2}+\frac{1}{k_3^2}\right)\right.\nonumber\\
&  - 60 \frac{(\bm{k}_2\cdot\bm{k}_3)^2(\bm{k}_2\cdot\bm{k}_4)}{k_2^2k_3^2k_4^2} + \bm{k}_3 \rightarrow\bm{k}_4 + \bm{k}_2 \rightarrow\bm{k}_4\left.\right]  P(k_2)P(k_3)P(k_4)
\end{align}
All the other relativistic galaxy bias solutions also satisfy the consistency relation. For $n=4$ we only check the dilation part of the consistency relation.


\us{\section{Conformal Fermi Coordinates and Comoving gauge}
\label{app:CFCconsistency}
In this Appendix, we show that the coordinate transformation \eqref{eq:coordtran}-\eqref{coordinatetr} used to compute the consistency relation, which eliminates the long mode in the local frame in comoving gauge, is equivalent to the coordinate transformation that goes to Conformal Fermi Coordinates (CFC) described in Refs.~\cite{Pajer:2013ana,Dai:2015rda}, up to first derivatives of the metric perturbations. This has already been shown in the synchronous gauge by~\cite{Umeh:2019jqg}.
We start writing the metric in comoving gauge at linear order in the initial curvature perturbation as  \cite{Boubekeur:2008kn}
\begin{equation}
ds^2 = a(\eta)^{2}\left\{-d\eta^2-\frac{4}{5aH}\partial_i\zeta d\eta dx^i+\left[(1+2\zeta)\delta_{ij}+\gamma_{ij}\right]dx^idx^j\right\}. \label{linearmetric}
\end{equation} 
Following the steps presented in \cite{Dai:2015rda}, the coordinate transformation between the global coordinate system and the CFC coordinates to quadratic order is given by
\begin{equation}
x^{\mu}(x^i_F) = x^{\mu}(P) + a_F(\eta_F)(e_i)^\mu x_F^i -\frac{1}{2} a_F^2(\eta_F)\left(\tilde{\Gamma}^{\mu}_{\alpha\beta}\right)_P(e_i)^\alpha(e_j)^\beta_Px_F^ix_F^j,
\end{equation}
where $P$ is the point on the central geodesic of a local observer with Fermi coordinates $\{\eta_F,\bm{0}\}$ and the Christoffel symbols are computed on the central geodesic with the conformal metric $\tilde{g}_{\mu,\nu}=a^{-2}(\eta)g_{\mu\nu}$. The tetrads in this case are written
\begin{align}
(e_0)^\mu&=u^\mu = \frac{1}{a}\left(1,v^i\right),\\ 
(e_i)^\mu  &= \left(0,\frac{1}{a}(1-\zeta)\delta_{ij}\right).
\end{align}
Here, $v^i$ is related to the off-diagonal part of the metric \eqref{linearmetric}  as $v^i = \frac{2}{5a^2H}\partial_i\zeta$. The observer is thus not at rest and the spatial part of the coordinates contains an extra piece $x^i(P)=a v^i(x_F^0-\eta_F)$. Thus, the coordinate transformation is
\begin{align}
x^0 (x_F)&= x_F^0 -\frac{2}{5aH}  \partial_i\partial_j \zeta x_F^ix_F^j, \\
x^i (x_F)&= \frac{a_F(\eta_F)}{a(\eta)}(1-\zeta)x^i + \frac{1}{2}\frac{a_F^2}{a^2}\left[\partial_{i} \zeta x^2 - 2\partial_j \zeta x^i x^j\right] +\frac{1}{5}\eta^2\partial_i \zeta.
\end{align}
Taking into account that in CFC coordinates the Hubble factor is defined from $H_F=\frac{1}{3}\nabla_\mu u^{\mu}$ to linear order in perturbations, the scale factor in CFC coordinates is related to the Hubble factor in the comoving gauge as
\begin{equation}
\frac{a_F(\eta_F)}{a(\eta)} = 1 -\frac{1}{3} \partial^2\zeta\int d\eta\left(\frac{2}{5aH}\right).
\end{equation}
The inverse coordinate transformation up to first derivatives on $\zeta$ is then given by \footnote{We take the limit of vanishing local curvature, therefore we ignore second derivatives on the initial perturbation $\zeta$ since physically they induce curvature \cite{Creminelli:2013cga}}
\begin{align}
x_F^0 (\eta,\bm{x})&= \eta, \\
x_F^i (\eta,\bm{x})&= (1+\zeta)x^i - \frac{1}{2} \partial_{i} \zeta x^2 +\partial_j \zeta x^i x^j -\frac{1}{5}\eta^2\partial_i \zeta.
\end{align}
We can see this is the same coordinate transformation obtained from the symmetries of the metric in comoving gauge. \eqref{eq:coordtran}-\eqref{coordinatetr}.
}

\section{Renormalization of operators}
\label{app:ernormOps}

We continue the work of section \ref{sec:renorm}, where we presented the renormalization of the terms proportional to $b_2^*$. We now turn to all the other operators.

\subsection*{Operator proportional to $b_1^*$}

Its initial conditions are given by
\be
\frac{b_1^*}{a}\delta_*(\bm{k}) = b_1^* \delta_\ell (\bm{k}) + b_1^* a_*^3 H_*^2 \int_{\bm{q}_1, \bm{q}_2} (2\pi)^3\delta_D(\bm{k} - \bm{q}_{12}) F_2^R(\bm{q}_1, \bm{q}_2) \delta_\ell(\bm{q}_1)\delta_\ell(\bm{q}_2)\,,
\ee
where the Newtonian non-linear terms have been ignored since they vanish in the early time limit $a \rightarrow 0$, but the relativistic corrections, given by eq.~\eqref{F2RSync}, survive in that limit. After renormalizing the background as discussed in \cite{Castiblanco:2018qsd}, the expectation value of this expression is zero as it should. Since cubic terms in the expansion vanish in the early time limit, no renormalization is necessary at the initial conditions, analogous to what happens in the Newtonian case. On the other hand, the non-linear evolution of this operator, dictated by eq.~\eqref{evolO}, gives non-zero correlators with long-wavelength perturbations at late time
\be
\lim_{k\rightarrow 0} \frac{b_1^*}{a}\la \delta_\ell(\bm{k})\delta(-\bm{k}, \eta) \ra = -\frac{40}{3} \frac{b_1^*}{a} a^5 H^2 \sigma^2_{-2} P(k)\,,
\ee
\begin{align}
\lim_{k_1\rightarrow 0} \lim_{k_2\rightarrow 0}\frac{b_1^*}{a}\left\langle \delta_\ell(\bm{k}_1) \delta_\ell(\bm{k}_2) \delta(\bm{k},\eta)\right\rangle' &=- \frac{b_1^*}{a} a^4 \left(\frac{58}{189} + \frac{101}{315}s^2(\bm{k}_1,\bm{k}_2) \right)\sigma^2 P(k_1) P(k_2) \nn \\
&\phantom{=} -\frac{b_1^*}{a} a^6 H^2 \left(\frac{3025}{63} + \frac{37}{210}s^2(\bm{k}_1,\bm{k}_2) \right)\sigma^2_{-2} P(k_1) P(k_2)\,,
\end{align}
which can be canceled by using the following counter-term kernels
\be
M_1^{\delta,c.t.}(\bm{k}) = \frac{40}{3} \frac{b_1^*}{a} a^5 H^2 \sigma^2_{-2}\,,
\ee
\be
M_2^{\delta,c.t.}(\bm{k}_1, \bm{k}_2) = \frac{b_1^*}{a} a^4 \left(\frac{58}{189} + \frac{101}{315}s^2(\bm{k}_1,\bm{k}_2) \right)\sigma^2 + \frac{b_1^*}{a} a^6 H^2 \left(\frac{3025}{63} + \frac{37}{210}s^2(\bm{k}_1,\bm{k}_2) \right)\sigma^2_{-2}\,.
\ee

\subsubsection*{Operator proportional to $b_{s^2}^*$}

Its initial conditions are given by
\begin{multline}
\frac{b_{s^2}^*}{a^2}(S^2_*)(\bm{k}) = b_{s^2}^* \int_{\bm{q}_1, \bm{q}_2} (2\pi)^3\delta_D(\bm{k} - \bm{q}_{12}) \left(\frac{(\bm{q}_1.\bm{q}_2)^2}{q_1^2 q_2^2} - \frac{1}{3}\right)\delta_\ell(\bm{q}_1)\delta_\ell(\bm{q}_2) \\
 + b_{s^2}^* a_*^3 H_*^2\int_{\bm{q}_1, \bm{q}_2, \bm{q}_3} (2\pi)^3\delta_D(\bm{k} - \bm{q}_{123}) M_3^{s^2,R}(\bm{q}_1, \bm{q}_2,  \bm{q}_3)\delta_\ell(\bm{q}_1)\delta_\ell(\bm{q}_2) \delta_\ell(\bm{q}_3)
\end{multline}
where $M_3^{s^2,R}$ is given in equation \eqref{eq:M3Rbk}, and is obtained by using equation \eqref{eq:Sij}. The expectation value of the bare operator is
\be
\frac{b_{s^2}^*}{a^2}\la (S^2_*) \ra = \frac{2}{3} b_{s^2}^* \sigma^2\,.
\ee
This leads to the renormalization $b_{s^2}^*/a^2 [(S^2)_*]_\Lambda = b_{s^2}^*/a^2 (S^2)_* - \frac{2}{3} b_{s^2}^* \sigma^2 + ...$. Next, we correlate with a long-wavelength perturbation
\be
\lim_{k\rightarrow 0}\frac{b_{s^2}^*}{a^2} \langle \delta_\ell(\bm{k}) (S^2)_*(\bm{k})\rangle' = \lim_{k\rightarrow 0}3\int_{\bm{q}} M^{s^2,R}_3(\bm{k},\bm{q},-\bm{q})P(q)P(k) = \left(-\frac{20}{3k^2}\sigma^2 + \frac{20}{9} \sigma_{-2}^2\right)a_*^3 H_*^2 b_{s^2}^* P(k)\,,
\ee
which is cancelled by writing
\be
\frac{b_{s^2}^*}{a^2}[(S^2)_*]_\Lambda = \frac{b_{s^2}^*}{a^2}(S^2)_* - \frac{2}{3} b_{s^2}^* \sigma^2  \us{+} \frac{8}{3}b_{s^2}^* \sigma^2 \zeta -\frac{20}{9} b_{s^2}^* a_*^3 H_*^2 \sigma_{-2}^2 \delta_\ell\,,
\label{eq:renormS2star}
\ee
where the term proportional to $\zeta$ is the same as discussed in section \ref{sec:renorm}, and combines with the first counter-term to form the same evaluated at the physical cutoff. We then use these as initial conditions for the solution of equation \eqref{evolO} to obtain operator $[(S^2)]$ evolved in time using equation \eqref{evolO} with initial conditions given by \eqref{eq:renormS2star}. One can check that it satisfies
$$
\la [(S^2)] \ra = 0\,,\quad \lim_{k\rightarrow 0} \la \delta_\ell(\bm{k}) [(S^2)](-\bm{k}) \ra = 0\,.
$$
However, there is still a cutoff dependence in the correlator
\be
\lim_{k_1\rightarrow 0} \lim_{k_2\rightarrow 0}\frac{b_{s^2}^*}{a^2}\left\langle \delta_\ell(\bm{k}_1) \delta_\ell(\bm{k}_2) [(S^2)](\bm{k},\eta)\right\rangle' = -\frac{5}{3} a^4 H^2 b_{s^2}^*\sigma^2_{-2}\left(5\frac{(\bm{k}_1.\bm{k}_2)^2}{k_1^2 k_2^2}+1\right)P(k_1) P(k_2)\,,
\ee
which has to be cancelled with an additional counter-term. The resulting counter-terms are
\be
M_0^{s^2, c.t.}(\bm{k}) = -\frac{2}{3} \frac{b_{s^2}^*}{a^2} a^2 \sigma^2(2\pi)^3 \delta_D(\bm{k})\,,
\ee
\be
M_1^{s^2, c.t.}(\bm{k}) = -\frac{2}{3} \frac{b_{s^2}^*}{a^2} a^3 \sigma^2  + \frac{20}{3k^2}\frac{b_{s^2}^*}{a^2} a^5H^2 \sigma^2 -\frac{20}{9} \frac{b_{s^2}^*}{a^2} a^5 H^2 \sigma_{-2}^2\,,
\ee
\begin{align}
M_2^{s^2, c.t.}(\bm{k}_1, \bm{k}_2) &= -\frac{1}{3} \frac{b_{s^2}^*}{a^2} a^4 \sigma^2\left(G_2^N(\bm{k}_1,\bm{k}_2) + \alpha(\bm{k}_1,\bm{k}_2)\right) \nn \\
&\phantom{=} + a^6 H^2 \frac{b_{s^2}^*}{a^2}\sigma^2_{-2} \left(-\frac{40}{9} + \frac{25}{6} s^2(\bm{k}_1, \bm{k}_2) -\frac{20}{9}\alpha(\bm{k}_1,\bm{k}_2)\right)\nn \\
&\phantom{=}-\frac{2}{3} \frac{b_{s^2}^*}{a^2} a^6 H^2 \sigma^2 \left(G_2^R(\bm{k}_1, \bm{k}_2) + 3 F_2^\psi(\bm{k}_1, \bm{k}_2) +\frac{15}{2} \frac{\bm{k}_1.\bm{k}_2}{k_1^2 k_2^2} - \frac{10}{k_2^2}\alpha(\bm{k}_1,\bm{k}_2) \right)\,.
\end{align}

\subsubsection*{Operator proportional to $b_3^*$}
The expression for this operator at initial time is
\begin{multline}
\frac{b_3^*}{6a^3} \delta^3_* = \frac{1}{6}b_3^*\int_{\bm{q}_1, \bm{q}_2,\bm{q}_3} \!\!\!(2\pi)^3 \delta_D(\bm{k} - \bm{q}_{123}) \delta_\ell(\bm{q}_1) \delta_\ell(\bm{q}_2) \delta_\ell(\bm{q}_3) \\
+ \frac{1}{6}b_3^*\int_{\bm{q}_1, \bm{q}_2,\bm{q}_3,\bm{q}_4} \!\!\!\!\!(2\pi)^3 \delta_D(\bm{k} - \bm{q}_{1234}) M_4^{\delta^3,R}(\bm{q}_1,\bm{q}_2,\bm{q}_3,\bm{q}_4) \delta_\ell(\bm{q}_1) \delta_\ell(\bm{q}_2) \delta_\ell(\bm{q}_3) \delta_\ell(\bm{q}_4)\,,
\end{multline}
where $M_4^{\delta^3,R}$ is given in equation \eqref{eq:M4Rb3}. The correlators give
\be
\frac{b_3^*}{6a^3} \lim_{k\rightarrow 0}\langle \delta_\ell(\bm{k}) \delta^3_*(\bm{k})\rangle' =  \frac{1}{2}\sigma^2b_3^* P(k)\,,
\ee
\begin{multline}
\lim_{k_1\rightarrow 0} \lim_{k_2 \rightarrow 0}\frac{b_3^*}{6a^3}\langle \delta_\ell(\bm{k}_1) \delta_\ell(\bm{k}_2) \delta^3_*(\bm{k})\rangle'  = \lim_{k_1\rightarrow 0}\lim_{k_2 \rightarrow 0}12\int_{\bm{q}} M^{\delta^3,R}_4(\bm{k}_1, \bm{k}_2, \bm{q},-\bm{q})P(q)P(k_1) P(k_2) \\
= \frac{1}{2}b_3^*a_*^3 H_*^2\left(2\sigma^2 F_2^R(\bm{k}_1, \bm{k}_2) -10\sigma^2\left(\frac{1}{k_1^2} + \frac{1}{k_2^2}\right) - 20 \sigma^2_{-2}\right)P(k_1)P(k_2)\,,
\end{multline}
which can be eliminated at second order by
\be
\frac{b_3^*}{6a^3}[\delta^3_*]_\Lambda = \frac{b_3^*}{6a^3} \delta^3_* -  \frac{1}{2}\frac{b_3^*}{a_*^3} a_*^2\sigma^2 \delta_* \us{+}  2\frac{b_3^*}{a_*^3} a_*^2\sigma^2 \delta_* \zeta + 10 b_3^* a_*^3 H_*^2 \sigma^2_{-2} \delta_\ell^2\,,
\ee
where $\delta_*$ is the dark matter solution at initial time, containing its linear and quadratic piece. We see that the counter-terms reshuffle into geometric operators of the initial time hypersurface as they should, and again we find the term proportional to $\zeta$ which corresponds to a rescaling of the cutoff. In this case there are no additional counter-terms needed for the evolved operator at the order at which we are working, and we find
\be
M_1^{\delta^3,c.t.}(\bm{k}) = -\frac{1}{2} \frac{b_3^*}{a^3} a^3 \sigma^2\,,
\ee
\be
M_2^{\delta^3,c.t.}(\bm{k}_1, \bm{k}_2) = -\frac{1}{2} \frac{b_3^*}{a^3} a^4 \sigma^2 \alpha(\bm{k}_1,\bm{k}_2)-\frac{1}{2}\frac{b_3^*}{a^3}a^6 H^2\left(\sigma^2 F_2^R(\bm{k}_1, \bm{k}_2) -5\sigma^2\left(\frac{1}{k_1^2} + \frac{1}{k_2^2}\right) - 10 \sigma^2_{-2}\right)\,.
\ee

\subsubsection*{Operator proportional to $b_{\delta s^2}^*$}

The expression for this operator at initial time is
\begin{multline*}
\frac{b_{\delta s^2}^*}{a^3} (S^2)_* \delta_*  = b_{\delta s^2}^*\int_{\bm{q}_1, \bm{q}_2,\bm{q}_3} \!\!\!(2\pi)^3 \delta_D(\bm{k} - \bm{q}_{123}) \left(\frac{(\bm{q}_1.\bm{q}_2)^2}{q_1^2 q_2^2} - \frac{1}{3}\right)\delta_\ell(\bm{q}_1) \delta_\ell(\bm{q}_2) \delta_\ell(\bm{q}_3) \\
+ b_{\delta s^2}^*\int_{\bm{q}_1, \bm{q}_2,\bm{q}_3,\bm{q}_4} \!\!\!\!\!(2\pi)^3 \delta_D(\bm{k} - \bm{q}_{1234}) M_4^{\delta s^2,R}(\bm{q}_1,\bm{q}_2,\bm{q}_3,\bm{q}_4) \delta_\ell(\bm{q}_1) \delta_\ell(\bm{q}_2) \delta_\ell(\bm{q}_3) \delta_\ell(\bm{q}_4)\,,
\end{multline*}
where $M_4^{\delta s^2,R}$ is given in equation \eqref{eq:M4Rbds2}. The correlators are
\be
\lim_{k\rightarrow 0}\frac{b_{\delta s^2}^*}{a^3} \langle \delta_\ell(\bm{k}) ((S^2)_*\delta_*)(\bm{k})\rangle' =  \frac{2}{3}\sigma^2b_{\delta s^2}^* P(k)\,,
\ee
\begin{multline}
\lim_{k_1\rightarrow 0} \lim_{k_2 \rightarrow 0}\frac{b_{\delta s^2}^*}{a^3}\langle \delta_\ell(\bm{k}_1) \delta_\ell(\bm{k}_2) ((S^2)_*\delta_*)(\bm{k})\rangle'  = \lim_{k_1\rightarrow 0}\lim_{k_2 \rightarrow 0}12\int_{\bm{q}} M^{\delta s^2,R}_4(\bm{k}_1, \bm{k}_2, \bm{q},-\bm{q})P(q)P(k_1) P(k_2) \\
= \frac{2}{3}b_{\delta s^2}^*a_*^3 H_*^2\left(2\sigma^2 F_2^R(\bm{k}_1, \bm{k}_2) -10\sigma^2\left(\frac{1}{k_1^2} + \frac{1}{k_2^2}\right) + 40\sigma^2_{-2}\right)P(k_1)P(k_2)\,,
\end{multline}
which can be eliminated at second order by
\be
\frac{b_{\delta s^2}^*}{a^3}[(S^2)_* \delta_*]_\Lambda = \frac{b_{\delta s^2}^*}{a^3}(S^2)_*\delta_* -  \frac{2}{3}\frac{b_{\delta s^2}^*}{a_*^3} a_*^2\sigma^2 \delta_* \us{+}  \frac{8}{3}\frac{b_{\delta s^2}^*}{a_*^3} a_*^2\sigma^2 \delta_* \zeta -\frac{40}{3} b_{\delta s^2}^* a_*^3 H_*^2 \sigma^2_{-2} \delta_*^2\,,
\ee
Evolving the operator renormalized at the initial hypersurface gives a new combination of terms that we call $[S^2 \delta ]$ for which all correlators with long-wavelength perturbations are zero, such that no new counter-terms are needed.
After reordering the terms, we finally obtain
\be
M_1^{\delta s^2,c.t.}(\bm{k}) = -  \frac{2}{3}\frac{b_{\delta s^2}^*}{a^3} a^3\sigma^2 \,,
\ee
\begin{multline}
M_2^{\delta s^2,c.t.}(\bm{k}_1, \bm{k}_2) = -\frac{2}{3} \frac{b_{\delta s^2}^*}{a^3} a^4 \sigma^2 \alpha^2(\bm{k}_1,\bm{k}_2)\\
-\frac{2}{3}\frac{b_{\delta s^2}^*}{a^3}a^6 H^2\left(\sigma^2 F_2^R(\bm{k}_1, \bm{k}_2) -5\sigma^2\left(\frac{1}{k_1^2} + \frac{1}{k_2^2}\right) - 20 \sigma^2_{-2}\right)\,.
\end{multline}

\subsubsection*{Operator proportional to $b_{s^3}^*$}

The calculation is analogous to the previous ones, and we find
$$
M_1^{s^3,c.t.}(\bm{k}) = 0\,, \quad
M_2^{s^3,c.t.}(\bm{k}_1, \bm{k}_2) =\frac{10}{3}\frac{b_{s^3}^*}{a^3}a^6 H^2\sigma^2_{-2} s^2(\bm{k}_1,\bm{k}_2)\,.
$$

\subsubsection*{Quartic terms}

The renormalization of quartic terms is straightforward at the order needed to compute the one-loop bispectrum. We need only require that the correlation with two long-wavelength perturbations at initial time cancel. We find
$$
M_2^{\delta^4,c.t.}(\bm{k}_1,\bm{k}_2) = -\frac{1}{4}\frac{b_4^*}{a^4}a^4\sigma^2\,,\quad M_2^{\delta^2s^2,c.t.}(\bm{k}_1, \bm{k}_2) =  -\frac{b_{\delta^2 s^2}^*}{a^4} a^4 \sigma^2 \left(s^2(\bm{k}_1,\bm{k}_2) + \frac{2}{3}\right)\,,
$$
$$
M_2^{\delta s^3,c.t.}(\bm{k}_1,\bm{k}_2) = 0\,,\quad M_2^{(s^2)^2,c.t.}(\bm{k}_1,\bm{k}_2) = -\frac{28}{15}\frac{b_{2,s^2}^*}{a^4}a^4\sigma^2 s^2(\bm{k}_1,\bm{k}_2)\,,
$$
$$
M_2^{s^4,c.t.}(\bm{k}_1,\bm{k}_2) = -\frac{14}{15}\frac{b_{s^4}^*}{a^4}a^4\sigma^2 s^2(\bm{k}_1,\bm{k}_2).
$$


\us{
\section{1-loop Galaxy Bispectrum }\label{app:G_BS}
The next-to-leading-order (or 1-loop) corrections are decomposed in 4 pieces
\begin{multline}
	B^g_{1-loop}(\bm{k}_1,\bm{k}_2,\bm{k}_3,\eta) = B^g_{222}(\bm{k}_1,\bm{k}_2,\bm{k}_3,\eta)+B_{321}^{g \; I}(\bm{k}_1,\bm{k}_2,\bm{k}_3,\eta) \\ +B_{321}^{g \, II}(\bm{k}_1,\bm{k}_2,\bm{k}_3,\eta)+B^g_{411}(\bm{k}_1,\bm{k}_2,\bm{k}_3,\eta) \, .
\end{multline}
Note that each part contains the 1-loop correlations between matter-matter-matter density contrast. We refer to \cite{Castiblanco:2018qsd} for full expressions for the matter bispectrum. Here we will focus on the correlations with the bias operators.  We decompose the galaxy perturbations as $\delta_g=\delta_m + \delta_b$ where $\delta_b$ only contains the part proportional to the bias parameters.}

\us{First we deal with $B^g_{222}$,
\begin{multline}
	B^g_{222}(\bm{k}_1,\bm{k}_2,\bm{k}_3,\eta) = \bigg[\frac{1}{3}B^{mmm}_{222}(\bm{k}_1,\bm{k}_2,\bm{k}_3,\eta) + B^{bmm}_{222}(\bm{k}_1,\bm{k}_2,\bm{k}_3,\eta) \\
    + B^{bbm}_{222}(\bm{k}_1,\bm{k}_2,\bm{k}_3,\eta) + B^{bbb}_{222}(\bm{k}_1,\bm{k}_2,\bm{k}_3,\eta)\bigg] + 2\,\text{cyclic perms.}\,,
\end{multline}
where
\begin{multline}
	B^{bmm}_{222}(\bm{k}_1,\bm{k}_2,\bm{k}_3,\eta) = 8 a^6(\eta) \sum _{\mathcal{O}}b_\mathcal{O}^\mathcal{L} \int_{\bm{q}} M_2^{\mathcal{O}} (\bm{q},\bm{k}_1-\bm{q}) F_2(\bm{k}_1-\bm{q},\bm{k}_2+\bm{q}) F_2(\bm{k}_2+\bm{q},-\bm{q}) \\
 \times P_{L}(q) P_{L}(|\bm{k}_1-\bm{q}|) P_{L}(|\bm{k}_2+\bm{q}|)\,,
\end{multline}
\begin{multline}
	B^{bbm}_{222}(\bm{k}_1,\bm{k}_2,\bm{k}_3,\eta) =8 a^6(\eta) \sum _{\mathcal{O},\mathcal{O'}}b_\mathcal{O}^\mathcal{L}b_\mathcal{O'}^\mathcal{L} \int_{\bm{q}} M_2^{\mathcal{O}} (\bm{q},\bm{k}_1-\bm{q}) M_2^{\mathcal{O'}}(\bm{k}_1-\bm{q},\bm{k}_3+\bm{q}) F_2(\bm{k}_3+\bm{q},-\bm{q}) \\
 \times P_{L}(q) P_{L}(|\bm{k}_1-\bm{q}|) P_{L}(|\bm{k}_2+\bm{q}|)\,,
\end{multline}
\begin{multline}
	B^{bbb}_{222}(\bm{k}_1,\bm{k}_2,\bm{k}_3,\eta) = 8 a^6(\eta) \sum _{\mathcal{O},\mathcal{O'},\mathcal{O''}}b_\mathcal{O}^\mathcal{L} b_\mathcal{O'}^\mathcal{L}b_\mathcal{O''}^\mathcal{L} \int_{\bm{q}} M_2^{\mathcal{O}} (\bm{q},\bm{k}_1-\bm{q}) M_2^{\mathcal{O'}}(\bm{k}_1-\bm{q},\bm{k}_2+\bm{q})  M_2^{\mathcal{O'''}}(\bm{k}_2+\bm{q},-\bm{q}) \\
 \times P_{L}(q) P_{L}(|\bm{k}_1-\bm{q}|) P_{L}(|\bm{k}_3+\bm{q}|)\,.
\end{multline}
Recall that $ F_2(\bm{k}_1,\bm{k}_2) = F_2^N(\bm{k}_1,\bm{k}_2) + H^2a^2 F_2^R(\bm{k}_1,\bm{k}_2)$ and $M_2^\mathcal{O}(\bm{k}_1,\bm{k}_2) = M_2^{\mathcal{O},N}(\bm{k}_1,\bm{k}_2) + a^2 H^2 M_2^{\mathcal{O},R}(\bm{k}_1,\bm{k}_2)$, so one has to be careful taking convolution products and keep only one relativistic operator per term.}

\us{
For $B_{321}^{g}$, we have
\begin{align}
	B_{321}^{g}(\bm{k}_1,\bm{k}_2,\bm{k}_3,\eta)&=  \frac{1}{6}B_{321}^{mmm}(\bm{k}_1,\bm{k}_2,\bm{k}_3,\eta) + B_{321}^{bbb}(\bm{k}_1,\bm{k}_2,\bm{k}_3,\eta)
	+ B_{321}^{bmm}(\bm{k}_1,\bm{k}_2,\bm{k}_3,\eta) \nonumber\\
	&\phantom{=} + B_{321}^{mbm}(\bm{k}_1,\bm{k}_2,\bm{k}_3,\eta) + B_{321}^{mmb}(\bm{k}_1,\bm{k}_2,\bm{k}_3,\eta) 
	+ B_{321}^{bbm}(\bm{k}_1,\bm{k}_2,\bm{k}_3,\eta) \nonumber\\
	&\phantom{=}+ B_{321}^{mbb}(\bm{k}_1,\bm{k}_2,\bm{k}_3,\eta) + B_{321}^{bmb}(\bm{k}_1,\bm{k}_2,\bm{k}_3,\eta) + 5\;\text{permutations}\,,
\end{align}
we then split each bispectrum into its two possible contractions $B_{321} = B_{321}^I + B_{321}^{II}$, where
\begin{align}
B_{321}^{bmm\; I}(\bm{k}_1,\bm{k}_2,\bm{k}_3,\eta)&= 6 P_{L}(k_3) \sum _{\mathcal{O}} b_\mathcal{O}^\mathcal{L} \int_{\bm{q}} M_3^{\mathcal{O}}(\bm{q},\bm{k}_2-\bm{q},\bm{k}_3) F_2(\bm{q},\bm{k}_2-\bm{q}) P_{L}(q) P_{L}(|\bm{k}_2-\bm{q}|)\,,\\
B_{321}^{mbm\; I}(\bm{k}_1,\bm{k}_2,\bm{k}_3,\eta)&= 
6 P_{L}(k_3) \sum _{\mathcal{O}} b_\mathcal{O}^\mathcal{L} \int_{\bm{q}} F_3(\bm{q},\bm{k}_2-\bm{q},\bm{k}_3) M_2^{\mathcal{O}}(\bm{q},\bm{k}_2-\bm{q}) P_{L}(q) P_{L}(|\bm{k}_2-\bm{q}|)\,,\\
B_{321}^{mmb\; I}(\bm{k}_1,\bm{k}_2,\bm{k}_3,\eta)&= 6 P_{L}(k_3) b_1^\mathcal{L} \int_{\bm{q}} F_3(\bm{q},\bm{k}_2-\bm{q},\bm{k}_3) F_2(\bm{q},\bm{k}_2-\bm{q}) P_{L}(q) P_{L}(|\bm{k}_2-\bm{q}|)\,,\\
B_{321}^{bbm\; I}(\bm{k}_1,\bm{k}_2,\bm{k}_3,\eta)& =
6 P_{L}(k_3) \sum_{\mathcal{O},\mathcal{O'}} b_\mathcal{O}^\mathcal{L} b_\mathcal{O'}^\mathcal{L} \int_{\bm{q}} M_3^{\mathcal{O}}(\bm{q},\bm{k}_2-\bm{q},\bm{k}_3) M_2^{\mathcal{O}}(\bm{q},\bm{k}_2-\bm{q}) P_{L}(q) P_{L}(|\bm{k}_2-\bm{q}|)\,,\\
B_{321}^{bmb\; I}(\bm{k}_1,\bm{k}_2,\bm{k}_3,\eta)& = 6 P_{L}(k_3) \sum _{\mathcal{O}} b_\mathcal{O}^\mathcal{L} b_1^\mathcal{L} \int_{\bm{q}} M_3^{\mathcal{O}}(\bm{q},\bm{k}_2-\bm{q},\bm{k}_3) F_2(\bm{q},\bm{k}_2-\bm{q}) P_{L}(q) P_{L}(|\bm{k}_2-\bm{q}|)\,,\\
B_{321}^{mbb\; I}(\bm{k}_1,\bm{k}_2,\bm{k}_3,\eta)& = 6 P_{L}(k_3) \sum _{\mathcal{O}} b_\mathcal{O}^\mathcal{L} b_1^\mathcal{L}\int_{\bm{q}} F_3(\bm{q},\bm{k}_2-\bm{q},\bm{k}_3) M_2^{\mathcal{O}}(\bm{q},\bm{k}_2-\bm{q}) P_{L}(q) P_{L}(|\bm{k}_2-\bm{q}|)\,,\\
B_{321}^{bbb\; I}(\bm{k}_1,\bm{k}_2,\bm{k}_3,\eta) &=  6 P_{L}(k_3) \sum _{\mathcal{O},\mathcal{O'}}b_1^\mathcal{L}  b_\mathcal{O}^\mathcal{L} b_\mathcal{O'}^\mathcal{L} \int_{\bm{q}} M_3^{\mathcal{O}}(\bm{q},\bm{k}_2-\bm{q},\bm{k}_3) M_2^{\mathcal{O}}(\bm{q},\bm{k}_2-\bm{q}) P_{L}(q) P_{L}(|\bm{k}_2-\bm{q}|)\,.
\end{align}
\begin{align}
B_{321}^{bmm\; II}(\bm{k}_1,\bm{k}_2,\bm{k}_3,\eta)&=
6\sum _{\mathcal{O}}b_\mathcal{O}^\mathcal{L} P_{L}(k_1) P_{L}(k_3)  F_2(\bm{k}_1,\bm{k}_3) \int_{\bm{q}} P_{L}(q)M_3^{\mathcal{O'}}(\bm{q},-\bm{q},\bm{k}_1)\, ,\\
B_{321}^{mbm\; II}(\bm{k}_1,\bm{k}_2,\bm{k}_3,\eta)&=
6\sum _{\mathcal{O}} b_\mathcal{O}^\mathcal{L} P_{L}(k_1) P_{L}(k_3)  M_2^{\mathcal{O}}(\bm{k}_1,\bm{k}_3) \int_{\bm{q}} P_{L}(q)F_3(\bm{q},-\bm{q},\bm{k}_1)\, ,\\
B_{321}^{mmb\; II}(\bm{k}_1,\bm{k}_2,\bm{k}_3,\eta)&= 6b_1^\mathcal{L} P_{L}(k_1) P_{L}(k_3)  F_2(\bm{k}_1,\bm{k}_3) \int_{\bm{q}} P_{L}(q)F_3(\bm{q},-\bm{q},\bm{k}_1)\, ,\\
B_{321}^{bbm\; II}(\bm{k}_1,\bm{k}_2,\bm{k}_3,\eta)&=
6\sum _{\mathcal{O},\mathcal{O'}} b_\mathcal{O}^\mathcal{L}b_\mathcal{O'}^\mathcal{L} P_{L}(k_1) P_{L}(k_3)  M_2^{\mathcal{O}}(\bm{k}_1,\bm{k}_3) \int_{\bm{q}} P_{L}(q)M_3^{\mathcal{O'}}(\bm{q},-\bm{q},\bm{k}_1)\,,\\
B_{321}^{bmb\; II}(\bm{k}_1,\bm{k}_2,\bm{k}_3,\eta)&= 6\sum _{\mathcal{O}}b_1^\mathcal{L} b_\mathcal{O}^\mathcal{L} P_{L}(k_1) P_{L}(k_3)  F_2(\bm{k}_1,\bm{k}_3) \int_{\bm{q}} P_{L}(q)M_3^{\mathcal{O}}(\bm{q},-\bm{q},\bm{k}_1)\,,\\
B_{321}^{mbb\; II}(\bm{k}_1,\bm{k}_2,\bm{k}_3,\eta)&=6 \sum _{\mathcal{O}}b_1^\mathcal{L} b_\mathcal{O}^\mathcal{L} P_{L}(k_1) P_{L}(k_3)  M_2^{\mathcal{O}}(\bm{k}_1,\bm{k}_3) \int_{\bm{q}} P_{L}(q)F_3(\bm{q},-\bm{q},\bm{k}_1)\,,\\
B_{321}^{bbb\; II}(\bm{k}_1,\bm{k}_2,\bm{k}_3,\eta) &= 6\sum _{\mathcal{O},\mathcal{O'}}b_1^\mathcal{L} b_\mathcal{O}^\mathcal{L}b_\mathcal{O'}^\mathcal{L} P_{L}(k_1) P_{L}(k_3)  M_2^{\mathcal{O}}(\bm{k}_1,\bm{k}_3) \int_{\bm{q}} P_{L}(q)M_3^{\mathcal{O'}}(\bm{q},-\bm{q},\bm{k}_1)\,.
\end{align}
}

\us{
Finally for $B^g_{411}$, we have
\begin{align}
B^g_{411}(\bm{k}_1,\bm{k}_2,\bm{k}_3,\eta) &=  \frac{1}{3}B^{mmm}_{411}(\bm{k}_1,\bm{k}_2,\bm{k}_3,\eta) + B^{bmm}_{411}(\bm{k}_1,\bm{k}_2,\bm{k}_3,\eta)  \nonumber \\ &\phantom{=} + 2B^{mmb}_{411}(\bm{k}_1,\bm{k}_2,\bm{k}_3,\eta) + 2B^{bbm}_{411}(\bm{k}_1,\bm{k}_2,\bm{k}_3,\eta) \nonumber \\ &\phantom{=} + B^{mbb}_{411}(\bm{k}_1,\bm{k}_2,\bm{k}_3,\eta) + B^{bbb}_{411}(\bm{k}_1,\bm{k}_2,\bm{k}_3,\eta) 
	+ 2\;\text{cyclic perms.} \, ,
\end{align}
where
\begin{align}
B^{bmm}_{411}(\bm{k}_1,\bm{k}_2,\bm{k}_3,\eta) &=  12P_{L}(k_2) P_{L}(k_3)\sum _{\mathcal{O}} b_\mathcal{O}^\mathcal{L} \int_{\bm{q}} M_4^{\mathcal{O}}(\bm{q},-\bm{q},-\bm{k}_2,-\bm{k}_3) P_{L}(q)\, ,\\
B^{mbm}_{411}(\bm{k}_1,\bm{k}_2,\bm{k}_3,\eta) &=  12 P_{L}(k_2) P_{L}(k_3) b_1^\mathcal{L} \int_{\bm{q}} F_4(\bm{q},-\bm{q},-\bm{k}_2,-\bm{k}_3) P_{L}(q)\, ,\\
B^{bbm}_{411}(\bm{k}_1,\bm{k}_2,\bm{k}_3,\eta)&= 12 P_{L}(k_2) P_{L}(k_3)\sum _{\mathcal{O}} b_\mathcal{O}^\mathcal{L}b_1^\mathcal{L} \int_{\bm{q}} M_4^{\mathcal{O}}(\bm{q},-\bm{q},-\bm{k}_2,-\bm{k}_3) P_{L}(q) \,,\\
B^{mbb}_{411}(\bm{k}_1,\bm{k}_2,\bm{k}_3,\eta)&= 12 P_{L}(k_2) P_{L}(k_3) (b_1^\mathcal{L})^2 \int_{\bm{q}} F_4(\bm{q},-\bm{q},-\bm{k}_2,-\bm{k}_3) P_{L}(q)\,,\\
B^{bbb}_{411}(\bm{k}_1,\bm{k}_2,\bm{k}_3,\eta)&=  12 P_{L}(k_2) P_{L}(k_3)\sum _{\mathcal{O}} b_\mathcal{O}^\mathcal{L} (b_1^\mathcal{L})^2 \int_{\bm{q}} M_4^{\mathcal{O}}(\bm{q},-\bm{q},-\bm{k}_2,-\bm{k}_3) P_{L}(q)\,.
\end{align}
}

\section{Cross correlation functions}\label{sec:newplts}
In this appendix we present the plots for the one-loop cross correlation functions when one of the density perturbations correspond to galaxy and the others to matter.
\subsection{Cross power spectrum}
\begin{figure}[H]
        \centering
        \begin{tabular}[t]{cc}
        \includegraphics[width=0.47\textwidth]{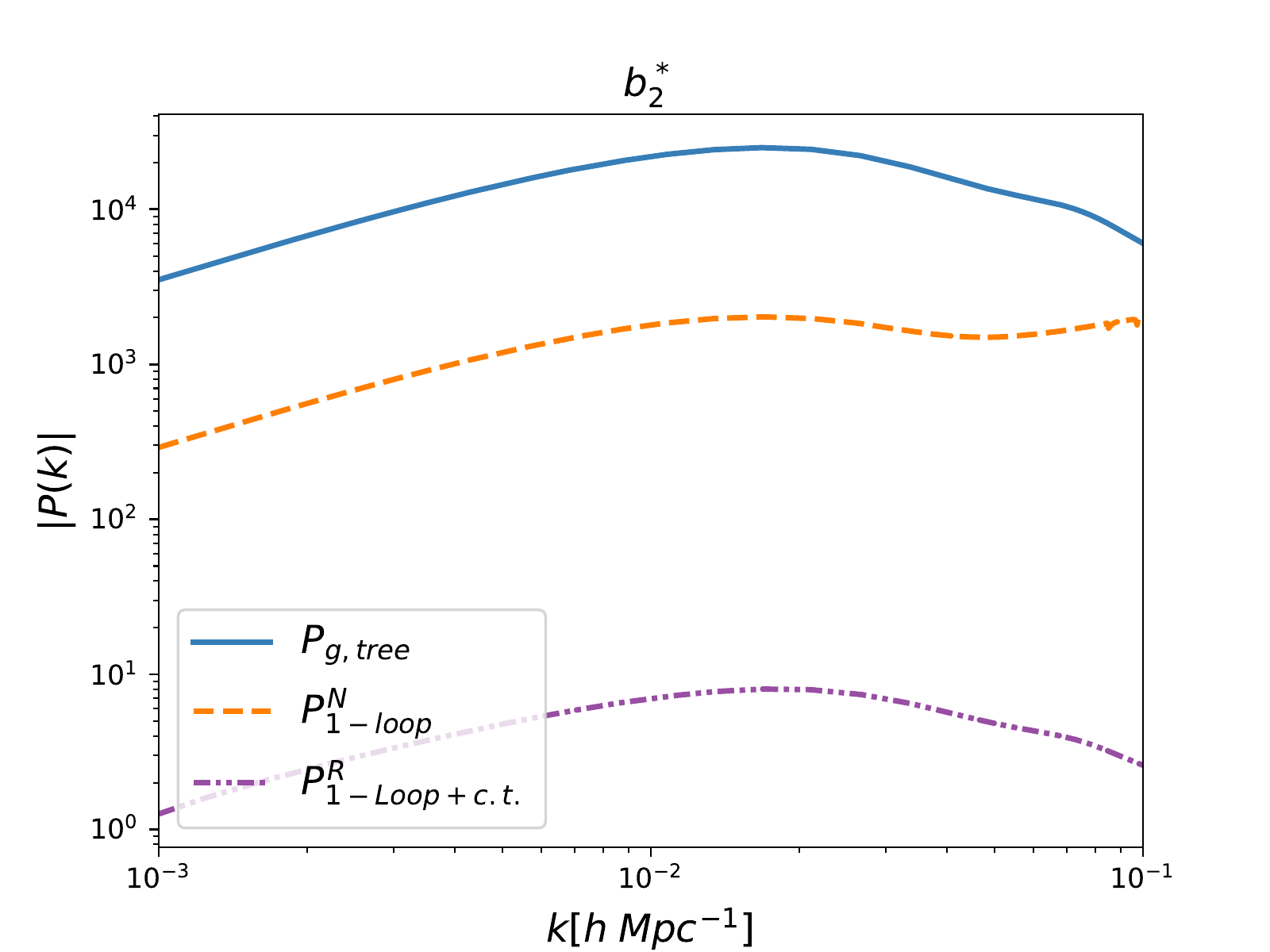} &
        \includegraphics[width=0.47\textwidth]{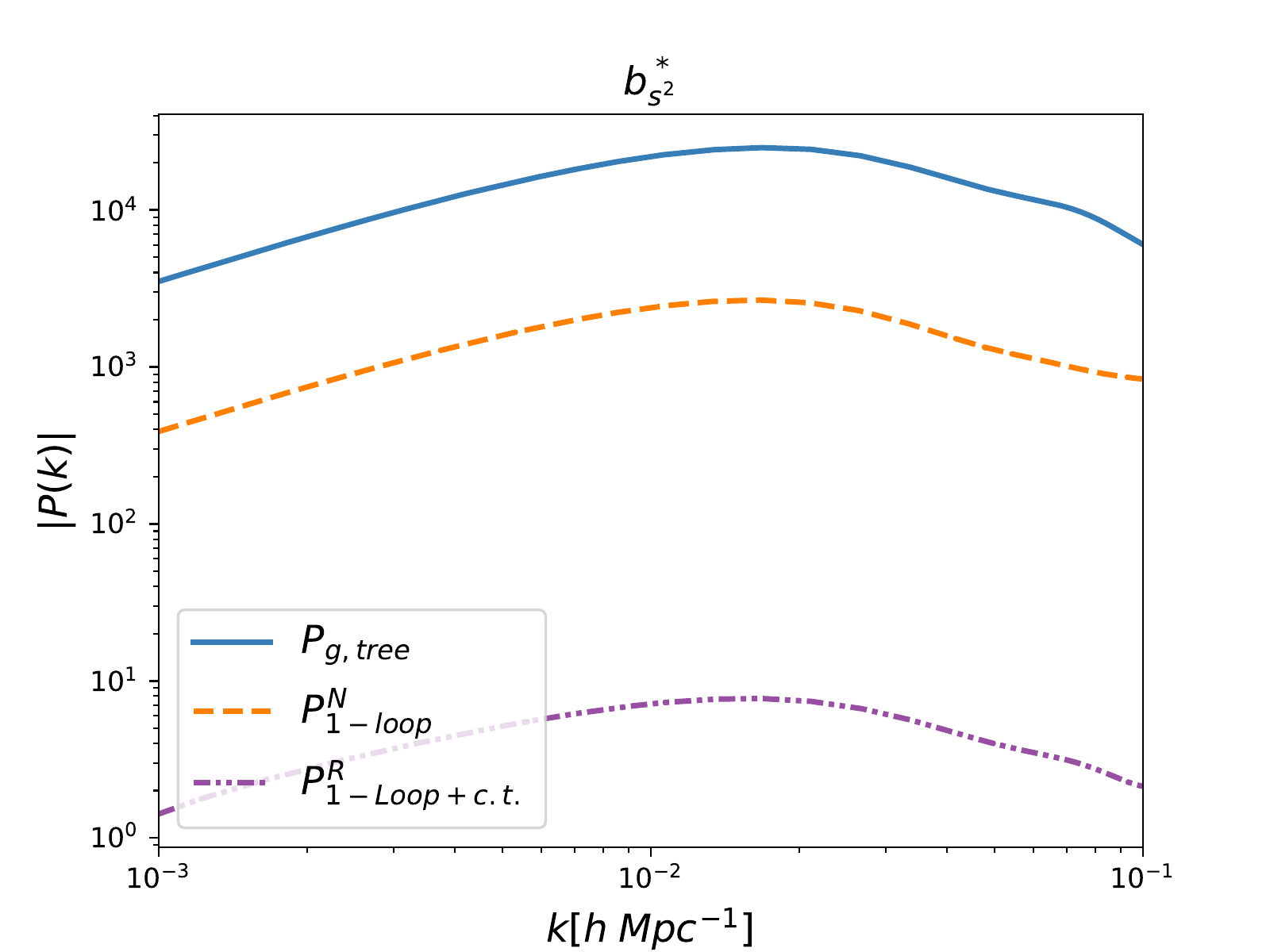}
        \end{tabular}
        \caption{\label{fig:PScross} One-loop  cross power spectrum compared with the galaxy tree level power spectrum computed from the operators proportional to $b_2^*$ and $b_{s^2}^*$ (left and right panels respectively) with all other bias parameters set to zero. We separated the Newtonian (N) and Relativistic (R) contributions}.
\end{figure}

In fig.~\ref{fig:PScross} we show the one-loop galaxy-matter power spectrum contribution for the quadratic operators proportional to $b^*_2$ and $b^*_s$.
\subsection{Cross bispectrum}
\begin{figure}[H]
        \centering
        \begin{tabular}[t]{cc}
        \includegraphics[width=0.48\textwidth]{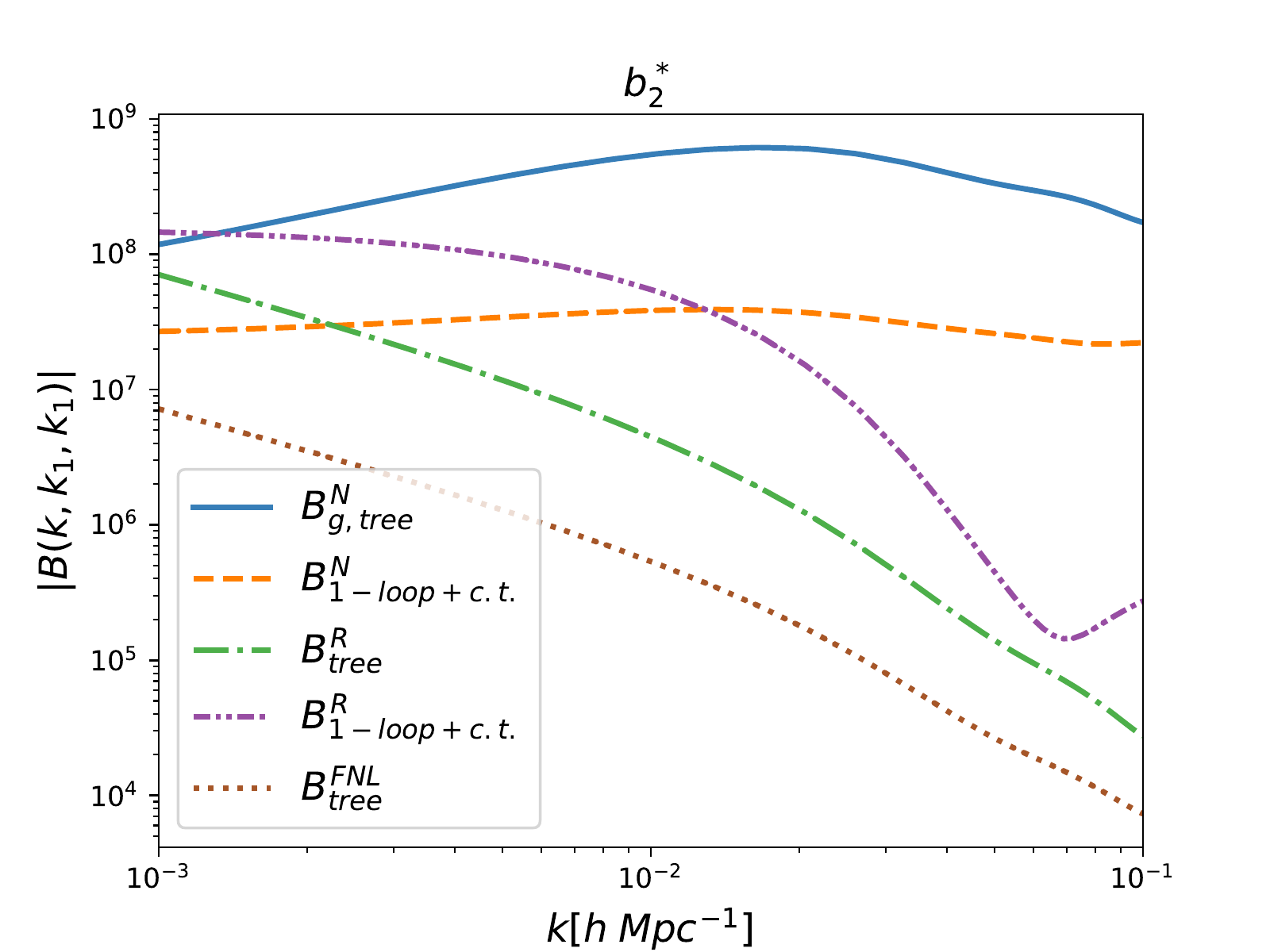} &
        \includegraphics[width=0.48\textwidth]{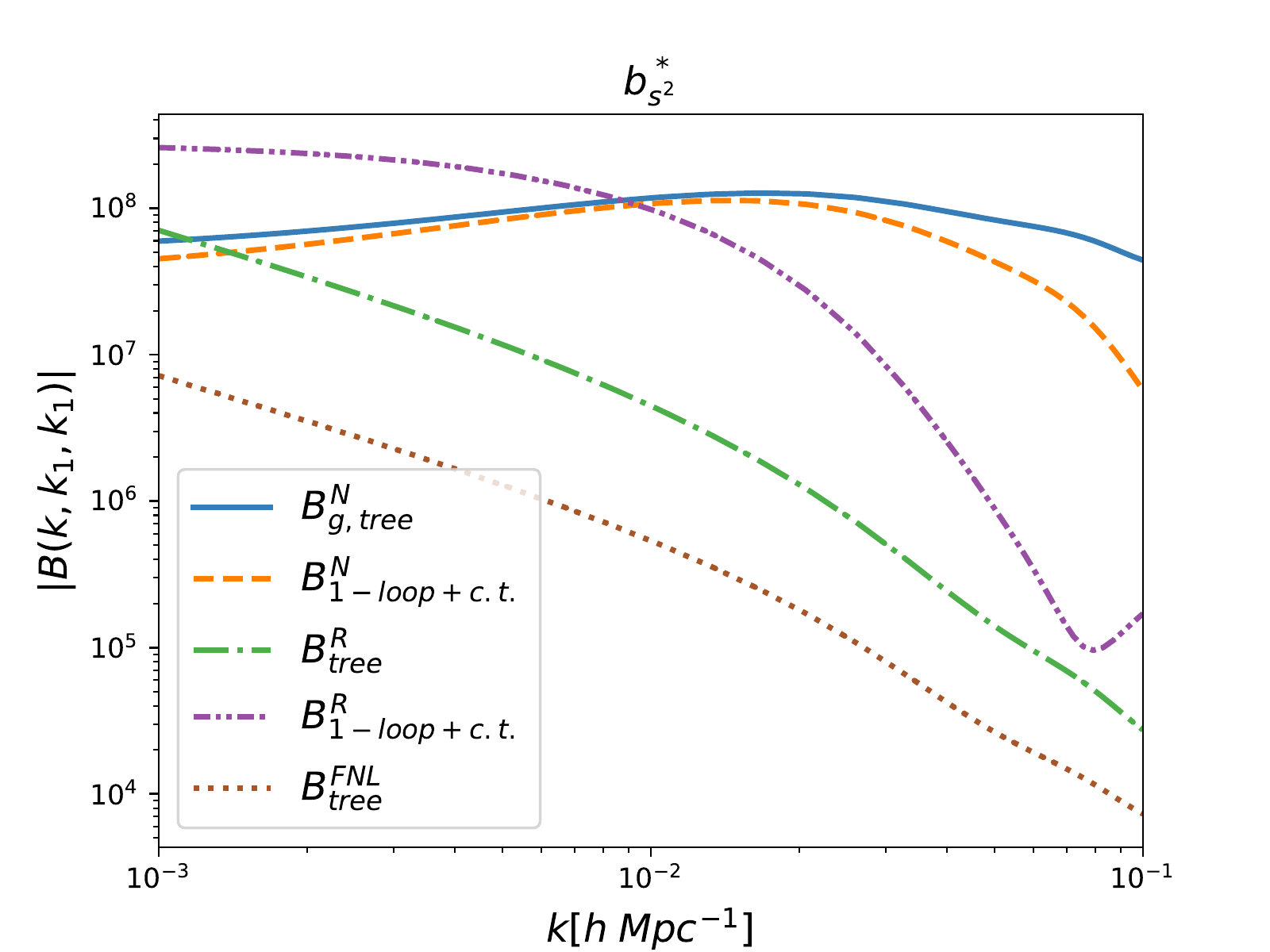}
        \end{tabular}
        \caption{\label{fig:squeezedCross1} Comparison between the galaxy tree-level bispectrum, the one-loop cross bispectrum, and a primordial bispectrum signal with local non-Gaussianity of $f_{NL} = 1$. We separated the Newtonian (N) and Relativistic (R) contributions. All lines are computed from the operators proportional to $b_2^*$ and $b_{s^2}^*$ (left and right panel). For each plot, we set all bias parameters to zero except for the one being studied. We fixed $k_1 = 0.1\,h\,\text{Mpc}^{-1}$  and $k$ is varied. The \textbf{galaxy} density contrast momentum $k$ is squeezed. All quantities are evaluated at redshift $z = 0$.}
\end{figure}

\begin{figure}[H]
        \centering
        \begin{tabular}[t]{cc}
        \includegraphics[width=0.49\textwidth]{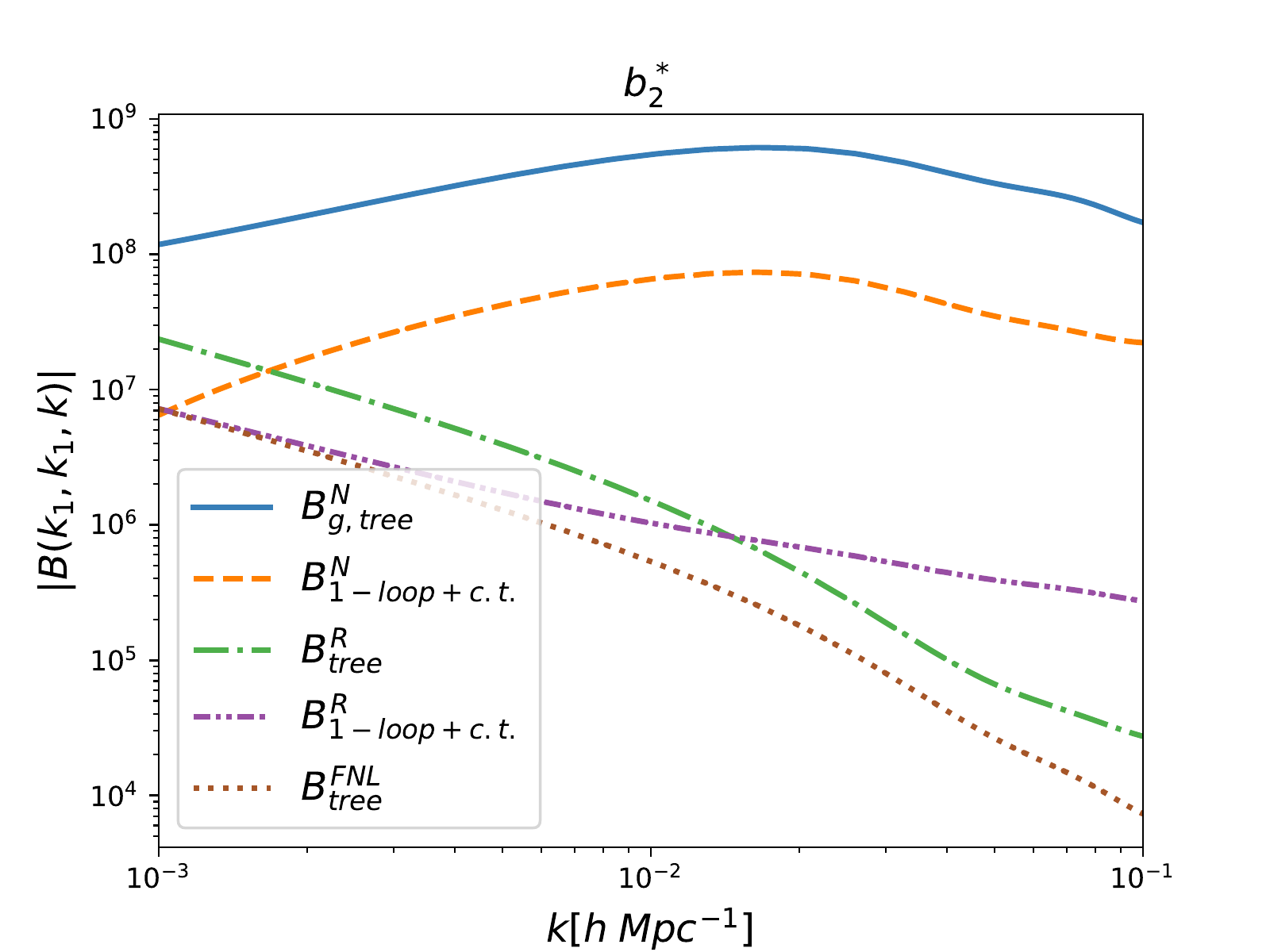}  &
        \includegraphics[width=0.49\textwidth]{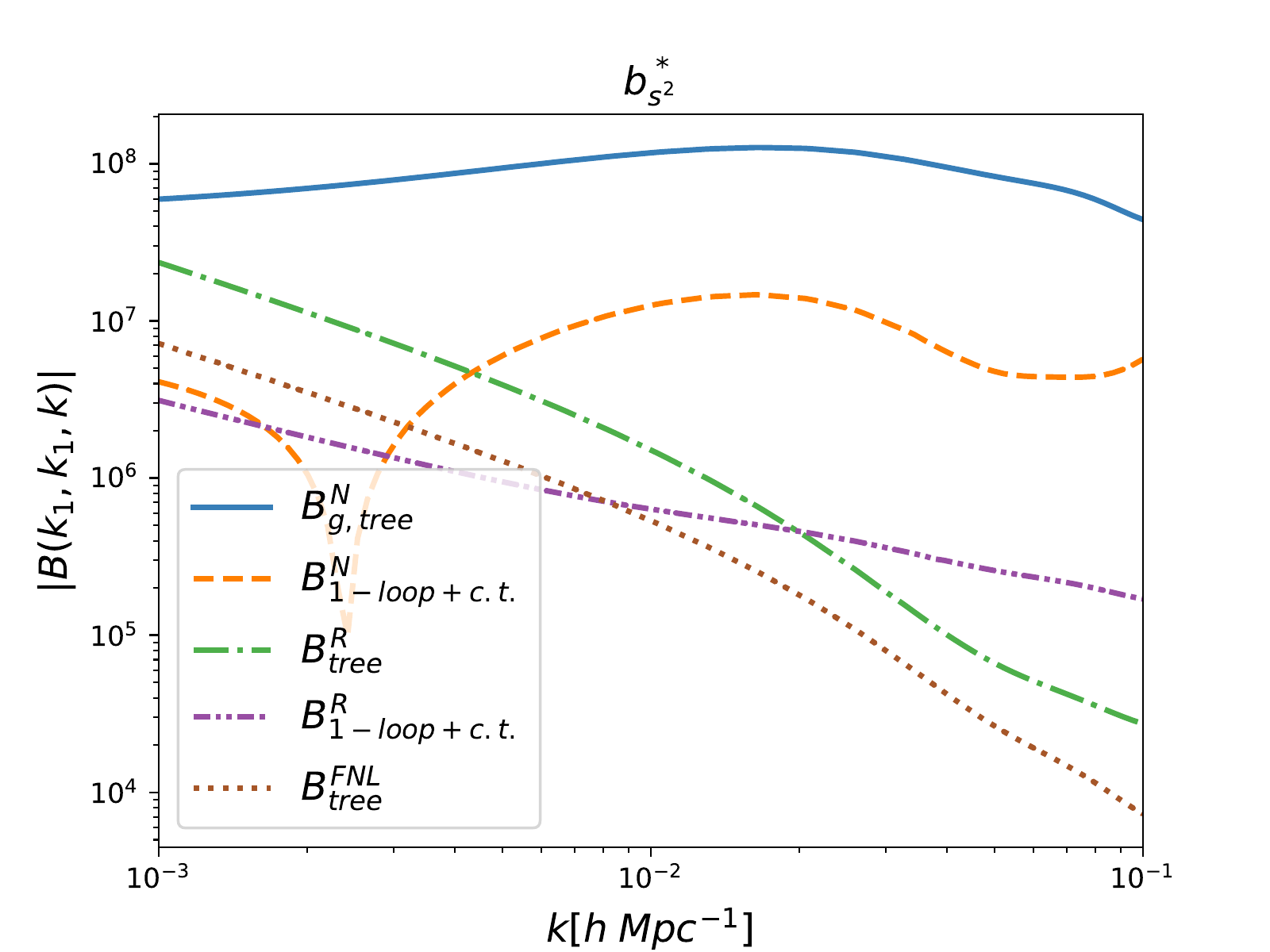} 
        \end{tabular}
        \caption{\label{fig:squeezedCross2}   Comparison between the galaxy tree-level bispectrum, the one-loop cross bispectrum, and a primordial bispectrum signal with local non-Gaussianity of $f_{NL} = 1$. We separated the Newtonian (N) and Relativistic (R) contributions. All lines are computed from the operators proportional to $b_2^*$ and $b_{s^2}^*$ (left and right panel). For each plot, we set all bias parameters to zero except for the one being studied. We fixed $k_1 = 0.1\,h\,\text{Mpc}^{-1}$  and $k$ is varied. The \textbf{matter} density contrast momentum is squeezed.All quantities are evaluated at redshift $z = 0$.}
\end{figure}

\begin{figure}[H]
        \centering
        \begin{tabular}[t]{cc}
        \includegraphics[width=0.48\textwidth]{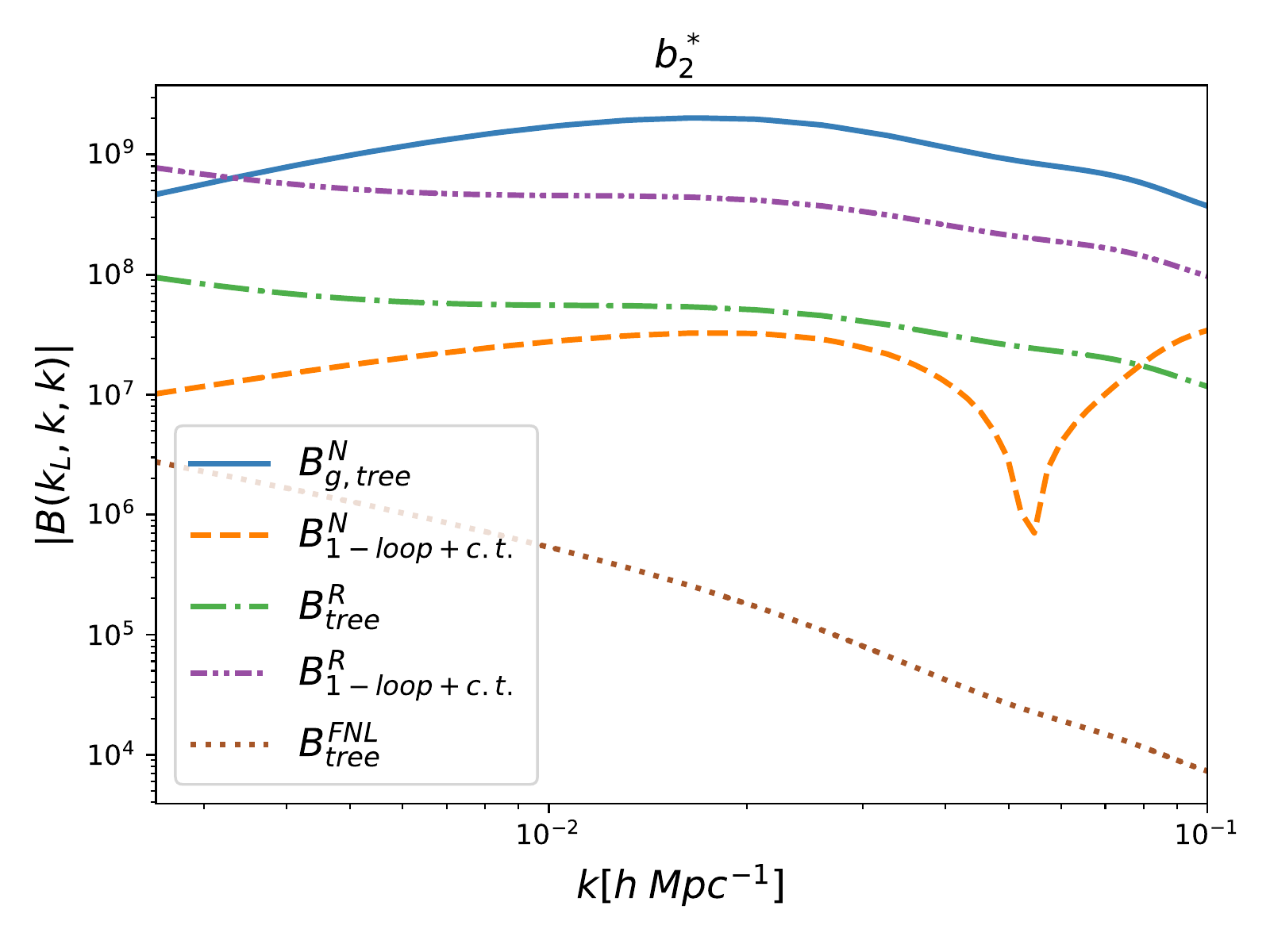} &
        \includegraphics[width=0.48\textwidth]{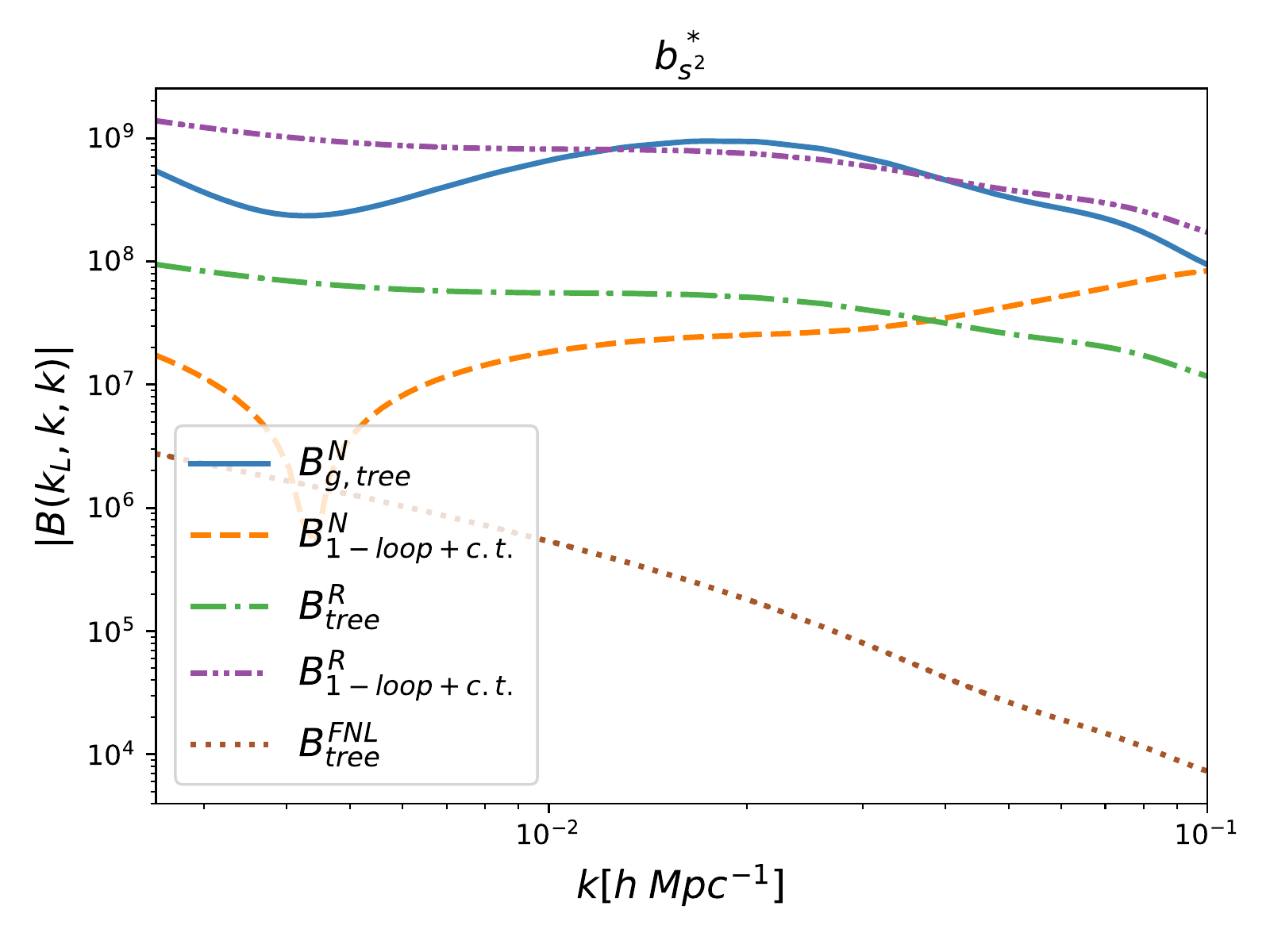} 
        \end{tabular}
        \caption{\label{fig:klongCross1}Comparison between the tree-level bispectrum, the one-loop bispectrum, and a primordial bispectrum signal with local non-Gaussianity of $f_{NL} = 1$. We separated the Newtonian (N) and Relativistic (R) contributions. All lines are computed from the operators proportional to $b_2^*$ and $b_{s^2}^*$ (left and right panel). For each plot, we set all bias parameters to zero except for the one being studied. The \textbf{galaxy} density contrast has momentum $k_L$. We fixed $k_L = 0.005\,h\,\text{Mpc}^{-1}$  and $k$ is varied.  All quantities are evaluated at redshift $z = 0$.}
\end{figure}

\begin{figure}[H]
        \centering
        \begin{tabular}[t]{cc}
        \includegraphics[width=0.49\textwidth]{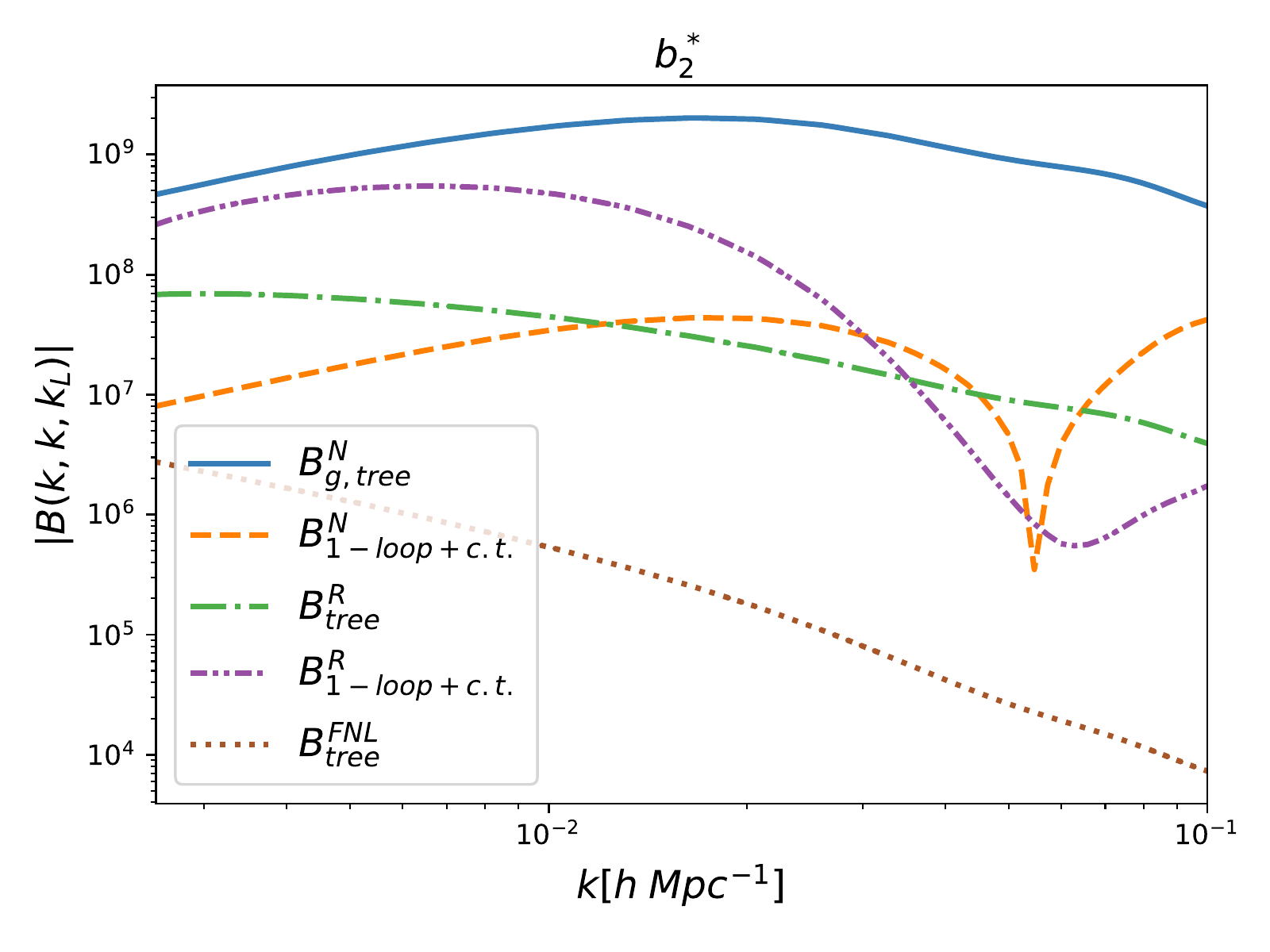}  &
        \includegraphics[width=0.49\textwidth]{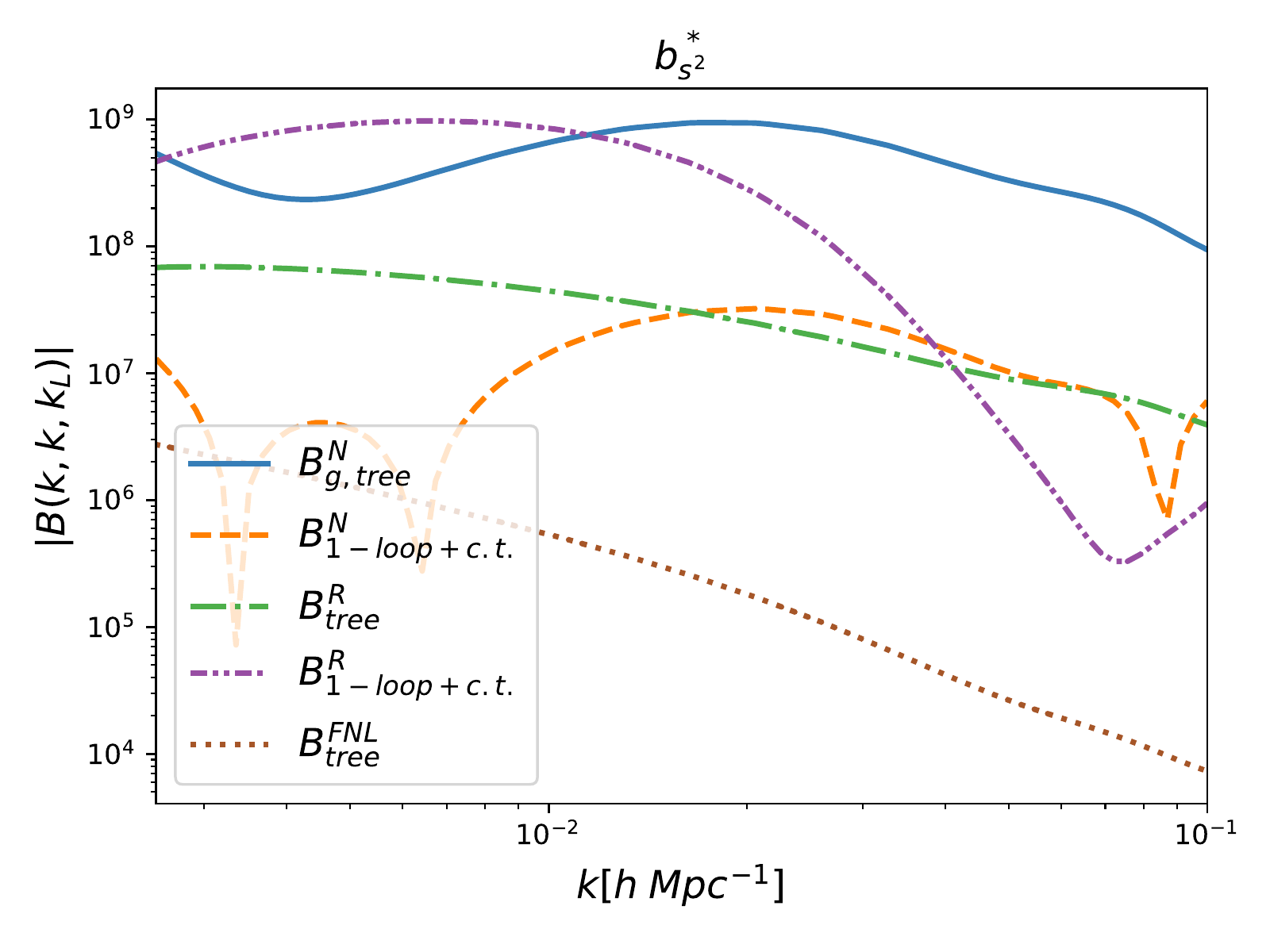}   
        \end{tabular}
        \caption{\label{fig:klongCross2} Comparison between the tree-level bispectrum, the one-loop bispectrum, and a primordial bispectrum signal with local non-Gaussianity of $f_{NL} = 1$. We separated the Newtonian (N) and Relativistic (R) contributions. All lines are computed from the operators proportional to $b_2^*$ and $b_{s^2}^*$ (left and right panel). For each plot, we set all bias parameters to zero except for the one being studied. One of the \textbf{matter} density contrast has momentum $k_L$. We fixed $k_L = 0.005\,h\,\text{Mpc}^{-1}$  and $k$ is varied.  All quantities are evaluated at redshift $z = 0$.}
\end{figure}

Figure \ref{fig:squeezedCross1}   shows the one-loop  galaxy-matter-matter bispectrum for the quadratic operators proportional to  $b^*_2$ and $b^*_s$, the galaxy density contrast momentum is squeezed. Figure  \ref{fig:squeezedCross2} shows the one-loop  galaxy-matter-matter bispectrum for the quadratic operators proportional to  $b^*_2$ and $b^*_s$, in this case one of the mater density contrast momentum is squeezed. Figures \ref{fig:klongCross1} and \ref{fig:klongCross2} present the one-loop  galaxy-matter-matter bispectrum for the quadratic operators proportional to  $b^*_2$ and $b^*_s$, here one the momentum is soft $k_L=0.005\,h\,\text{Mpc}^{-1}$. In figure \ref{fig:klongCross1} the galaxy density contrasts has momentum $k_L$ while in figure \ref{fig:klongCross2} one of the matter density contrast has momentum $k_L$.
 

\bibliographystyle{JHEP.bst}
\bibliography{bibbiasRG}
\end{document}